\documentclass[namedreferences,hyperref,optionalrh]{spr-sola}
\usepackage{graphicx}       % For eps figures, newer & more powerfull
\usepackage[section]{placeins}
\usepackage{lscape}
\usepackage{appendix}
\usepackage{color}          % For color text: \color command
\usepackage{lmodern}        % Prevents errors of font shapes `OT1/cmr/m/n' and
\chardef\us=`\_

\begin{document}

\begin{frontmatter}
\title{Magnetic Field-Constrained Ensemble Image Segmentation of Coronal Holes in Chromospheric Observations}

\author[addressref={aff1,aff2,aff3},corref,email={jalanderos@ucsd.edu}]
{\inits{J.A.}\fnm{Jaime A.}~\snm{Landeros}\orcid{0009-0008-9297-5820}}
\author[addressref=aff1]
{\inits{M.S.}\fnm{Michael S.}~\snm{Kirk}\orcid{0000-0001-9874-1429}}
\author[addressref=aff1]
{\inits{C.N.}\fnm{C. Nick}~\snm{Arge}\orcid{0000-0001-9326-3448}}
\author[addressref=aff4]
{\inits{L.E.}\fnm{Laura E.}~\snm{Boucheron}\orcid{0000-0002-8187-1566}}
\author[addressref=aff5]
{\inits{J.}\fnm{Jie}~\snm{Zhang}\orcid{0000-0003-0951-2486}}
\author[addressref={aff1,aff6}]
{\inits{V.M.}\fnm{Vadim M.}~\snm{Uritsky}\orcid{0000-0002-5871-6605}}
\author[addressref=aff4]
{\inits{J.A.}\fnm{Jeremy A.}~\snm{Grajeda}\orcid{0009-0008-3189-8200}}
\author[addressref=aff5]
{\inits{M.}\fnm{Matthew}~\snm{Dupertuis}}

\address[id=aff1]{NASA Goddard Space Flight Center, Greenbelt, MD 20771, USA}
\address[id=aff2]{California State Polytechnic University Pomona, Pomona, CA 91768, USA}
\address[id=aff3]{ADNET Systems Inc, Bethesda, MD 20817, USA}
\address[id=aff4]{Klipsch School of Electrical and Computer Engineering, New Mexico State University, Las Cruces, NM 88003, USA}
\address[id=aff5]{George Mason University, Fairfax, VA 22030, USA}
\address[id=aff6]{Catholic University of America, Washington, DC, 20064, USA}

\runningauthor{J.A. Landeros et al.}
\runningtitle{Magnetic-Field Constrained Segmentation of Coronal Holes}

\begin{abstract}
Coronal Holes (CHs) are large-scale, low-density regions in the solar atmosphere which may expel high-speed solar wind streams that incite hazardous, geomagnetic storms. Coronal and solar wind models can predict these high-speed streams and the performance of the coronal model can be validated against segmented CH boundaries. We present a novel method named Sub-Transition Region Identification of Ensemble Coronal Holes (STRIDE-CH) to address prominent challenges in segmenting CHs with Extreme Ultraviolet (EUV) imagery. Ground-based, chromospheric He I 10830 Å line imagery and underlying Fe I photospheric magnetograms are revisited to disambiguate CHs from filaments and quiet Sun, overcome obscuration by coronal loops, and complement established methods in the community which use space-borne, coronal EUV observations. Classical computer vision techniques are applied to constrain the radiative and magnetic properties of detected CHs, produce an ensemble of boundaries, and compile these boundaries in a confidence map that quantifies the likelihood of CH presence throughout the solar disk. This method is science-enabling towards future studies of CH formation and variability from a mid-atmospheric perspective.
\end{abstract}
\keywords{Coronal Holes; Chromosphere; Magnetic fields, Photosphere}
\end{frontmatter}
%-------------------------------------------------

\section{Introduction}
     \label{S-Introduction} 

The ambient solar wind originates primarily from Coronal Holes (CHs) and comprises the environment through which the greatest space weather hazards propagate. CHs are observationally defined as regions of low density in the solar atmosphere, while being theoretically defined as regions of photospheric magnetic field lines that are open to the Heliosphere (for a review, \citealp{2009LRSP....6....3C}). The detailed, 3-dimensional coronal field structure has not yet lent itself to routine measurement \citep{2012LRSP....9....6M}, but the reduced emission present in Extreme Ultraviolet (EUV) solar imagery may be used to help identify the footpoints of the permanently open field. Accurate and reliable segmentation of CHs has become a common goal in the solar physics community, as it both facilitates fundamental solar physics research and modeling and operational improvements. Basic research into the open flux problem \citep{2017SoPh..292...18L,2017ApJ...848...70L,2019SoPh..294...19W,2024ApJ...964..115A} and CH formation and variability \citep{2014SoPh..289.1349T,2017ApJ...835..268H} are enabled by the calculation of physical characteristics and uncertainties within the detected boundaries of this solar feature population. Coronal models may be validated through comparison between derived and observationally detected CHs \citep{2005ASPC..346..251T}, with the potential to provide feedback on input maps and model parameters to improve operational space weather forecasts. The Wang-Sheeley-Arge (WSA) coronal model \citep{2000JGR...10510465A, 2004JASTP..66.1295A} may be calibrated in this manner to augment ongoing efforts in calibrating predictions with observed in situ quantities \citep{2020ApJ...891..165R,2023SpWea..2103555I,2024Samara} and coronal field structure \citep{2017ApJ...844...93J}.

\begin{figure}    %%%%%%%%%%%%%%%%%% Showcase on Sarnoff camera observation
\centerline{
    \hspace{-0.015\textwidth}
    \includegraphics[width=0.43\textwidth]{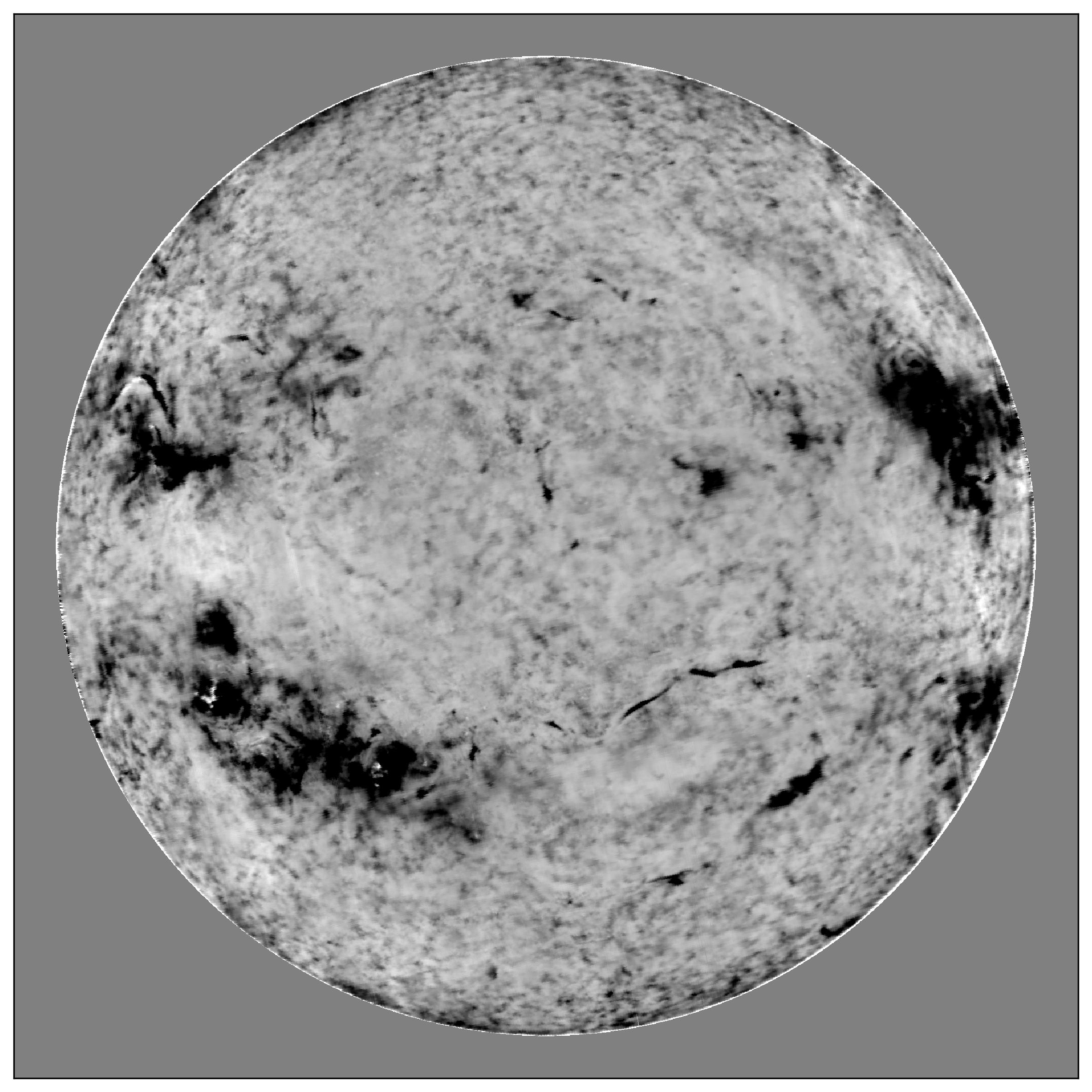}
    \hspace{0.12\textwidth}
    \includegraphics[width=0.43\textwidth]{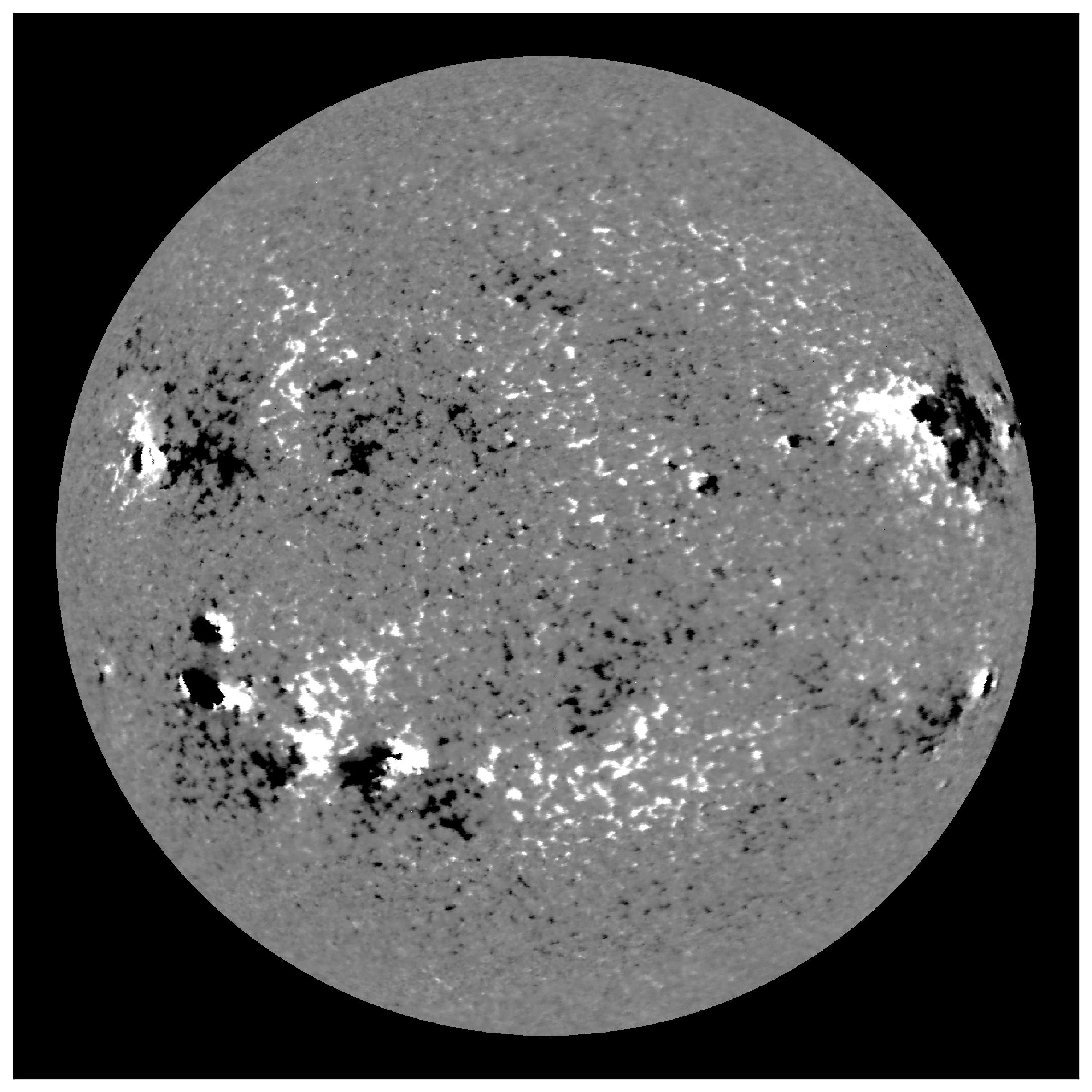}
    }
\vspace{-0.415\textwidth}   % Shift close to the panel top 
\centerline{\Large \bf      % Labels
    \hspace{0 \textwidth}       \color{white}{(a)}
    \hspace{0.48\textwidth}    \color{white}{(b)}
    \hfill
    }
\vspace{0.375\textwidth}    % Shift back to the panel bottom 
\centerline{
    \hspace*{-0.025\textwidth}
    \includegraphics[width=0.59\textwidth]{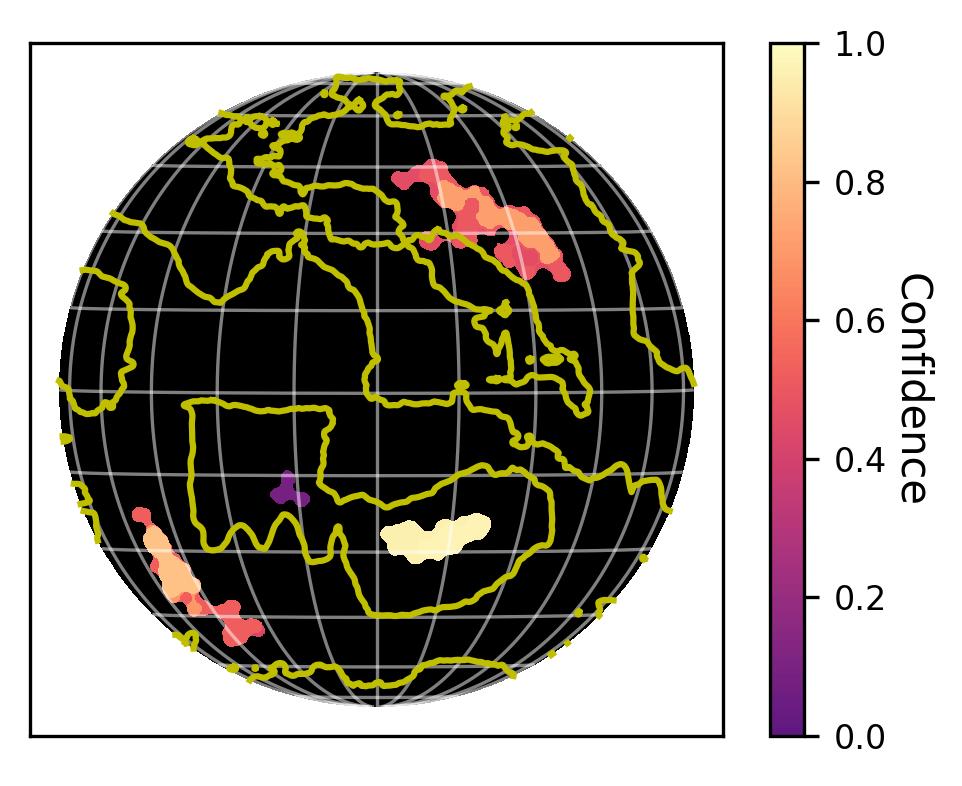}
    \hspace{-0.025\textwidth}
    \raisebox{0.07\height}{\includegraphics[width=0.43\textwidth]{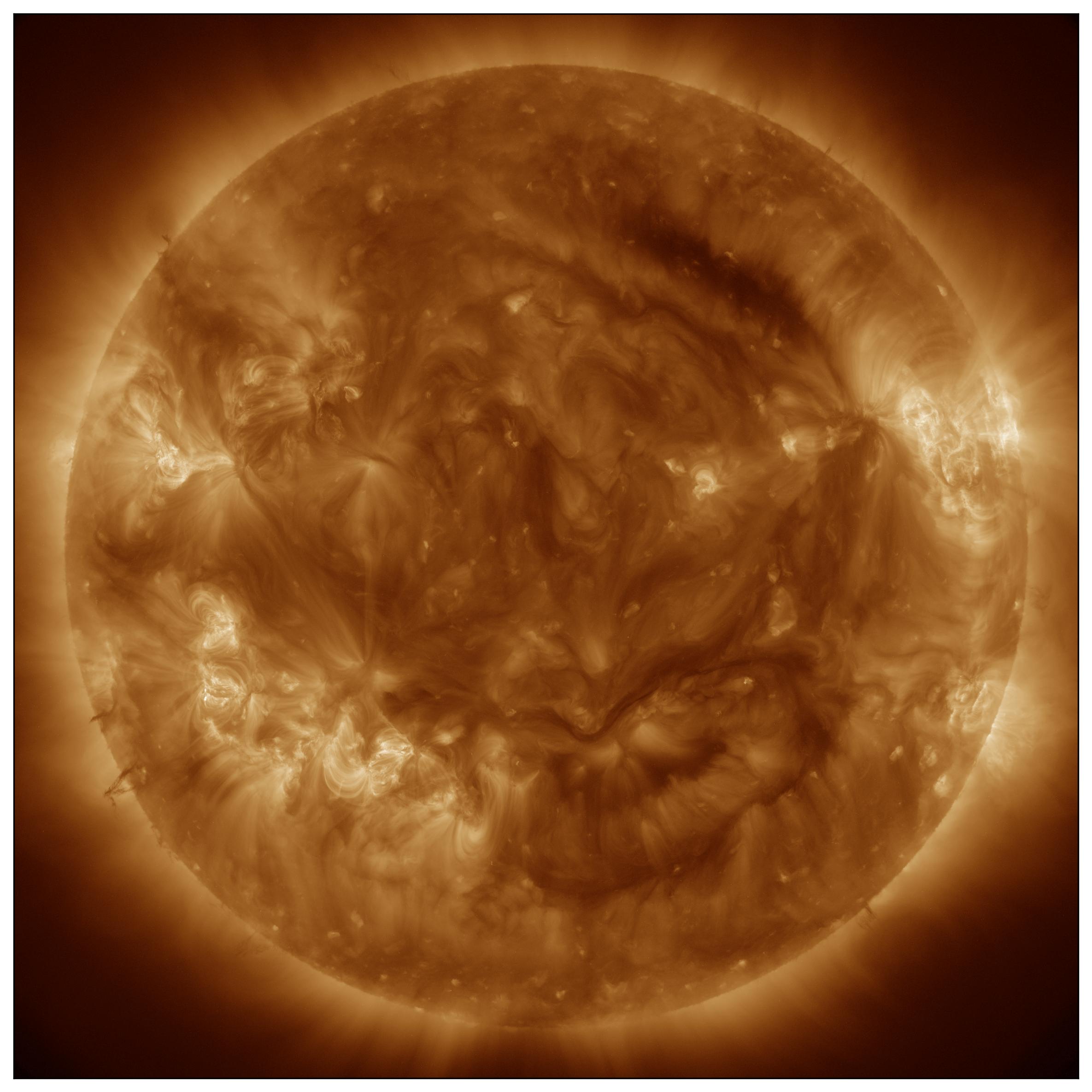}}
}
\vspace{-0.435\textwidth}   % Shift close to the panel top 
\centerline{\Large \bf      % Labels
    \hspace{0 \textwidth}     (c)
    \hspace{0.48\textwidth}  \color{white}{(d)}
    \hfill
    }
\vspace{0.385\textwidth}    % Shift back to the panel 
              
\caption{The newly presented automated CH detection method, STRIDE-CH, applied to SOLIS observations on June 11, 2012 in the rising phase of Solar Cycle 24. A He I 10830 Å spectroheliogram (a) and a line of sight magnetogram (b), both from the SOLIS VSM Sarnoff era camera, are ingested by STRIDE-CH to produce an ensemble CH segmentation map (c). An SDO AIA 193 Å EUV observation is included for reference (d). The He I spectroheliogram is saturated between ± 100 mÅ, the magnetogram is saturated between ± 50 G, and yellow contours on the segmentation map denote the locations of macroscopic polarity inversion lines.
        }
\label{F-sarnoff}
\end{figure}

   % {\bf --- CH property characterization} \\
A precursor to the development of a robust segmentation method is a proper characterization of the properties of CHs in solar imagery. CHs are predominantly magnetically unipolar, as observable in magnetograms of the photospheric field. They exhibit a dark appearance in Extreme Ultraviolet (EUV) imagery due to weak emission from low coronal plasma densities \citep{2009LRSP....6....3C}, as demonstrated by the two CHs west of disk center in Figure~\ref{F-sarnoff}d. CHs in He I 10830 Å full disk images have an opposite character, with a bright and smooth appearance that may be observed in the same two CHs in Figure~\ref{F-sarnoff}a. The bright appearance is due to continuum emission only being weakly absorbed by the low densities in the neutral, chromospheric orthohelium species \citep{1975ApJ...199L..63Z,1987SvAL...13..255P}, while the apparent smoothness is a result of the chromospheric network within CHs exhibiting reduced contrast as compared to that across the quiet Sun \citep{2002SoPh..211...31H}. The unique and complex He I 10830 Å spectral line lies in the near infrared range and is the only known CH diagnostic that may be observed by ground-based observatories with high spatial resolution \citep{1997ApJ...489..375A}. Furthermore, large-scale CHs appear coherently shaped without thin extrusions, exhibit long evolutionary time scales, and cover a minority of the solar surface with an extent between 5 - 17\% in surface area during solar maximum and minimum respectively \citep{2014ApJ...783..142L,2024ApJ...964..115A}.

   % {\bf --- CH disambiguation from other solar features} \\
Observational characteristics that may be used to uniquely identify CHs from other large scale, persistent structures appearing on the solar disk are now described, as summarized in Table~\ref{T-solar-feature-properties}. Quantification of the properties in the first column with multiple “No” entries under non-CH features is of particular interest for aiding in the task of CH segmentation. It is important to note that these observational appearances arise from the thermodynamic and magnetic physical properties of these solar features. Active regions in the first column may be clearly differentiated from CHs in the final column through their brightness in EUV (strong emission) and darkness in He I (strong absorption). Filaments are closed field regions that lie along polarity inversion lines and appear similarly dark to CHs in 193 and 195 Å EUV imagery, which are commonly used to identify CHs \citep{2018SoPh..293...71H}. This dark appearance is due to their cool, dense plasma which weakly emits EUV. A filament is visible in Figure~\ref{F-sarnoff}d as a dark structure surrounding the previously mentioned dark, southwestern CH. Filaments appear differently from CHs in He I and magnetograms, however, as they have a dark appearance and approximately bipolar magnetic field. Polarity inversion lines lying along filament and non-filament regions are visualized in Figure~\ref{F-sarnoff}c by yellow contours, with the visualization method described in Section~\ref{S-Methods}. The broader quiet Sun appears substantially brighter than CHs in EUV, but the contrast becomes marginal along the CH boundary such that there is often ambiguity in its precise location. Regions of quiet Sun appear with similar brightness and variable degrees of smoothness as compared to CHs in He I, making them the most difficult region type to differentiate from CHs in this spectral line. For example, the bright, equatorial, quiet Sun regions near the east and west limbs in Figure~\ref{F-sarnoff}a appear similarly bright to CHs, but they may be distinguished through their lack of unipolarity. 

\begin{table}
\caption{Properties of large scale, persistent, on-disk solar features and their observational appearances. Features are ascribed a ``Yes'', ``No'', ``Marginal'', or ``Variable'' entry for qualitative physical properties that are characteristic of CHs. The He I 10830 Å network contrast property has ``-'' entries for solar features where this property is irrelevant for CH disambiguation as they may be readily disambiguated by the degree of He I 10830 Å absorption.
}
\label{T-solar-feature-properties}
\begin{tabular}{lcccc}     % define the column alignment
                           % l: left, c: center, r: right
\hline                     % horizontal line
Physical Property   & Active    & Filament  & Quiet     & Coronal \\
(Appearance)        & Region    &           & Sun       & Hole\\
\hline
Weak EUV Emission   & No        & Yes       & Marginal  & Yes \\
(Dark)              &  &  &  &  \\
Weak He I Absorption& No        & No        & Yes       & Yes \\
(Bright)            &  &  &  &  \\
Low He I Network    & -         & -         & Variable  & Yes \\
Contrast (Smooth)   &  &  &  &  \\
Predominant         & No        & No        & No        & Yes \\
Unipolarity         &  &  &  &  \\

\hline
\end{tabular}
\end{table}

   % {\bf --- Lack of  CH ground truth} \\
The challenges in solar feature disambiguation are exacerbated by both the variability in appearances when observed in different optical filters and the biases of each algorithm that aims to extract CH boundaries. Thus, CH detection is inherently an ambiguous segmentation task \citep{2021arXiv210316265C} with no widely accepted ground truth in CH boundary location \citep{2021ApJ...918...21L}. Limb effects complicate matters, including brightening from stray light in EUV and darkening from emission becoming foreshortened far from disk center in He I and in continuum emission. In addition, the dense plasma in large, extruding coronal loops may obscure entire CHs in EUV \citep{2023ApJ...958...43C}. This is particularly the case near the limb where projection effects are greatest \citep{2018JSWSC...8A...2G}, and is exemplified by the southeastern active region in Figure~\ref{F-sarnoff}d whose coronal loops obscure a CH lying further east of disk center. This obscured CH becomes readily apparent as a bright region in the He I observation in Figure~\ref{F-sarnoff}a, as well as in later EUV observations after the Sun has rotated to bring the CH closer to disk center.

   % {bf --- CH detection methods review} \\
A CH detection method developed by \citet{2005ASPC..346..261H} stands amongst the earliest efforts in this area of research, creating boundary estimates using ground observed He I 10830 Å spectroheliograms and Fe I 6301.5 Å photospheric magnetograms. Their primary motivation was to automate manual CH detection to facilitate CH variability studies, as pioneered by \citet{2002SoPh..211...31H}. As routine, space-borne EUV observations became available, a variety of CH detection methods were developed with the advantage of improved intensity contrast over ground-based He I observations \citep{2012SoPh..281..793R,2014A&A...561A..29V,2016SoPh..291.2353B,2018MNRAS.481.5014I,2019SoPh..294..144H}. Challenges in CH detection have been addressed in several ways such as using magnetogram information to help resolve the CH-filament ambiguity \citep{2008SoPh..248..425S,2014AGUFMSH21A4089H,2015JSWSC...5A..23R,2022arXiv220710070J}. Segmentation of multispectral observations, primarily in EUV and X-ray imagery, has confronted the lack of ground truth in CHs across optical filters \citep{2005ASPC..346..251T,2009SoPh..256...87K,2018SoPh..293...71H,2018JSWSC...8A...2G,2021A&A...652A..13J}, and a specialized perimeter tracing technique has confronted the challenge of polar CH detection due to projection effects \citep{2009SoPh..257...99K}.

   % {\bf --- Introduction to STRIDE-CH and road map of article} \\
Our newly developed CH segmentation method, Sub-Transition Region Identification of Ensemble Coronal Holes (STRIDE-CH), confronts the ambiguous segmentation task by pairing He I and magnetogram observations to complement approaches using EUV and X-ray solar imagery. Although now out of operations, the method developed by \citet{2005ASPC..346..261H} used a common set of input data to STRIDE-CH and this work will compare the two. The approach employed involves classical image processing and segmentation techniques with tunable design variables, or parameters defining the design of STRIDE-CH, that target specific CH properties as observed in the aforementioned observations below the transition region. Data sources are overviewed in Section~\ref{S-Observations}, with the ensuing pre-processing, segmentation, design variable selection, and extraction of CH physical properties detailed in Section~\ref{S-Methods}. Cases demonstrating the method’s strengths and weaknesses are provided in Section~\ref{S-Results}, followed by discussion in Section~\ref{S-Discussion} and conclusions in Section~\ref{S-Conclusion}.

\section{Observations}
    \label{S-Observations}
   % {\bf --- Data sources} \\
The National Solar Observatory (NSO) has collected He I 10830 Å spectroheliograms and Fe I 6301.5 Å line of sight magnetograms with the Kitt Peak Vacuum Telescope (KPVT) \citep{1976ApOpt..15...33L} from 1974 to 2003 and with the Synoptic Optical Long-term Investigations of the Sun (SOLIS) Vector Spectromagnetograph (VSM) \citep{2003ASPC..307...13K} from 2004 to 2015. Care must be taken in the treatment of diverse observational sources, as data products that build on top of them magnify their differences. Different NSO telescopes and cameras produce observations with not only quantitative calibration differences in He I, but systematic biases at the poles as well that may be identified qualitatively, as detailed in Section~\ref{S-Discussion}. Full-disk KPVT spectromagnetograph (SPM) spectroheliograms and magnetograms captured from 1999 to 2003 are openly available at ~\url{https://nispdata.nso.edu/ftp/kpvt/daily/raw/} and SOLIS data is available on the SOLIS information website \href{https://solis.nso.edu/0/vsm/VSMDataSearch.php?stime=1059717600&etime=1701647999&thumbs=0&pagesize=150&obsmode[]=1083i&sobsmode=1&sobstype=&display=1}{\nolinkurl{https://solis.nso.edu/0/vsm/VSMDataSearch.php}}. These data are available in the Flexible Image Transport System (FITS) file format at a near daily cadence. Instruments that will provide He I observations in the future are highlighted in Section~\ref{S-Conclusion}.

   % {\bf --- Equivalent width and CH as bright, positive regions} \\
The data array within the NSO He I FITS files that is relevant for the task of segmentation is labeled as equivalent width. While this data is only a proxy for the true equivalent width, which is a scalar quantification of the strength of a spectral line that may be expressed in units of mÅ \citep{2003Rutten}, it quantifies He I absorption throughout the full solar disk and will be referred to as a He I observation herein. Negatively valued, dark regions correspond to strongly absorbing active regions and filaments, whereas positively valued, bright regions correspond to weakly absorbing CHs, quiet Sun, and polarity inversion lines \citep{1996SoPh..163...79B,2002SoPh..211...31H}. Thus, it is a necessary but insufficient condition for the identification of weakly absorbing CHs that they consist of primarily positively valued regions.

\section{Methods} 
      \label{S-Methods}   

   % {\bf --- Open-source packages} \\
Data in the FITS file format is extracted with the FITS file reading capability from the astropy package \citep{2022ApJ...935..167A} and converted into GenericMap objects with the Sunpy package \citep{sunpy_community2020}. Image operations are in large part performed with the scikit-image \citep{scikit-image} and numpy \citep{harris2020array} libraries, while visualizations are achieved with matplotlib \citep{Hunter:2007}. The relevant code is openly available at \url{https://github.com/jalanderos/STRIDE-CH} and a fixed version has been published \citep{my_zenodo}. Pre-processing and boundary detection are discussed below.

\subsection{Data Preparation} 
  \label{Data-Preparation}
\begin{figure}    %%%%%%%%%%%%%%%%%% He I observation histograms
\centerline{
         \includegraphics[width=0.5\textwidth]{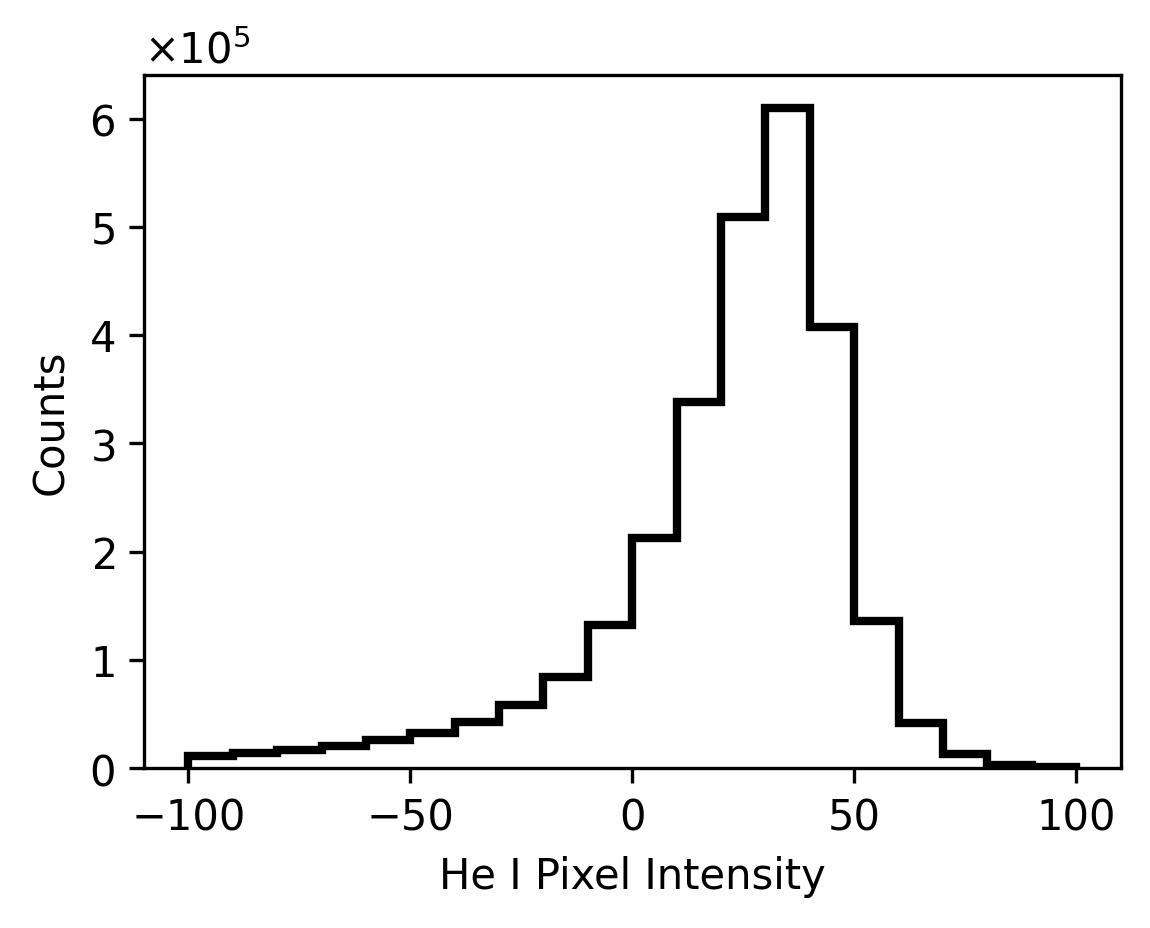}
         \includegraphics[width=0.5\textwidth]{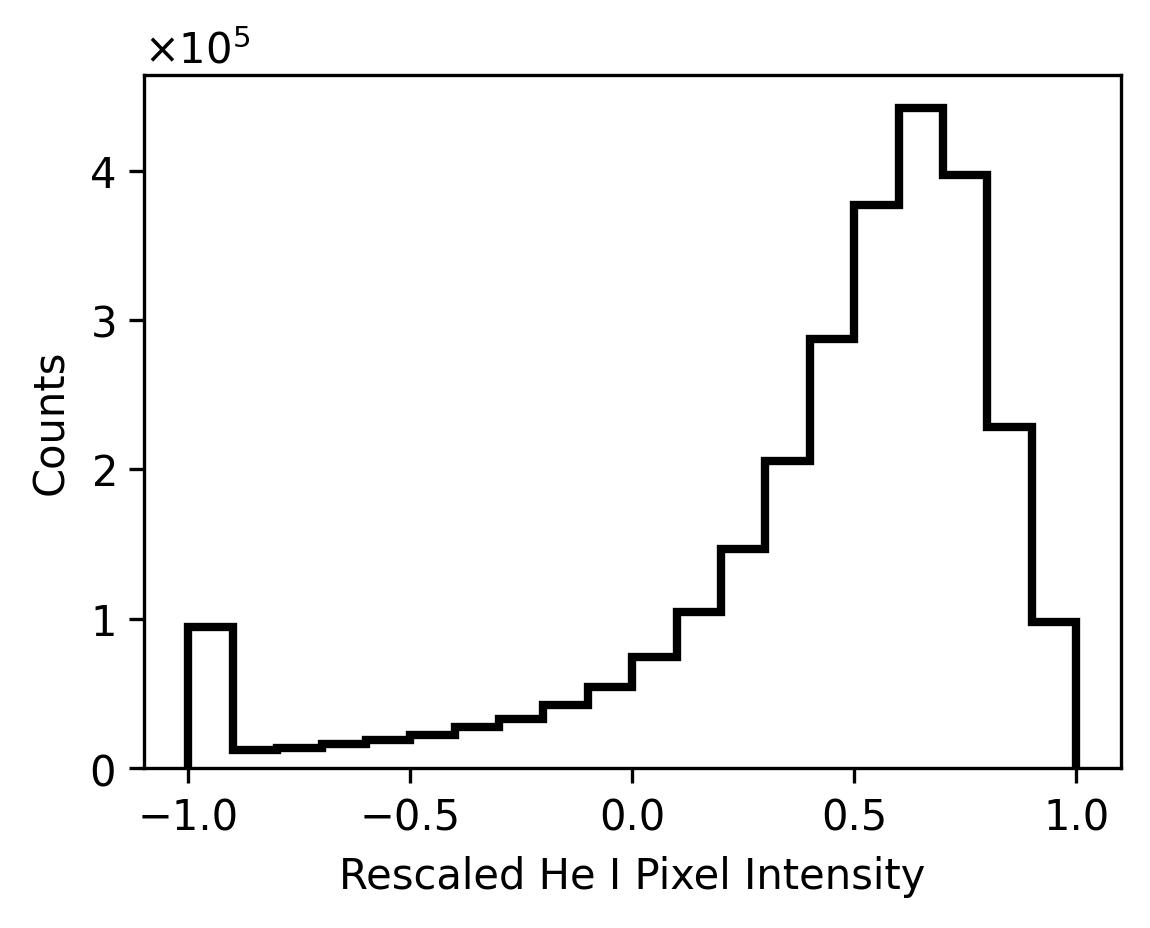}
        }
\vspace{-0.37\textwidth}   % Shift close to the panel top 
\centerline{\bf
    \hspace{0.055 \textwidth}    (a)
    \hspace{0.445\textwidth}      (b)
    \hfill
   }
\vspace{0.32\textwidth}    % Shift back to the panel bottom 
              
\caption{
    Histograms before (a) and after (b) data preparation steps of the SOLIS He I observation of Figure~\ref{F-sarnoff} on June 11, 2012 in the rising phase of Solar Cycle 24. The histogram of the observation (a) excludes the zero-valued pixels in the image background and only includes pixel values within ± 100 mÅ. A high number of counts may be found in the leftmost bin of the histogram of the rescaled observation (b) as a result of mapping pixels with intensities in the negatively-valued tail of the distribution to the minimum rescaled value.
        }
\label{F-he-hist}
\end{figure}

Pre-processing is applied to He I observations to standardize contrast amongst images gathered over time. The pixel intensities are linearly rescaled such that the 2$^{\rm nd}$ to the 98$^{\rm th}$ percentiles of the intensities present in the image are mapped to lie $\in [-1,1]$, with a resulting histogram depicted in Figure~\ref{F-he-hist}b. The pixel intensities below the 2$^{\rm nd}$ percentile and above the 98$^{\rm th}$ percentile are assigned values of -1 and 1 respectively. The off-disk background is then filled with Not a Number (NaN) entries to remove them from consideration. Magnetograms, on the other hand, need only be aligned with He I observations via differential rotation to account for differences in observational times. This alignment is achieved with Sunpy.

\subsection{Segmentation} %%%%%%%%%%%%%%
  \label{S-Segmentation}

   % {\bf --- Target physical properties and threshold} \\
The physical properties of brightness and shape coherence that characterize CHs in He I are used in the segmentation procedure, along with the unipolarity in underlying magnetograms. Pre-processed He I observation pixel intensities above a percentage-based threshold $I_{\rm thresh}$ are assigned a value of one in a binary mask to isolate bright regions. The threshold is defined such that it adapts to the challenging nature of the poor contrast in He I, as exemplified by the unimodal histogram in Figure~\ref{F-he-hist}a. Whereas a bimodal distribution would lend itself well to a threshold in the local minima between modes of CH and quiet sun intensities, histograms of He I observations are consistently unimodally distributed and thus lack a clear location for thresholding. To allow for tuning relative to the pixel intensity at which the mode occurs, which is variable over time, the threshold is therefore defined as a percentage from the zero value up to the mode of the histogram. Selection of $I_{\rm thresh}>100 \%$ would correspond to an intensity threshold past the location of the mode. Discussion on tuning of the threshold level and further influential design variables will be discussed at the end of this section.

   % {\bf --- Morphological operations} \\
The resulting binary mask is then operated upon with morphological open and close operations \citep{gonzalez2018digital}, which are classical image processing methods applied here to reshape isolated, bright regions into coherently-shaped candidate CHs with smoothed boundaries. These operations use a Structuring Element (SE), which is chosen to be a circular disk, to modify an image. An accompanying design variable is the SE disk radius $r_{\rm SE}$, which is specified in units of Mm as measured at the center of the solar disk to maintain independence from image resolution. The application of a decomposed sequence of smaller elements reduces computational cost, while well approximating the operations performed by a full-sized disk \citep{ParkDECOMPOSITIONOS}.

   % {\bf --- Fill small holes and remove small regions} \\
Minor operations of filling small holes within candidate CHs and removing small, detected regions are then applied to focus on detecting primarily large, consolidated CHs in the first instantiation of this routine. The fill operation is non-parametric while the small object removal is defined with a minimum size of 3$\times10^{9}$ km$^{2}$, which has been determined in consideration of the absolute minimum CH size of 1$\times10^{9}$ km$^{2}$ of two supergranules defined by \citet{2002SoPh..211...31H} in an effort to remove spurious detections.

   % {\bf --- Ensemble and aggregation by unipolarity} \\
This primary segmentation procedure is applied with multiple sets of design variables $I_{\rm thresh},r_{\rm SE}$ to form a set of preliminarily segmented CH maps. The selected size of this set will be discussed at the end of this section. These segmented maps are then aggregated and judged by unipolarity to provide greater confidence in unipolar regions. We define unipolarity $U \in [0,1]$ as the ratio of average signed to average unsigned pseudo-radial magnetic field within a region:

\begin{equation}  \label{Eq-U-def}
U = \frac{| {\rm avg}(B_r^{(i)}) |} {{\rm avg}(| B_r^{(i)} |)}
\end{equation}

   % {\bf --- Unipolarity definiton} \\
This definition yields values of zero and one for perfectly bipolar and unipolar behavior, respectively, serving as the complement of that presented by \citet{2014ApJ...787..121K}. The pseudo-radial magnetic field of each pixel $B_r^{(i)}$ underlying a candidate CH is obtained by correcting the line-of-sight magnetic field observed in the respective magnetogram pixel for the effect of spherical projection with the local longitude $\phi^{(i)}$, latitude $\theta^{(i)}$, and global B-angle $\theta_{B_0}$ as follows:

\begin{equation}  \label{Eq-B-r-i-def}
B_r^{(i)} = \frac{B_{\rm LOS}^{(i)}} {\cos{ \phi^{(i)} } \cos{ ( \theta^{(i)} - \theta_{B_0} ) } }
\end{equation}

   % {\bf --- Unipolarity threshold and PIL visualization} \\
A final design variable is applied in the form of a threshold on the minimum unipolarity of a candidate to be accepted $U_{\rm thresh} \in [0,1)$. Each accepted detection is then assigned a confidence value by linearly rescaling its unipolarity $U\in [U_{\rm thresh},1]$ to $\mathrm{Confidence}\in[0,1]$. The confidence defined herein refers to a heuristic, yet automated, manner for highlighting regions that are more likely to be CHs, similar to that used in \citet{2023SoPh..298..133G}. Accepted candidate regions are then compiled into an ensemble segmentation map. Macroscopic polarity inversion lines are visualized in all ensemble maps as yellow contours, such as in Figure~\ref{F-sarnoff}c. These contours play no role in the segmentation of CHs or the assignment of confidence, but they serve to aid in highlighting non-CH regions. These lines are obtained by a simple smoothing of the aligned magnetogram and plotting the zero level contours. The uniform filter applied in smoothing is sized at 10\% of the image size.

   % {\bf --- Extracted physical properties} \\
Further CH physical properties to be extracted include center of mass coordinates. The area of each pixel $A^{(i)}$ is computed with Equation~\ref{Eq-A-i-def} by taking the product of distance scales per pixel on disk center, $\Delta x,\Delta y$, and correcting for projection as in Equation~\ref{Eq-B-r-i-def}. The center of mass longitude $\phi_{\rm CM}$ and latitude $\theta_{\rm CM}$ are computed for individual detected regions with their unique number of pixels $M$.

\begin{equation}  \label{Eq-A-i-def}
A^{(i)} = \frac{\Delta x \Delta y} {\cos{ \phi^{(i)} } \cos{ ( \theta^{(i)} - \theta_{B_0} ) } }
\end{equation}

\begin{equation}  \label{Eq-lon-cm-def}
\phi_{\rm CM} = \frac{ \sum_{i = 0}^{M} \phi^{(i)} A^{(i)} } { \sum_{i = 0}^{M} A^{(i)} }
\end{equation}

\begin{equation}  \label{Eq-lat-cm-def}
\theta_{\rm CM} = \frac{ \sum_{i = 0}^{M} \theta^{(i)} A^{(i)} } { \sum_{i = 0}^{M} A^{(i)} }
\end{equation}

\begin{landscape}
\begin{figure}    %%%%%%%%%%%%%%%%%% Method Progression on Sarnoff camera observation
\includegraphics[width=0.355\textwidth,clip=]{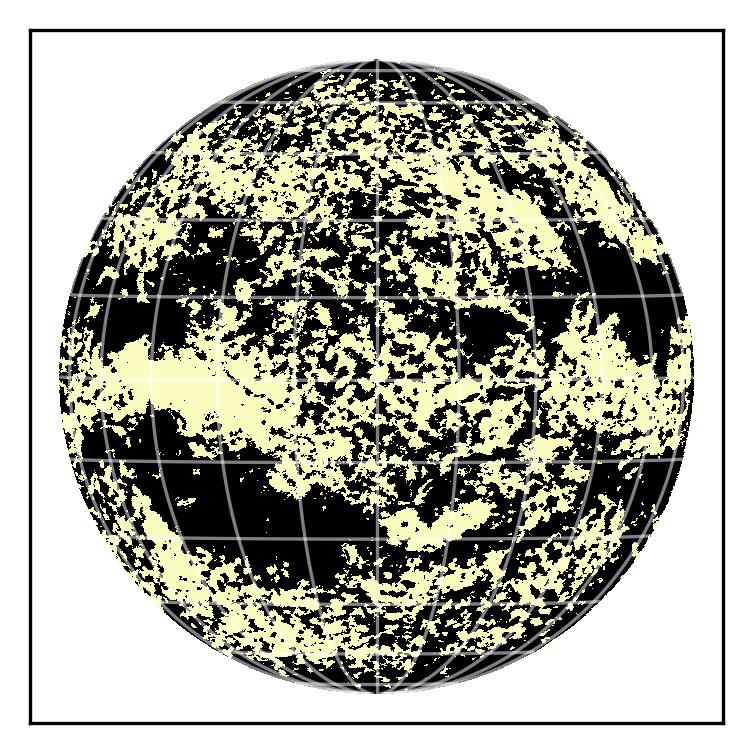}
\hspace{-0.015\textwidth}
\includegraphics[width=0.355\textwidth,clip=]{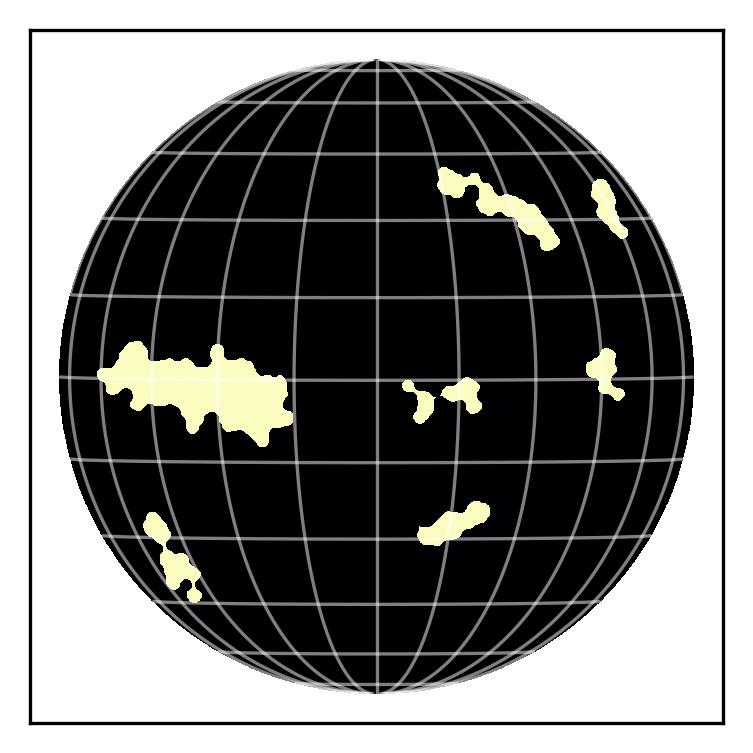}

\vspace{-0.345\textwidth}   % Shift close to the panel top 
\textbf{
    \hspace{0 \textwidth}   (a)
    \hspace{0.3\textwidth}  (b)
}
\vspace{0.315\textwidth}    % Shift back to the panel bottom

\includegraphics[width=0.355\textwidth,clip=]{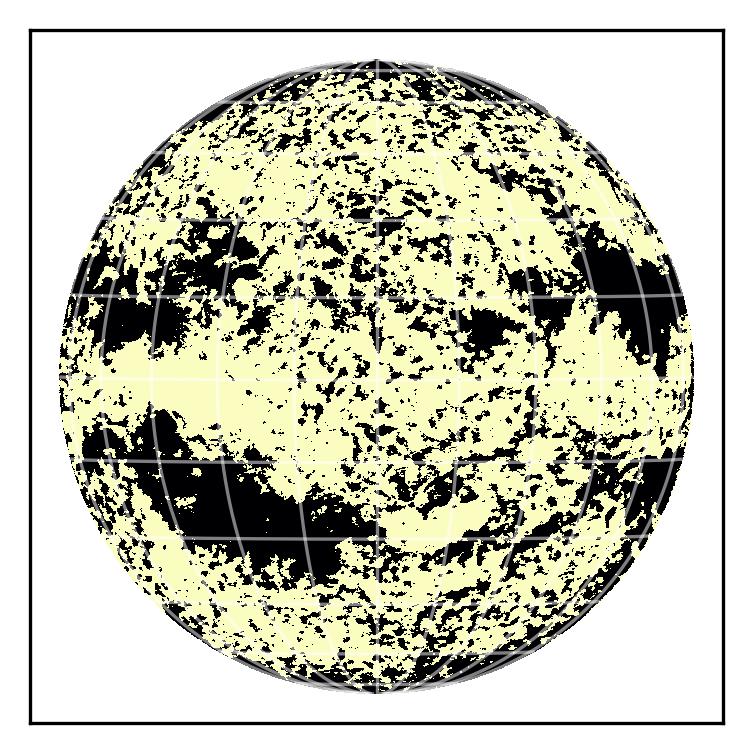}
\hspace{-0.015\textwidth}
\includegraphics[width=0.355\textwidth,clip=]{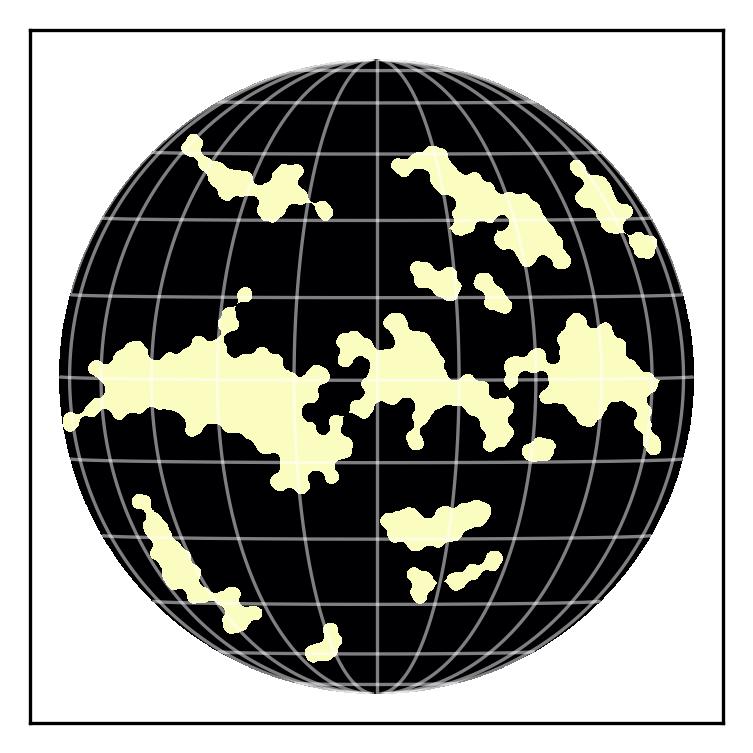}

\vspace{-0.345\textwidth}   % Shift close to the panel top 
\textbf{
    \hspace{0 \textwidth}   (c)
    \hspace{0.3\textwidth}  (d)
}
\vspace{0.3\textwidth}    % Shift back to the panel bottom

\vspace{-0.54\textwidth}
\hspace{0.7\textwidth}
\raisebox{0.54\height}{
    \includegraphics[width=0.46\textwidth,clip=]{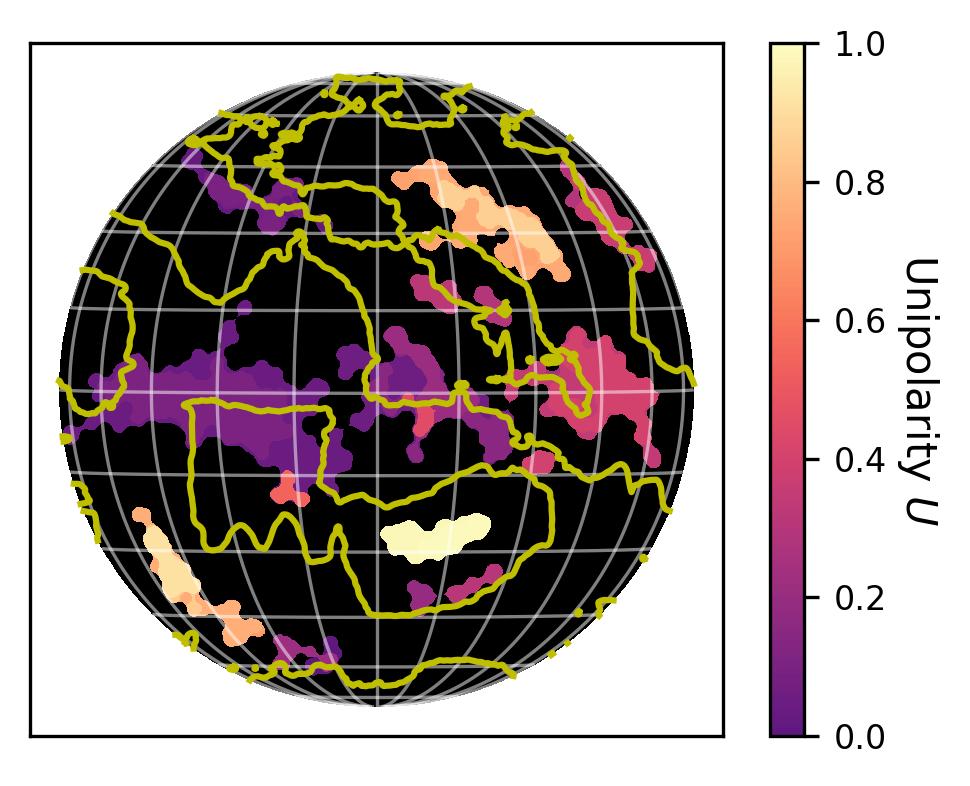}
    \hspace{-0.455\textwidth}
    \raisebox{0.32\textwidth}{\textbf{(e)}}
    \hspace{0.345\textwidth}
}
\raisebox{0.54\height}{
    \includegraphics[width=0.46\textwidth,clip=]{images/sarnoff-stride.jpeg}
    \hspace{-0.455\textwidth}
    \raisebox{0.32\textwidth}{\textbf{(f)}}
}

\caption{Progression through segmentation steps in STRIDE-CH, as applied to the observations of Figure~\ref{F-sarnoff} on June 11, 2012 in the rising phase of Solar Cycle 24. After the execution of data preparation steps on a He I observation and magnetogram, the He I observation is thresholded (a), boundaries are reshaped with morphological operations, small holes within candidate detections are filled, and small candidates are filtered out (b). A second member among the ensemble of segmentations is developed in (c) and (d) with more aggressive detection primarily due to a reduced brightness threshold $I_{\rm thresh}$. Two other preliminary segmentation masks as in (b) and (d) are developed and compiled into an ensemble with candidate regions colored by the unipolarity $U$ in their underlying magnetic field (e). A threshold on unipolarity $U_{\rm thresh}$ is applied to discard bipolar regions, yielding a final ensemble segmentation map with unipolarity-derived confidence (f). Yellow contours in (e) and (f) denote the locations of macroscopic polarity inversion lines to highlight boundaries which CHs should not span across.
        }
\label{F-sarnoff-flowchart}
\end{figure}
\end{landscape}

   % {\bf --- Design variable selection} \\
The values of the design variables $I_{\rm thresh},r_{\rm SE}$ have been selected to find a balance between excessive conservatism and aggression, wherein only a small subset of CHs or a majority of the entire solar disk would be detected, respectively. This was achieved by variable tuning to yield approximate consistency with CH sizes that were visually assessed in EUV 193 Å. STRIDE-CH is not restricted to recapitulate results from the multiplicity of EUV-based CH detection methods, however. The detected sizes and boundary locations of individual CHs may vary significantly from those observed in EUV, as demonstrated in Section~\ref{S-Results}. Figures~\ref{F-sarnoff-flowchart}a and~\ref{F-sarnoff-flowchart}b depict a relatively conservative design with design variables of $I_{\rm thresh} = 90 \%, r_{\rm SE} = 13$ Mm, while Figures~\ref{F-sarnoff-flowchart}c and~\ref{F-sarnoff-flowchart}d depict a relatively aggressive design detecting greater area with $I_{\rm thresh} = 70 \%, r_{\rm SE} = 15$ Mm. The latter design also demonstrates greater complexity in the boundaries of detected CHs, which is an important indicator of the physical processes that occur at the interface between the open and closed fields \citep{2022ApJ...937L..19M}. Thus, the ensemble size was selected to include four designs to combine low and high levels of detected area and visually assessed boundary complexity. The two remaining ensemble members are defined by $I_{\rm thresh} = 80 \%, r_{\rm SE} = 13$ Mm and $I_{\rm thresh} = 70 \%, r_{\rm SE} = 17$ Mm. The unipolarity threshold has been selected at $U_{\rm thresh} = 0.5$ across all ensemble members as it presents the the exact middleground between predominantly bipolar regions $U \in [0,0.5)$ and predominantly unipolar regions $U \in [0.5,1]$. The threshold may be placed elsewhere, but this value is both physically intuitive and reliably demarcates large-scale CHs from large-scale, non-CH regions, as demonstrated in Section~\ref{S-Results}.

\section{Results} 
      \label{S-Results}      

\begin{figure}    %%%%%%%%%%%%%%%%%% Center of mass histograms
\centerline{
       \includegraphics[height=0.41\textwidth]{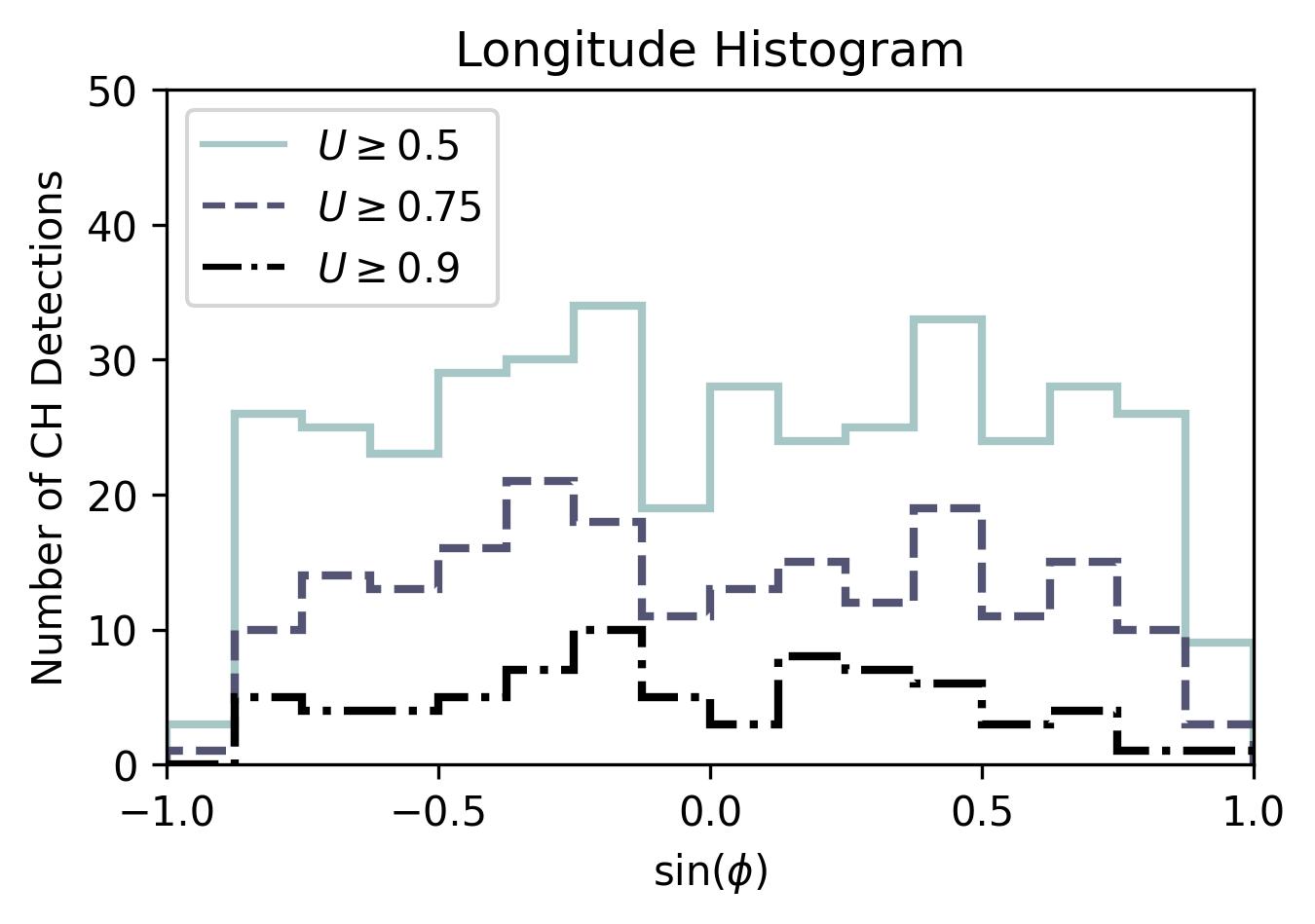}
         \hspace*{-0.03\textwidth}         \includegraphics[height=0.41\textwidth]{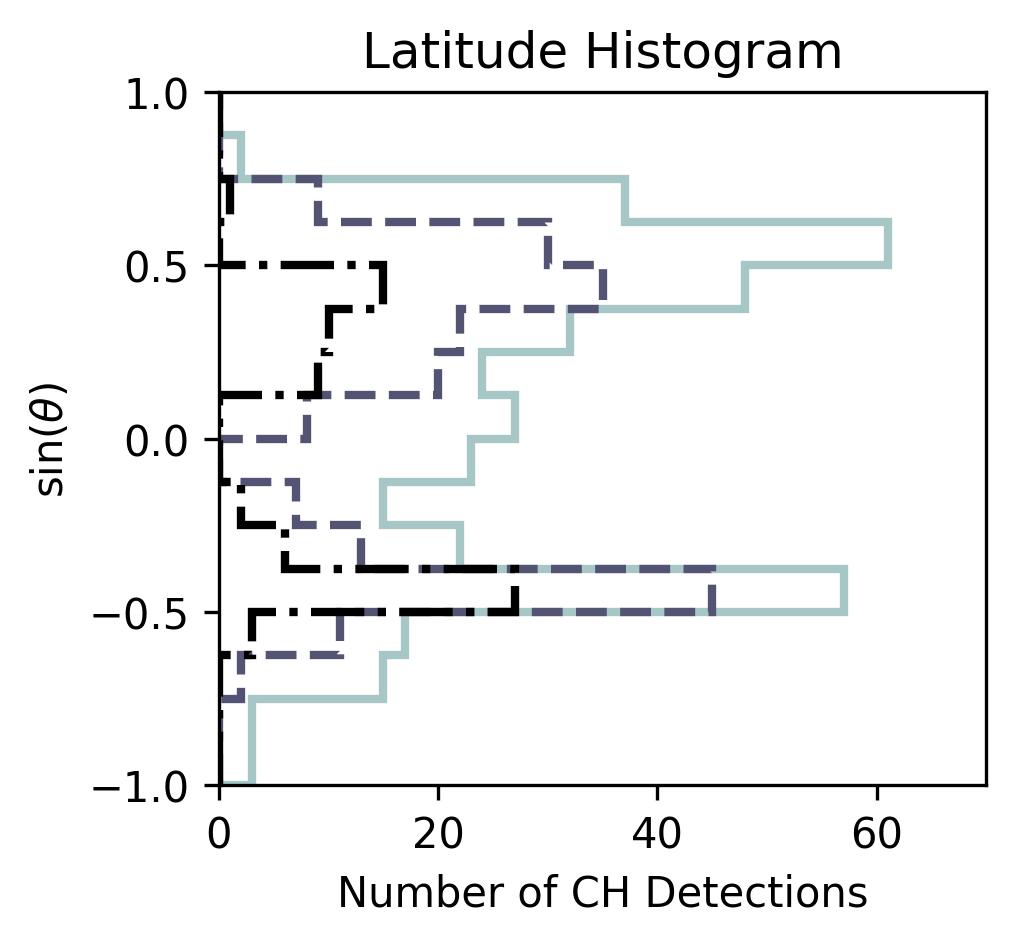}
        }
\vspace{-0.375\textwidth}   % Shift close to the panel top 
\centerline{\bf
    \hspace{0.49 \textwidth} (a)
    \hspace{0.38\textwidth} (b)
    \hfill
   }
\vspace{0.32\textwidth}    % Shift back to the panel bottom 
              
\caption{Histograms of CH center of mass coordinates as detected from April to August 2012 in the rise of solar cycle 24, stratified by unipolarity $U$. Longitude $\phi$ lies on the x-axis (a) and latitude $\theta$ lies on the y-axis (b), with sine of each angle being presented to mimic projection effects on the solar disk.
        }
\label{F-center-of-mass-hists}
\end{figure}

   % {\bf --- Longitude and latitude detection dropoffs} \\
The first period of interest lies in the rising phase of solar cycle 24 from April to August of 2012 and includes the case highlighted in Figures~\ref{F-sarnoff} and~\ref{F-sarnoff-flowchart}. The STEREO spacecraft \citep{2008SSRv..136....5K} were in quadrature with the Earth during this period such that near global imaging coverage of the Sun was achieved. While not immediately relevant for this work, analysis of this period is conducive to future studies aiming to constrain global coronal models as global synchronic detection maps may be developed \citep{2016ApJ...823...53C}. The SOLIS VSM captured 103 daily pairs of He I observations and magnetograms with its Sarnoff cameras during this period, constituting a 68.7\% duty cycle. A histogram of CHs detected by STRIDE-CH in this period is plotted against center of mass in sine-heliographic coordinates in Figure~\ref{F-center-of-mass-hists}, revealing relatively uniform detections in longitude as CHs rotate across the solar disk and a bimodal distribution in latitude, centered about ±26° latitude for the highest confidence detections. The centers of mass show distinct drop-offs past ±61° longitude and ±49° latitude, but portions of the detected CHs do extend farther towards the limb.

\begin{figure}    %%%%%%%%%%%%%%%%%% Rockwell camera 2009 case
\centerline{
    \raisebox{0.065\height}{
        \includegraphics[width=0.29\textwidth]{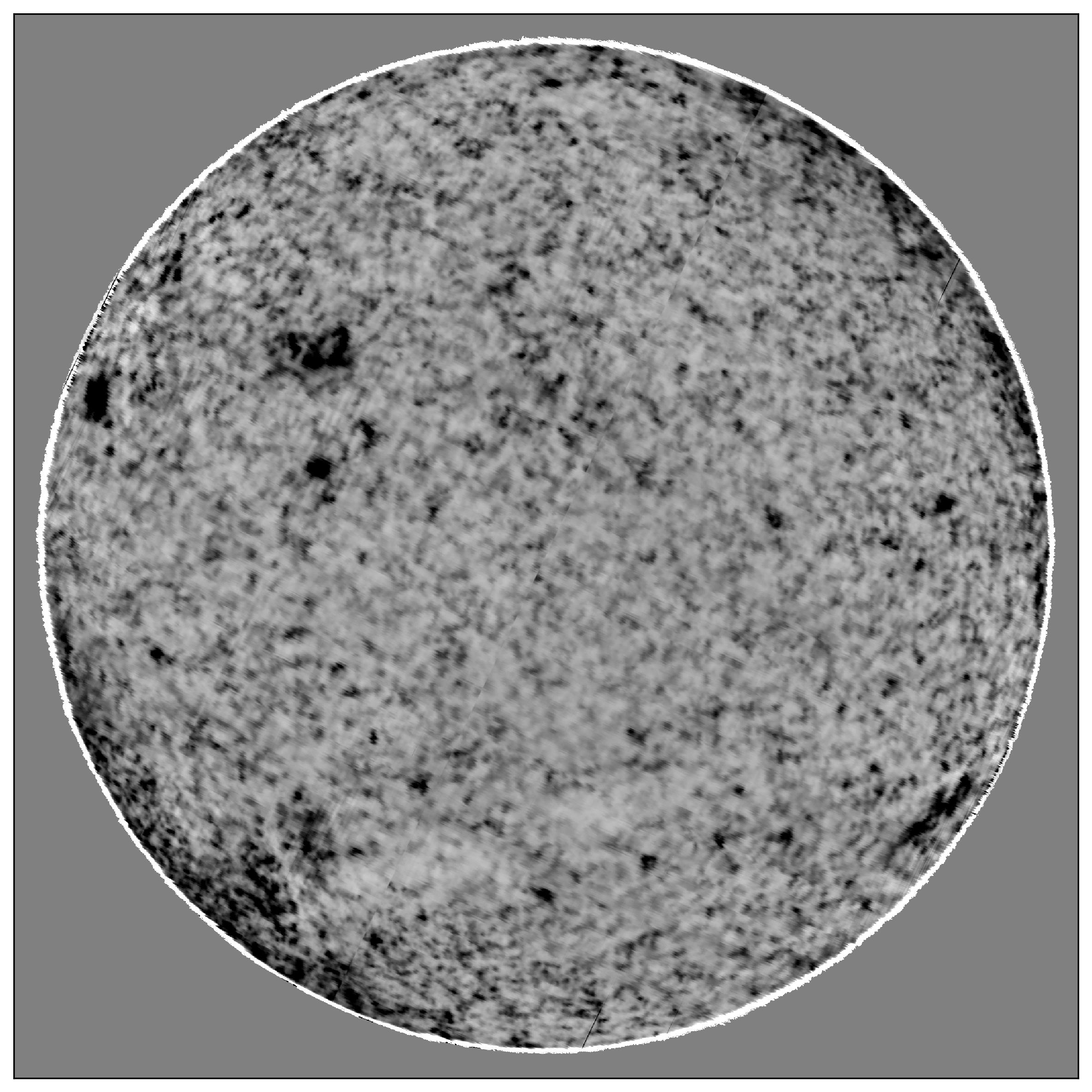}
    }
    \hspace{-0.02\textwidth}
    \includegraphics[width=0.397\textwidth]{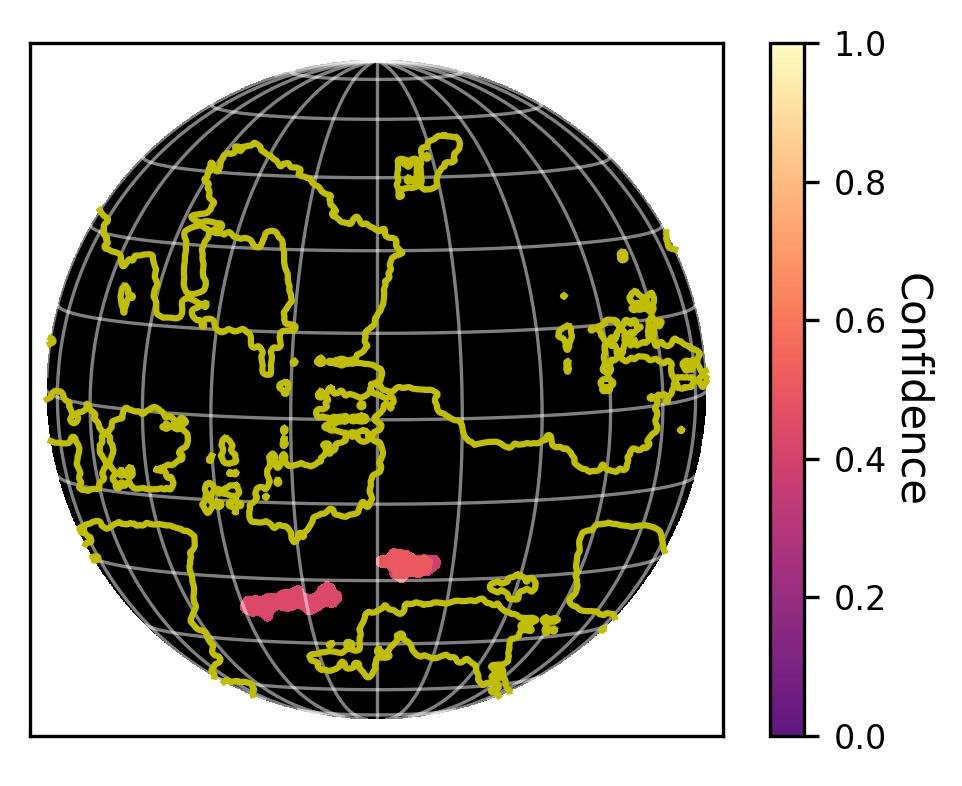}
    \hspace{-0.02\textwidth}
    \raisebox{0.065\height}{
        \includegraphics[width=0.29\textwidth]{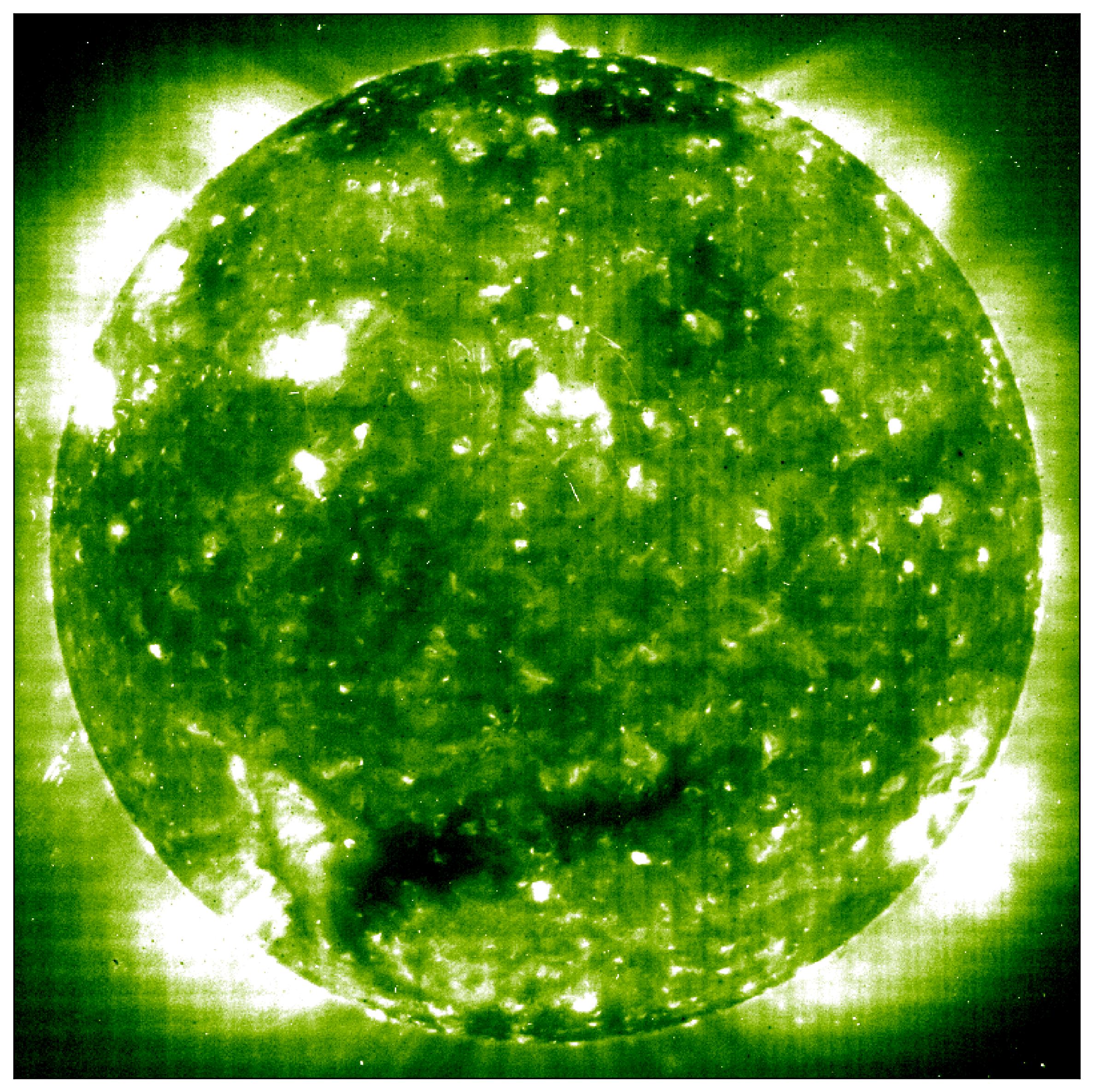}
    }
}
\vspace{-0.305\textwidth}     % Shift close to the panel top 
\centerline{\bf             % Includes the labels
    \hspace{0 \textwidth}       \color{white}{(a)}
    \hspace{0.245 \textwidth}   \color{black}{(b)}
    \hspace{0.33 \textwidth}    \color{white}{(c)}
    \hfill
    }
\vspace{0.25\textwidth}     % Shift back to the panel bottom

\small
        \caption{STRIDE-CH applied to SOLIS VSM Rockwell camera observations on October 22, 2009 in the minimum after Solar Cycle 23. The He I observation, saturated within ± 100 mÅ (a) and ensemble segmentation map (b) are juxtaposed with a SOHO EIT 195 Å observation (c) for reference. Yellow contours on the segmentation maps denote the locations of macroscopic polarity inversion lines.
                }
\label{F-rockwell-2009}
\end{figure}

\begin{figure}    %%%%%%%%%%%%%%%%%% Rockwell camera 2004 case
\centerline{
    \raisebox{0.065\height}{
        \includegraphics[width=0.29\textwidth]{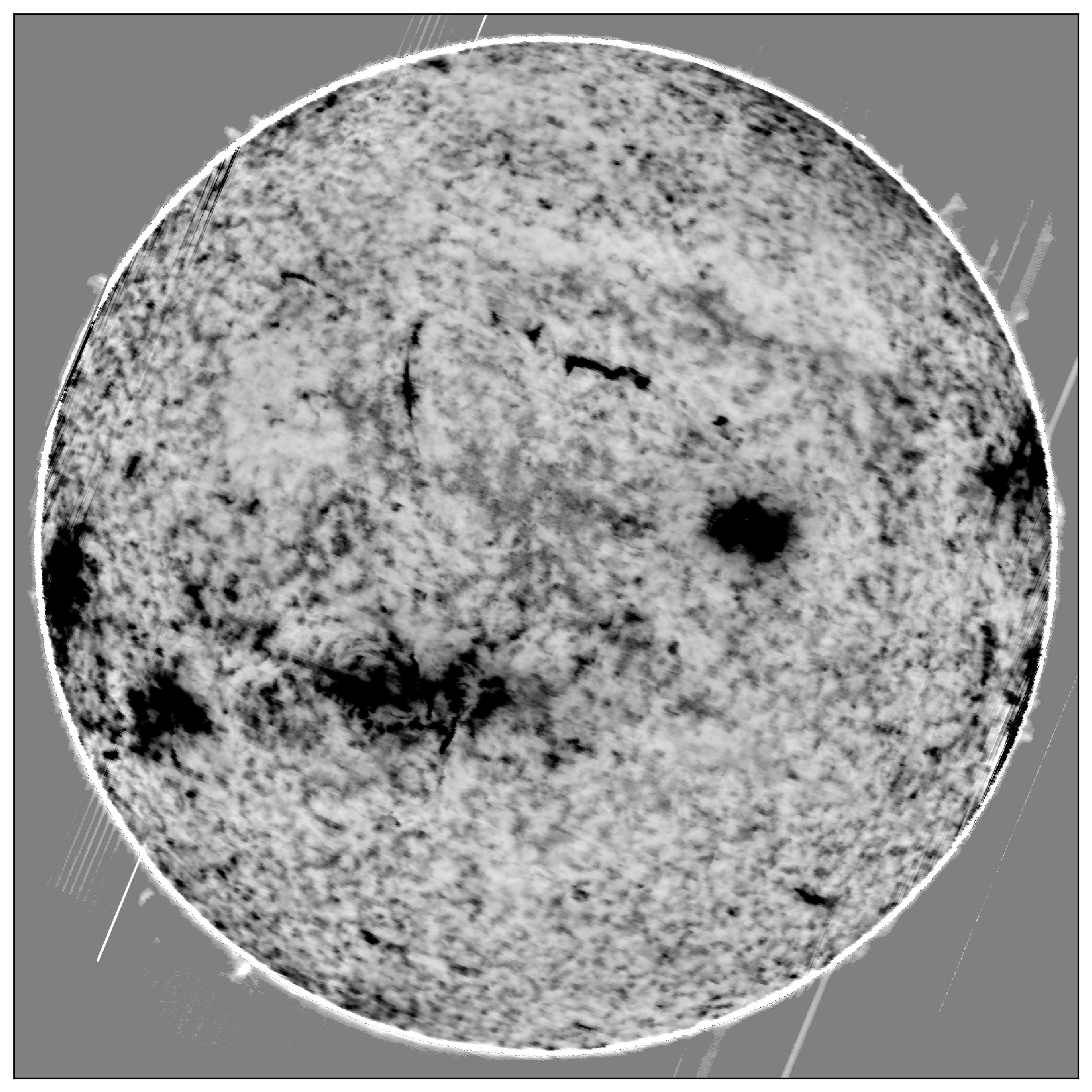}
    }
    \hspace{-0.02\textwidth}
    \includegraphics[width=0.397\textwidth]{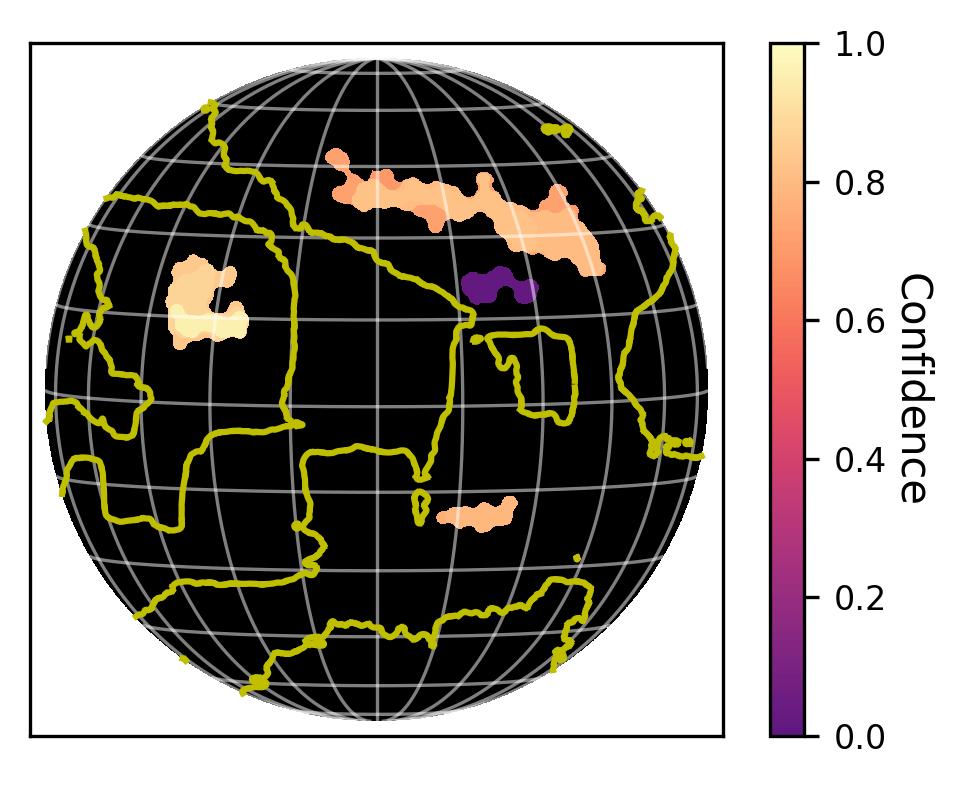}
    \hspace{-0.02\textwidth}
    \raisebox{0.065\height}{
        \includegraphics[width=0.29\textwidth]{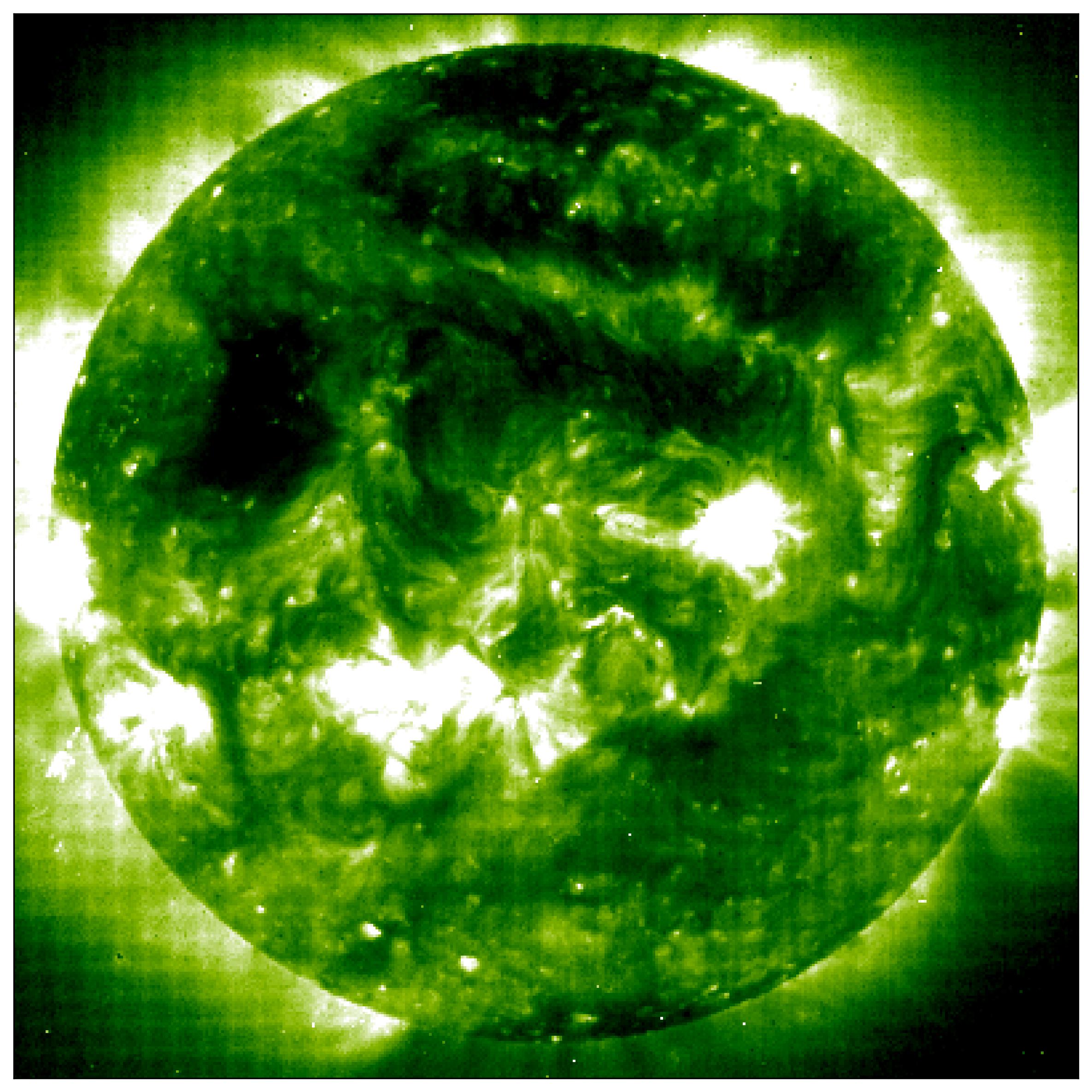}
    }
}
\vspace{-0.305\textwidth}     % Shift close to the panel top 
\centerline{\bf             % Includes the labels
    \hspace{0 \textwidth}       \color{white}{(a)}
    \hspace{0.245 \textwidth}   \color{black}{(b)}
    \hspace{0.33 \textwidth}    \color{white}{(c)}
    \hfill
    }
\vspace{0.25\textwidth}     % Shift back to the panel bottom

\small
    \caption{
    STRIDE-CH applied to SOLIS VSM Rockwell camera observations on November 13, 2004 in the declining phase of Solar Cycle 23. The He I observation, saturated within ± 100 mÅ (a) and ensemble segmentation map (b) are juxtaposed with a SOHO EIT 195 Å observation (c) for reference. Yellow contours on the segmentation maps denote the locations of macroscopic polarity inversion lines.
    }
\label{F-rockwell-2004}
\end{figure}

   % {\bf --- VSM Rockwell camera cases} \\
Earlier cases are presented to inspect the quality of STRIDE-CH across phases of the solar cycle and across cameras of the SOLIS instrument. Two cases lie within the Rockwell camera era of the SOLIS VSM, in the solar minimum following solar cycle 23 on October 22, 2009 and earlier in the declining phase on November 13, 2004. Application of STRIDE-CH to the solar minimum case in Figure~\ref{F-rockwell-2009} results in the detection of two southern hemisphere CHs, a failure to detect both northern and southern polar CHs that are easily identifiable in EUV, and successful null detections elsewhere on the disk. The declining phase case in Figure~\ref{F-rockwell-2004} depicts a successfully detected northeastern CH with substantively different boundaries between the chromospheric He I and coronal EUV observations. A northern CH extending from the meridian towards the western limb is well detected. Below it, a filament appears equally as dark in EUV, but is successfully undetected by STRIDE-CH.

\begin{figure}    %%%%%%%%%%%%%%%%%% KPVT case
\centerline{
    \includegraphics[width=0.29\textwidth]{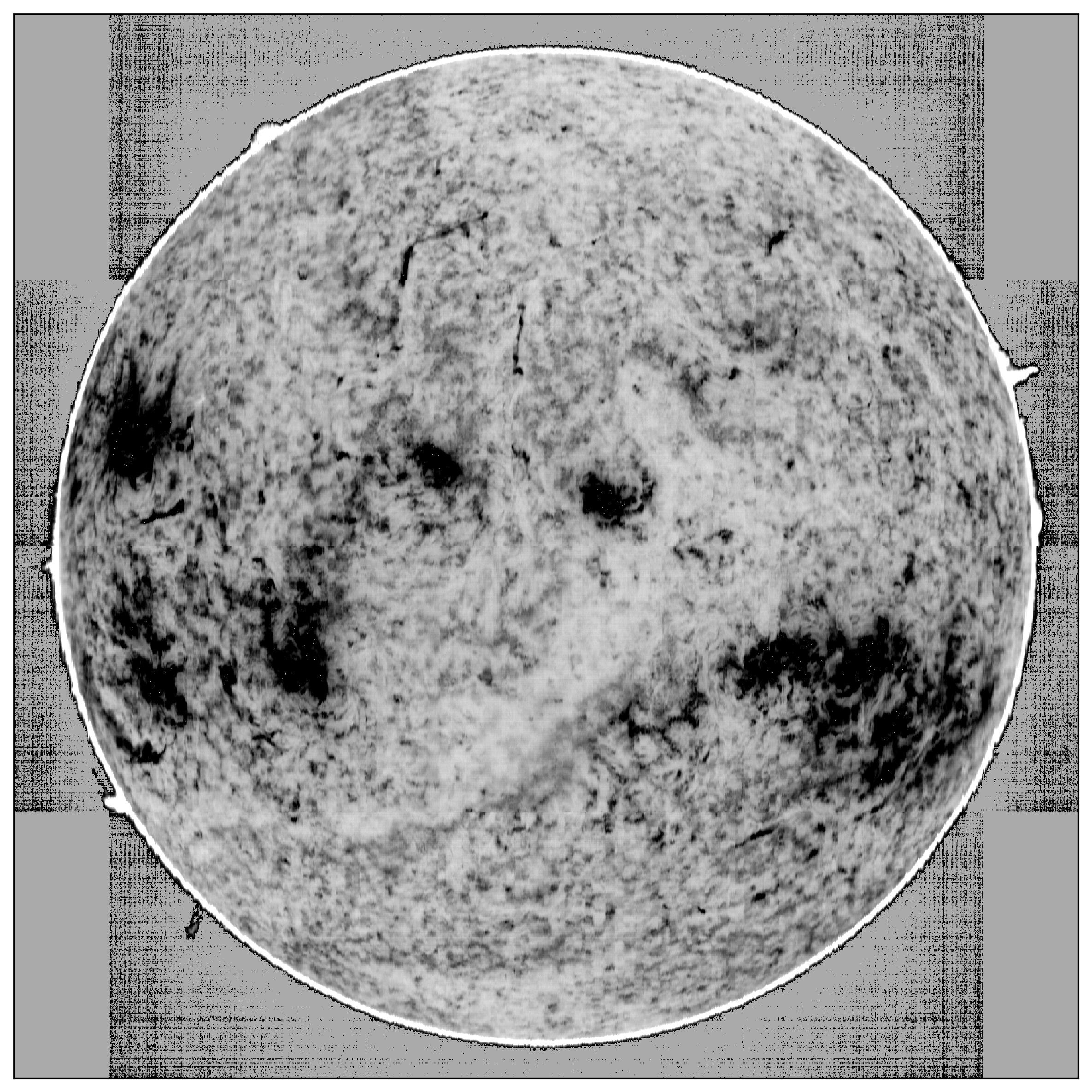}
    % \hspace*{0.095\textwidth}
    \includegraphics[width=0.29\textwidth]{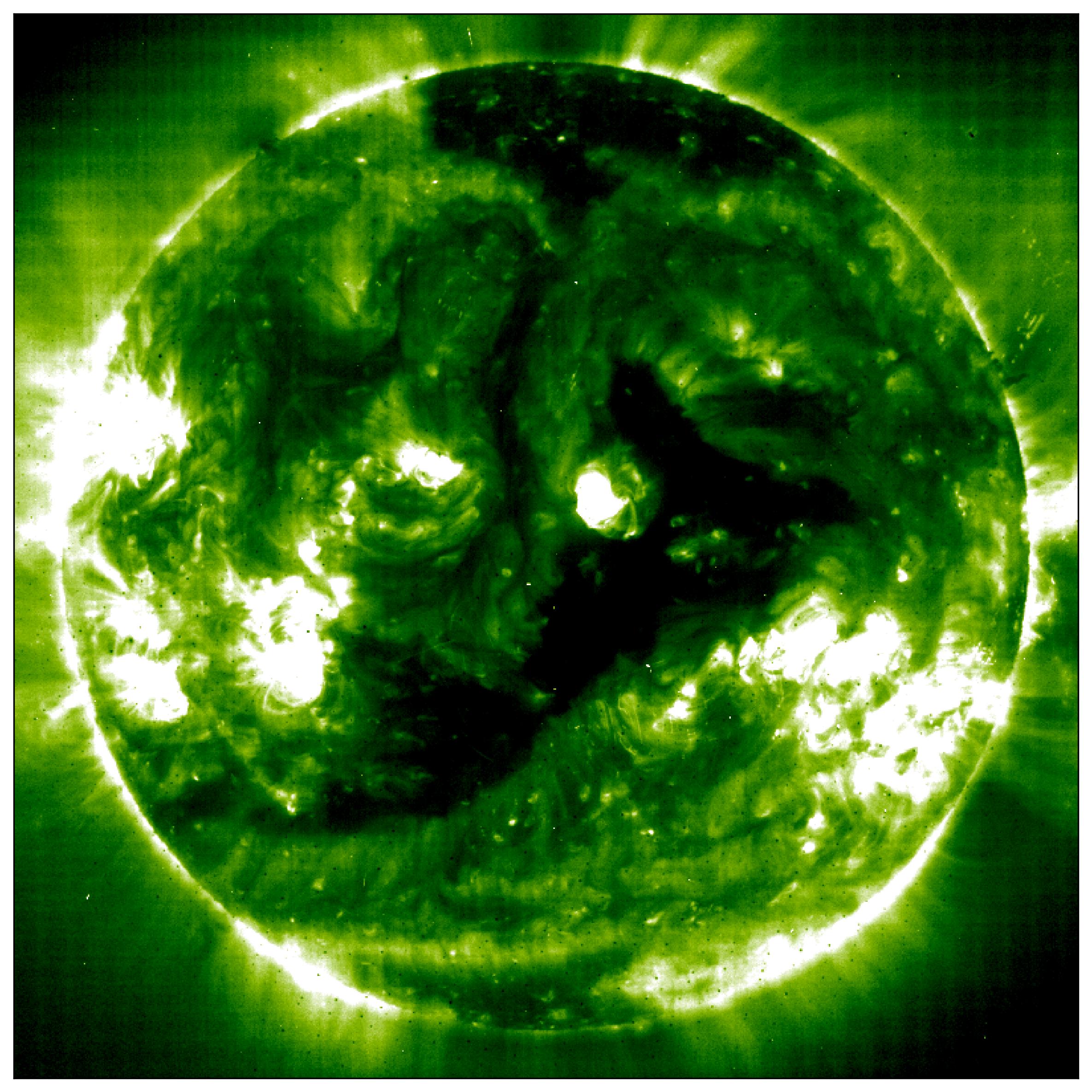}
    }
\vspace{-0.29\textwidth}   % Shift close to the panel top 
\centerline{\bf
    \hspace{0.205 \textwidth}   \color{white}{(a)}
    \hspace{0.235\textwidth}    \color{white}{(b)}
    \hfill
}
\vspace{0.26\textwidth}    % Shift back to the panel bottom 

\centerline{
    \hspace{0.01\textwidth}
    \raisebox{0.1\height}{
    \includegraphics[width=0.277\textwidth]{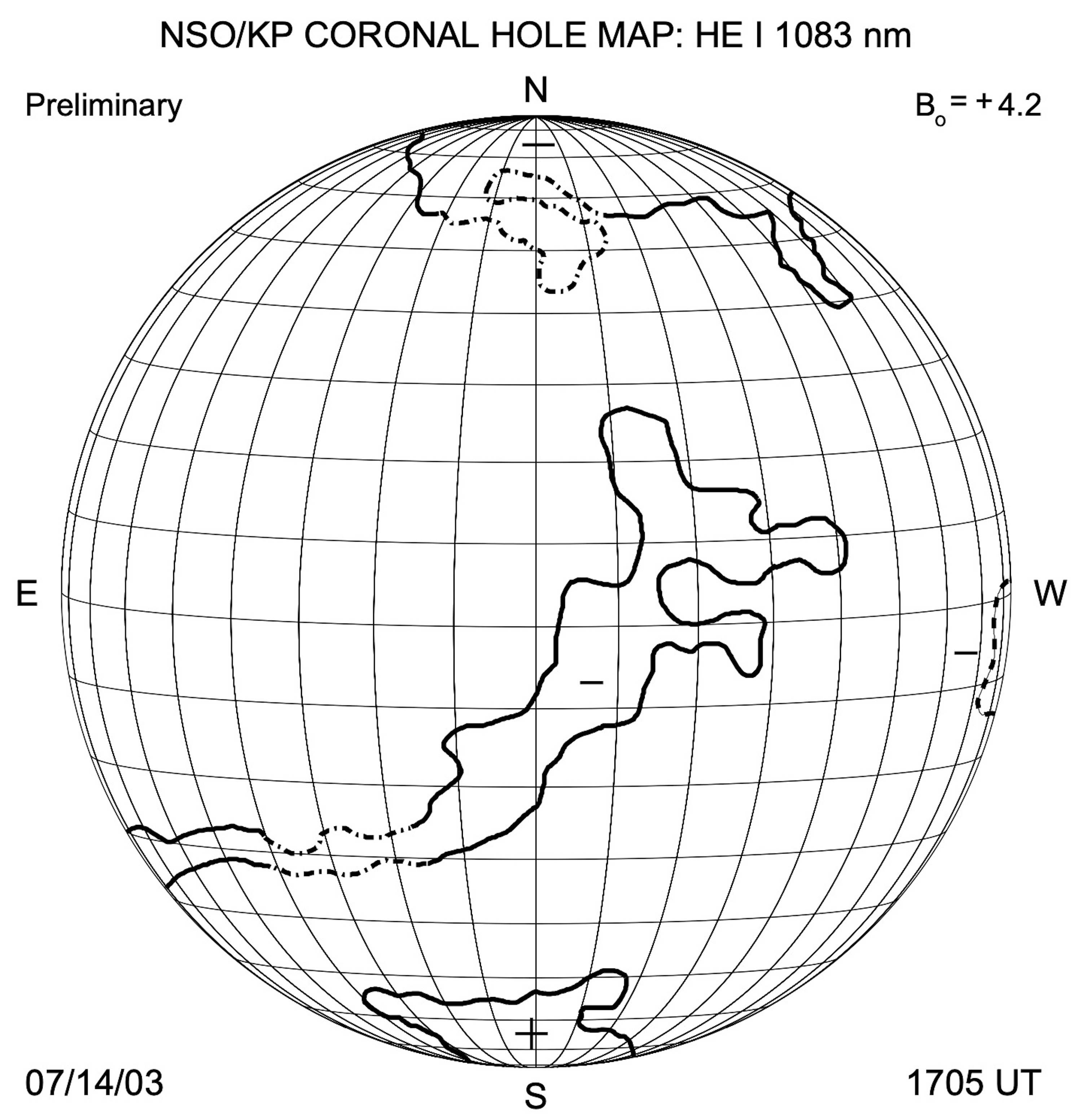}
    }
    \hspace{-0.03\textwidth}
    \includegraphics[width=0.322\textwidth]{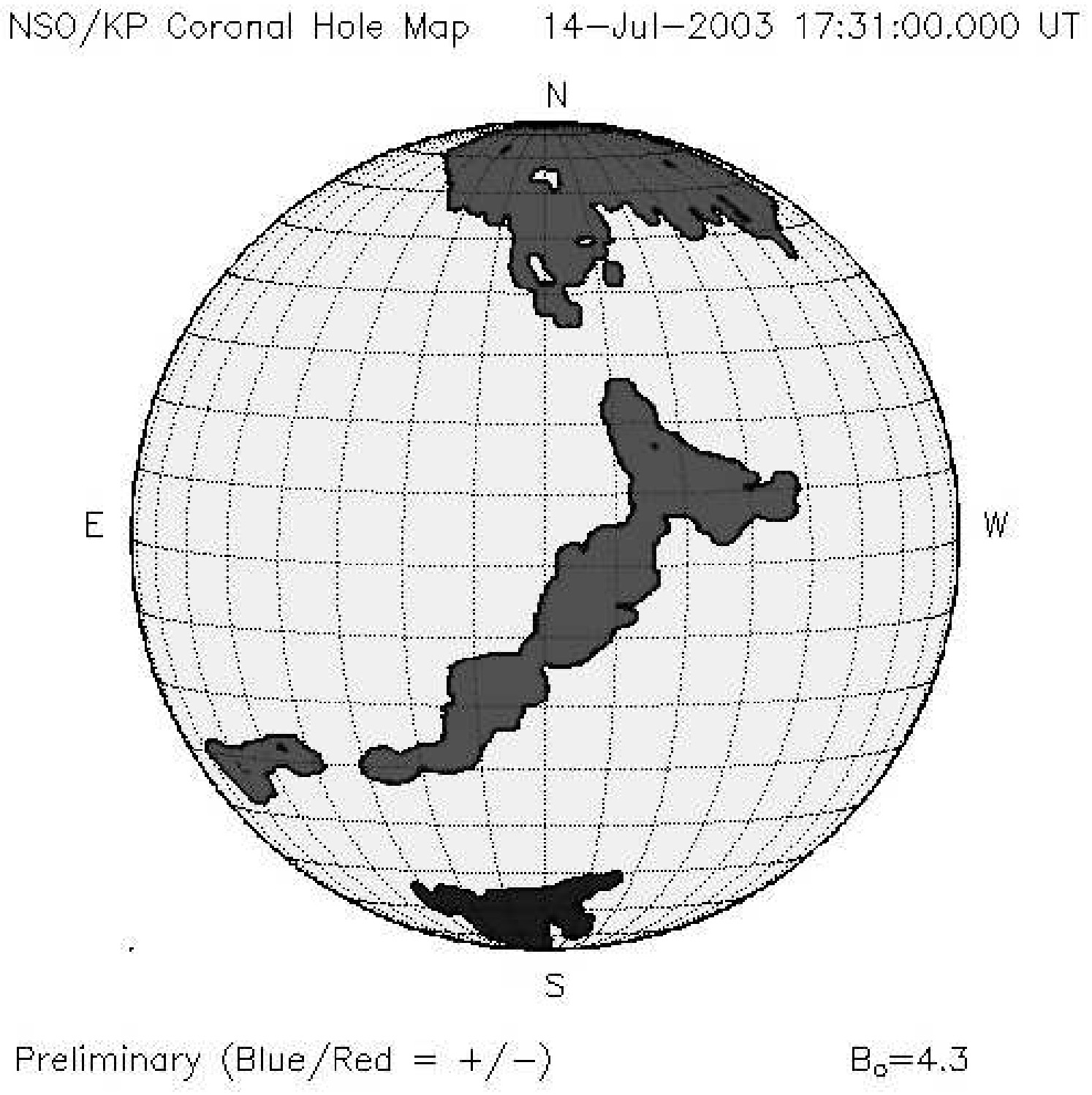}
    \hspace{-0.03\textwidth}
    \raisebox{0.01\height}{
    \includegraphics[width=0.39\textwidth]{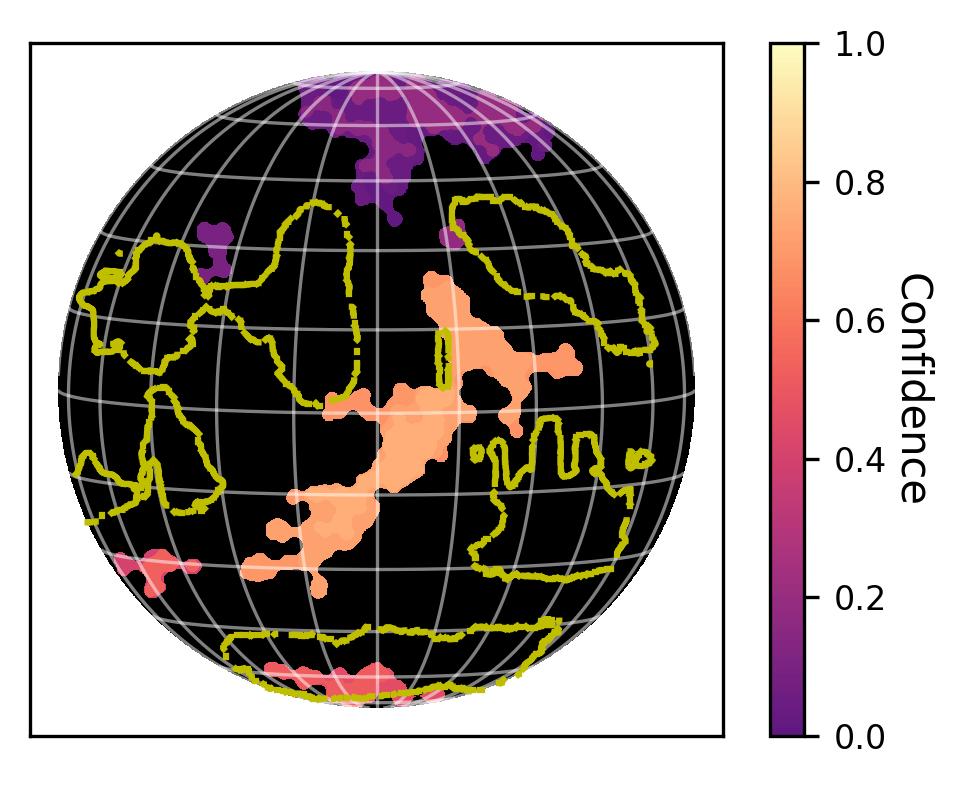}
    }
}
\vspace{-0.293\textwidth}   % Shift close to the panel top 
\centerline{\bf      % Labels
    \hspace{0\textwidth}        (c)
    \hspace{0.25\textwidth}     (d)
    \hspace{0.245\textwidth}     (e)
    \hfill
    }
\vspace{0.253\textwidth}    % Shift back to the panel bottom 

\small
        \caption{Prominent He I CH detection methods applied to KPVT observations on July 14, 2003, in the early decline of Solar Cycle 23. The He I observation, saturated within ± 100 mÅ (a) is juxtaposed with a SOHO EIT 195 Å observation (b) for reference. CH detection methods are demonstrated in the lower row, including a hand drawn map developed by \citet{2002SoPh..211...31H} (c), an automated detection map developed by \citet{2005ASPC..346..261H} (d), and a STRIDE-CH ensemble segmentation map with threshold levels modified for KPVT observations (e). Images (c) and (d) are reproduced from \citet{2005ASPC..346..261H}. Yellow contours on the STRIDE-CH map denote the locations of macroscopic polarity inversion lines.
                }
\label{F-kpvt}
\end{figure}

   % {\bf --- KPVT case and Henney \& Harvey comparison} \\
The final case is depicted in Figure~\ref{F-kpvt} and is the earliest considered, lying in the declining phase of solar cycle 23 on July 14, 2003, as observed by the KPVT SPM. As the KPVT SPM is a significantly different instrument from the SOLIS VSM used thus far, new design variables were required to produce stable results in an extended range of observations analyzed from this period. The He I pixel intensity threshold $I_{\rm thresh}$ was raised uniformly across all ensemble members  by 15\% and the unipolarity threshold was dropped to $U_{\rm thresh} = 0$ to reveal candidate CHs that have been measured as predominantly bipolar. The reasoning for the latter alteration is discussed in Section~\ref{S-Discussion}.

\section{Discussion} 
      \label{S-Discussion}

   % {\bf --- Sarnoff showcase: ability to stride past coronal loops} \\
The robustness of STRIDE-CH is evidenced by the results in Figures~\ref{F-sarnoff},~\ref{F-rockwell-2009},~\ref{F-rockwell-2004}, and~\ref{F-kpvt} which sample different phases of the solar cycle while maintaining fixed ensemble of design variable values for a given instrument. These design variables include threshold levels $I_{\rm thresh}$, SE disk radii for morphological operations $r_{\rm SE}$, and a unipolarity threshold $U_{\rm thresh}$. Figure~\ref{F-sarnoff} demonstrates several strengths of the newly developed algorithm when applied to SOLIS VSM Sarnoff camera observations in the rising phase of solar cycle 24. The three largest detections in Figure~\ref{F-sarnoff}c are all true CHs as they appear bright and smooth in He I while having high unipolarity, $U\ge0.5$. While the two western CHs may be readily observed in the corresponding EUV image, the southeastern CH on the other hand may only be observed in EUV at a later date when the CH has rotated towards disk center. This is an excellent example of the benefit of observing CHs in He I 10830 Å as its identification in EUV is difficult due to the extruding coronal loops in the neighboring active region, but it is readily apparent in He I and successfully detected by STRIDE-CH. 

   % {\bf --- Sarnoff showcase: ability to disambiguate PILs} \\
STRIDE-CH is also capable of eliminating initially detected regions that are not CHs based on the underlying unipolarity. The large equatorial regions of quiet Sun on the east and west limb which appear bright in the He I observation of Figure~\ref{F-sarnoff} are detected in the preliminary segmentation masks of Figure~\ref{F-sarnoff-flowchart}b and~\ref{F-sarnoff-flowchart}d. However, they have low unipolarity, appear bright in this EUV observation and in those taken at neighboring times, and thus are not CHs. These regions are sufficiently bright, coherently shaped, and large to be admitted by the preliminary segmentation procedure, but are insufficiently unipolar to be determined as CHs and are correctly eliminated. Their bright appearance in 10830 Å indicates that the underlying continuum emission is being weakly absorbed such that the chromospheric plasma present in these regions likely hosts a low density in the He I population that absorbs this wavelength.

   % {\bf --- Potential criticisms of unipolarity-derived confidence} \\
A limitation in unipolarity-derived confidence as a quantification of uncertainty is the effect of reduced confidence in CHs that host a greater density of mixed polarities \citep{2019A&A...629A..22H}. CHs that are measured to have imperfect unipolarity will directly result in STRIDE-CH presenting reduced confidence, regardless of whether the boundary is accurately located with respect to the boundary which would be segmented by a human expert in He I. This prediction of uncertainty is therefore not calibrated well as it would not be highly correlated with segmentation error. Well-calibrated uncertainty quantification is a desirable trait of image segmentation methods as predicting high uncertainty a priori can alert researchers and practitioners to anticipate error in boundary location and not rely on the automated results \citep{2021arXiv210316265C}. Another potential limitation in this definition of confidence lies in the fact that magnetograms incur substantial error beyond 60° from disk center \citep{2014ApJ...787..121K}, such that confidence might be artificially reduced near the limb. However, the derived unipolarity appears to remain elevated far from disk center, as evidenced by the southeastern and northwestern CHs in Figure~\ref{F-sarnoff} that are detected with high unipolarity and thus high confidence. This demonstrates the quality of SOLIS magnetograms and provides evidence of the legitimacy of unipolarity-derived confidence. For discussion on alternatives for deriving confidence, see the Appendix.

   % {\bf --- Lack of polar CH detections} \\
A lack of polar CH detections is demonstrated in the reduced latitudinal range of detected centers of mass in Figure~\ref{F-center-of-mass-hists}. Although polar CHs are important for understanding the topology and dynamics of the corona, it is well known that accurately segmenting these regions is challenging. Beyond the projection effect that foreshortens the observed area of near limb CHs, this shortcoming may be attributed to an instrumental effect present in SOLIS He I observations of noise at the poles. This instrumental effect is distinct from the traditional limb darkening observed in continuum emission as it appears to be enhanced at the poles rather than uniformly at the limb and is evident in Figures~\ref{F-sarnoff}a and~\ref{F-rockwell-2009}a. Additionally, while spectral fringes were present in the Rockwell camera era prior to 2010 \citep{2005AGUSMSP51A..02J}, they were enhanced after the replacement with Sarnoff cameras and prevented finely tuned limb darkening and zero offset corrections. This change in data preparation unfortunately rendered the \citet{2005ASPC..346..261H} routine unreliable in operation \citep{2016NSO}. The cases presented in this work demonstrate that STRIDE-CH performs satisfactorily on this SOLIS data away from the poles, where specialized techniques may be applied for the detection of polar CHs \citep{2009SoPh..257...99K}. The two latitudinal concentrations in Figure~\ref{F-center-of-mass-hists} reflects the rising phase of the solar cycle, as active regions abound at lower latitudes and vacate the equatorial region from a significant presence of CHs.

   % {\bf --- Rockwell: solar minimum} \\
Solar minimum, as sampled in Figure~\ref{F-rockwell-2009}, poses a challenge as mid to low latitude CHs tend to be sparse, requiring CH detection algorithms to be capable of not detecting any CHs at low latitudes when none are present. Additionally, the He I observations in this period coincided with the final year of Rockwell camera use prior to their replacement, which raised concerns on potential imagery degradation, as warranted by the grainier texture as compared to other He I observations displayed in this work. Despite these challenges and aside from the aforementioned lack of polar CH detection, reasonable CHs are segmented. 

   % {\bf --- Rockwell: altered boundaries in chromosphere and corona} \\
CHs observed in underlying chromospheric He I are typically expected to appear with smaller size and as a subset of those observed in coronal emission due to the superradial expansion of their magnetic field lines as they rise in altitude \citep{1997SoPh..176..107M}. However, the northeastern CH in the declining phase case in Figure~\ref{F-rockwell-2004} demonstrates an eastern boundary in EUV that flares significantly more towards the east limb than in He I, hinting towards a more complex relationship. Future studies of He I imagery may therefore reveal insights into the mid-atmospheric magnetic field structure of CHs, as they evolve from photospheric flux tubes up to their expanding forms in the corona.

   % {\bf --- Henney \& Harvey comparison and KPVT case: polar CH capturing} \\
The final application to KPVT SPM observations in Figure~\ref{F-kpvt} yields a favorable comparison between the elongated, disk center CH detected by the newly developed ensemble detection, the manual detection developed by \citet{2002SoPh..211...31H}, and the automated detection developed by \citet{2005ASPC..346..261H}. All methods achieve a true negative detection along the filament visible in EUV with a vertical span in the northern hemisphere, as well as true positive detections in the northern and southern polar CHs. The success achieved by STRIDE-CH in polar detection may be ascribed to the high quality of KPVT He I observations, which lack the instrumental polar noise effect that persists in observations made by SOLIS. However, these regions host lower unipolarity as a result of the weak signal to noise ratio in the polar regions of the KPVT magnetogram. This serves as a further testament to the quality of SOLIS magnetograms, which overcome this challenge and provide reliable unipolarity measurements from disk center.
   
\section{Conclusion} 
\label{S-Conclusion} 

   % {\bf --- Strengths} \\
STRIDE-CH accounts for multiple properties of CHs including weak absorption in the chromospheric He I spectral line, coherent morphology, and predominant unipolarity in the underlying photospheric field. Similar to the method developed by \citet{2005ASPC..346..261H}, the ingested He I observations allow the algorithm to stride past extruding coronal loops that obscure CHs while magnetograms allow it to disambiguate CHs from filaments and quiet Sun. STRIDE-CH builds upon it, however, with the following added capabilities:

\begin{enumerate}
  \item The ensemble of segmentations provides added flexibility in CH boundary assignment. Confidence in each of the candidate boundaries for a CH is then derived in a physically motivated manner from the unipolarity of the enclosed region.
  \item The novel method succeeds on observations after 2010 in the SOLIS VSM Sarnoff camera era where \citet{2005ASPC..346..261H} became unreliable in operation.
  \item The relevant code has been made available in an open-source, easily maintainable Python project.
\end{enumerate}

   % {\bf --- Quantified uncertainty; sacrifices} \\
The quantification of uncertainty through  the assigned confidence is crucial in this challenging segmentation task and aligns with recent community efforts towards uncertainty quantification of automatically detected CH boundaries \citep{2021ApJ...913...28R,2024ApJS..271....6R}. The method’s weaknesses are of importance as well, whether they are inherited from observations or introduced by the procedure and design variable selection. Polar CHs are not well detected in He I observations made by SOLIS, in large part due to an instrumental effect. This is a technological challenge rather than a fundamental limitation of the He I line, however, as evidenced by the clarity of polar CHs in KPVT He I observations and successful detection in Figure~\ref{F-kpvt}. A further limitation is that transitioning from research to operational maturity would require improvements to time series consistency.

    % {\bf --- Future He I observations} \\
Despite these tradeoffs, He I spectroheliograms provide valuable CH boundary information. Future instruments observing He I in a synoptic, full-disk fashion would therefore be vital. The Chromospheric Magnetometer (ChroMag) \citep{2023BAAS...55c.088D} will provide full-disk He I observations in a research capacity from the Mauna Loa Solar Observatory and, should the NSO’s critically important next generation Ground-based solar Observing Network (ngGONG) be funded, derivatives of ChroMag will be implemented to operationally observe He I \citep{2023BAAS...55c.320P}. As STRIDE-CH has been made available as an open-source tool, it may be implemented in a pipeline for either observatory. The quality of KPVT He I observations, which clearly display polar CHs and enable their detection by STRIDE-CH, should be sought after while the polar noise effect and the absence of limb and zero-offset corrections in SOLIS He I observations should be avoided. The improved signal to noise ratio of SOLIS magnetograms is desirable on the other hand, as it enables CHs far from disk center to be detected with high unipolarity, and thus confidence by STRIDE-CH.

   % {\bf --- Fusion of CHs from He I and EUV} \\
The ground-based, chromospheric perspective on CHs offered by near infrared, He I observations complements well the space-borne, coronal perspective provided by EUV observations. While the boundaries of CHs that are fully exposed to the line of sight may appear with greater sharpness and contrast, He I enables the detection of CHs obscured in EUV by greater emissive scale heights and coronal loops. The fusion of near side CHs detected by STRIDE-CH and an EUV-based segmentation method may thus better represent the distribution of CHs across the disk than any individual method. The improved Active Contours Without Edges (ACWE) algorithm for EUV-based segmentation presented by \citet{2023SoPh..298..133G} would be a particularly fitting complement to STRIDE-CH as it also predicts pixel-wise confidence in CH presence rather than binary decisions. The integration of STRIDE-CH and ACWE CHs into a fused confidence map would be a form of soft fusion within the object/decision level fusion classification (for a review, \citealp{2010IJIDF...1....5Z}). The litany of methods available for such a task is well summarized by \citet{2006Zeng}, but operations from fuzzy logic are especially promising. They have been applied for CH detection in a single EUV wavelength \citep{2014A&A...561A..29V} and in multiple EUV wavelengths, including 171 Å with formation in both the chromosphere and corona, through a soft fusion framework \citep{2007Barra}. Such work is foundational for future soft fusion with He I segmentations that do not suffer from the obscuration by coronal emission present in 171 Å.

   % {\bf --- Global, synchronic maps} \\
Furthermore, global, synchronic detection maps may be developed by merging near side CH maps, either from STRIDE-CH alone or as fused with EUV-based CHs, with those detected from STEREO Extreme Ultraviolet Imager (EUVI) observations \citep{2004SPIE.5171..111W}. These maps may be developed for periods in which the STEREO spacecraft are in quadrature with the Earth, such as during the 2012 period analyzed herein. They would benefit from the strengths of He I and magnetograms on the near-side as well as the availability of EUV observations away from the Sun-Earth line of sight. The global, instantaneous CH distributions would enable the refinement of answers towards research questions such as the rigid rotation of CHs \citep{2006ApJ...642L..69L} and the open flux problem \citep{2024ApJ...964..115A}. They would also support the application of model selection techniques (e.g. \citealt{Brunton_Kutz_2022}) to rigorously select the coronal models which best match the available segmented CH data. Such techniques may identify optimal magnetic input maps and model parameters automatically for eventual application in an operational forecast setting.

\begin{acks}

 This work utilizes SOLIS data obtained by the NSO Integrated Synoptic Program (NISP), managed by the National Solar Observatory, which is operated by the Association of Universities for Research in Astronomy (AURA), Inc. under a cooperative agreement with the National Science Foundation. With respect to the comparison in Figure~\ref{F-kpvt}, the coronal hole data used here was compiled by K. Harvey and F. Recely using NSO KPVT observations under a grant from the NSF.
\end{acks}

% Available additional data environments:
% required: authorcontribution, fundinginformation, dataavailability
% optional: materialsavailability, codeavailability
\begin{authorcontribution}
J.A.L. developed the algorithm, obtained the results, and wrote the manuscript. M.S.K. and C.N.A. guided and provided mentorship for this project and its implementation. All authors advised in the development of the algorithm, aided in the interpretation of results,  and reviewed the manuscript.
\end{authorcontribution}

\begin{fundinginformation}
This work was funded by the National Aeronautics and Space Agency (NASA) Internal Scientist Funding Model (ISFM) awards to the Wang-Sheeley-Arge (WSA) model team and the Center for Helioanalytics (CfHA). J.Z. and M.D. acknowledge the support from NSF multi-messenger astronomy supplement funding to NSO (award No. AST-0946422).
\end{fundinginformation}

\begin{dataavailability}
Websites for KPVT SPM and SOLIS VSM full-disk spectroheliograms and magnetograms are referenced in Section~\ref{S-Observations}. The FITS files for the ensemble CH segmentation maps in Figures~\ref{F-sarnoff}c,~\ref{F-rockwell-2009}b,~\ref{F-rockwell-2004}b, and~\ref{F-kpvt}e have been published \citep{my_zenodo}.
\end{dataavailability}

\begin{codeavailability}
The repository at \url{https://github.com/jalanderos/STRIDE-CH} contains the relevant code and a fixed version has been published \citep{my_zenodo}.
\end{codeavailability}

\begin{ethics}
\begin{conflict}
The authors declare that they have no conflicts of interest.
\end{conflict}
\end{ethics}

\appendix   

\section{Methods Considered for Deriving Confidence}
After candidate regions have been detected from He I observations, as demonstrated in Figure~\ref{F-sarnoff-flowchart}e, a method for assigning confidence to candidates must be defined to capture intra-algorithm uncertainty. The properties of CHs that were evaluated as bases for deriving confidence include smoothness in He I, or low chromospheric network contrast, and unipolarity in photospheric magnetograms. Table~\ref{T-solar-feature-properties} indicates that these properties distinguish CHs from some, if not all, other large scale solar features for deriving confidence. While CHs are both smooth and unipolar, the candidate regions detected by STRIDE-CH with smooth textures do not coincide completely with those exhibiting unipolarity. The following question then arises: \textit{Should confidence in true CH status be derived from unipolarity or smoothness?}

\subsection{Differences in Unipolar and Smooth Candidate Regions}
   % {\bf --- Smoothness metric} \\
Smoothness must first be quantified to provide evidence for the claim that differences exist between unipolar and smooth candidates, after which the question of confidence allocation may be addressed. The lack of smoothness within a candidate region in a He I observation may be quantified by the gradient median $g_\mathrm{med}$, calculated as follows.

\begin{enumerate}
    \item Pixels outside of the candidate region in question are masked.
    \item An approximation of the image gradient is obtained by application of the Sobel operator to the masked array \citep{CAVE_0396}.
    \item The gradient median $g_\mathrm{med}$ of the pixel-wise image gradient magnitudes $g$ within the candidate region is obtained.
\end{enumerate}

The scalar gradient median $g_\mathrm{med}$ quantifying roughness for each candidate is converted into a percentile to characterize texture relative to $n$ candidates under consideration as well as inverted to instead quantify smoothness. This results in the smoothness percentile $S\in [0,100]$. % small lie, really [0,100)

   % {\bf --- Unipolarity vs smoothness table} \\
The distinction between unipolar candidate CHs detected by STRIDE-CH and smooth candidates is evidenced by Table~\ref{T-unipolarity}, which consists of results from the 2012 period detailed in Section~\ref{S-Results}. Without the application of the unipolarity threshold as in Figure~\ref{F-sarnoff-flowchart}e, 1064 candidate boundaries were detected in the initial ensemble CH maps produced by STRIDE-CH. This does not reflect the number of distinct CHs even for a single date, as multiple candidate boundaries may characterize the same candidate region. Candidates are categorized as predominantly unipolar here if they have a unipolarity $U \ge 0.5$. In place of an absolute metric for distinguishing unipolar and bipolar categories, smoothness is judged relative to the $n$ candidates in the sample obtained from this period. High smoothness candidates host a smoothness percentile $S \ge 50$ such that half of the regions fall in this category. While one might expect these candidates to either be true CHs hosting both high smoothness and unipolarity or false CHs hosting neither, 41.6\% do not fall in either category as they achieve high marks in only one of the two properties in question. Thus, unipolarity and smoothness do not coincide entirely among candidate CHs.

An example of the smooth, yet bipolar category is the bright and smooth equatorial region to the east of disk center that appears in the He I observation of Figure~\ref{F-sarnoff}a and is discussed in Section~\ref{S-Discussion}. The majority of candidates are predominantly bipolar, at 73.6\%, and are thresholded away in the final construction of STRIDE-CH with unipolarity-derived confidence.

\begin{table}
\caption{Categorization of candidate regions by unipolarity and smoothness as detected from April to August 2012 in the rise of solar cycle 24. The sample is further described in Section~\ref{S-Results}.
}
\label{T-unipolarity}
\begin{tabular}{lcc}     % define the column alignment
                           % l: left, c: center, r: right
\hline                     % horizontal line
        & Predominantly     & Predominantly \\
        & Bipolar ($U<0.5$) & Unipolar ($U \ge 0.5$) \\
\hline
Smoothness Below        & 41.0\%    & 9.0\%\\
the Median ($S<50$)     &           &       \\
Smoothness Above        & 32.6\%    & 17.4\%\\
the Median ($S \ge 50$) &           &       \\

\hline
\textbf{All Candidates} & 73.6\% & 26.4\% \\
\end{tabular}
\end{table}

\subsection{Selection of Unipolarity over Smoothness for Deriving Confidence}
Unipolarity-derived confidence was ultimately selected as it provided the ability to disambiguate CHs from quiet Sun, a physical motivation to confidence, and increased time series consistency. As the evolution of CHs on the time scale of days is often modest, consistency in CHs detected over time is a desirable characteristic of segmentation algorithms.

   % {\bf --- Time series consistency} \\
Consistency may be quantified to a first approximation by the autocorrelation in the time series of detected area $f$. It should be noted that relative increases demonstrate improved consistency, but a value of 1 is undesirable as it would indicate that the detected area is constant over time and the evolution of CHs has not been captured. The global detected area for a given date is computed by summation over $N$ detected pixels across the disk of the pixel-wise area $A^{(i)}$, which was defined in Equation~\ref{Eq-A-i-def}.

\begin{equation}  \label{Eq-A-def}
A = \sum_{i = 0}^{N} A^{(i)}
\end{equation}

Autocorrelation is then computed as the Pearson’s correlation coefficient of the time series $f$ with itself upon applying an index lag $n=1$, yielding a quantity $\in[0,1]$:

\begin{equation}  \label{Eq-autocorr-def}
(f * f)[n=1] = \frac{{\rm cov}(f[m],f[m+1])} {{\rm Var}(f)}
\end{equation}

   % {\bf --- Time series consistency improvements} \\
The time series consistency provided by distinct methods for deriving confidence is displayed in Table~\ref{T-consistency}. The explored methods include:

\begin{enumerate}
    \item Single Preliminary Mask: The previously detailed steps of data preparation, brightness threshold, morphological operations, filling of small holes in candidates, and small object removal are applied to derive a single segmentation. Examples are depicted in Figure~\ref{F-sarnoff-flowchart}b and~\ref{F-sarnoff-flowchart}d.
    \item Area-Derived Confidence: A collection of preliminary detection masks form the members of an ensemble as in Figure~\ref{F-sarnoff-flowchart}e, but with confidence corresponding to low detected area. 100\% confidence is assigned to the most conservative global mask with the least detected area. Confidence assigned to masks with greater area uniformly descends to 0\%, which is assigned to the mask with greatest detected area.
    \item Smoothness-Derived Confidence: An ensemble of boundaries is compiled as in Figure~\ref{F-sarnoff-flowchart}e, but with confidence corresponding to the smoothness percentile $S$.
    \item Unipolarity-Derived Confidence: This method refers to the final construction of STRIDE-CH. Confidence in each candidate of the ensemble corresponds to the unipolarity $U$ of the magnetic field data within the region and a threshold is applied on unipolarity, as in Figure~\ref{F-sarnoff-flowchart}f.
\end{enumerate}

\begin{table}[H]
\caption{Comparison of the consistency between methods as quantified by the autocorrelation in the time series of detected area $(f * f)[n=1]\in[0,1]$, with a greater value indicating improved consistency. The methods are evaluated on observations from April to August 2012 in the rise of solar cycle 24. The sample is further described in Section~\ref{S-Results}.
}
\label{T-consistency}
\begin{tabular}{lcccc}     % define the column alignment
                           % l: left, c: center, r: right
\hline                     % horizontal line
Confidence  & Single        & Area          & Smoothness    & Unipolarity \\
Level       & Preliminary   & Derived       & Derived       & Derived \\
            & Mask          & Confidence    & Confidence    & Confidence \\
\hline
0\%         & 0.05              & 0.10          & 0.34              & 0.46 \\
50\%        & -                 & 0.12          & 0.27              & 0.46 \\
80\%        & -                 & 0.23          & 0.31              & 0.65 \\

\hline
\end{tabular}
\end{table}

   % {\bf --- Consistency improvement} \\
The addition of key elements in the algorithm results in marked improvements in the time series consistency of detections, as quantified in Table~\ref{T-consistency}. The gains realized by the implementation of even the most naïve ensemble method, with area-derived confidence that simply assigns greater confidence to conservatively detected small regions, doubles the autocorrelation in the lowest confidence level, with increasing values at higher confidence levels. This improved consistency demonstrate the benefits of ensemble boosting that have been developed by the machine learning community \citep{2012Zhou}. A lone preliminary segmentation may be considered a weak CH classifier with poor time series consistency but the aggregation of an ensemble may boost these segmentations into a unified, stronger classifier with improved consistency. Further rewards may be reaped by taking the radiative property of smoothness in the He I absorption into account, but the derivation of confidence from unipolarity results in the greatest autocorrelation values across confidence levels, adding a reason for the ultimate selection of unipolarity over smoothness as the confidence metric in ensemble map production.

In summary, unipolarity provides the following benefits over the smoothness criterion for allocating confidence towards candidate regions initially segmented by STRIDE-CH:

\begin{enumerate}
    \item The ability to resolve the CH-quiet Sun ambiguity is provided. Candidates that consist of quiet Sun will be predominantly bipolar, but may appear smooth and lead to false positive detections if judged by smoothness.
    \item Confidence becomes physically motivated. Unipolarity $U$ is derived from rigorous measurements of the magnetic field enclosed by a boundary and is an absolute quantity $\in$[0,1]. While the proper location of the unipolarity threshold may be a matter of debate, the selected value carries physical meaning. The smoothness percentile $S$, or any other such smoothness metric, is instead an emergent, visual property that is less comprehensively understood and is defined relative to the textures present amongst a sample of observations. Thus, a potential threshold would be less physically meaninigful.
    \item Unipolarity-derived confidence is shown in Table~\ref{T-consistency} to yield greater consistency over time in the total CH area detected by STRIDE-CH.
\end{enumerate}
  
%%% BIBLIOGRAPHY %%%%%%%%%%%%%%%%%%%%%%%%%%%%%%%%%%%%%%%%%%%%%%%%%%%%%%%%%%%

     % format of references provided by the journal (.bst)
\bibliographystyle{spr-mp-sola}
     % name your Bibtex file containing your references (.bib)
\bibliography{sola_bibliography}  

\begin{thebibliography}{71}
% BibTex style file: spr-mp-sola.bst, v2.06, 2023-07-27
\ifx\bisbn     \undefined \def\bisbn  #1{ISBN #1}\fi
\ifx\binits    \undefined \def\binits#1{#1}\fi
\ifx\bauthor   \undefined \def\bauthor#1{#1}\fi
\ifx\batitle   \undefined \def\batitle#1{#1}\fi
\ifx\bjtitle   \undefined \def\bjtitle#1{\textit{#1}}\fi
\ifx\bvolume   \undefined \def\bvolume#1{\textbf{#1}}\fi
\ifx\byear     \undefined \def\byear#1{#1}\fi
\ifx\bissue    \undefined \def\bissue#1{#1}\fi
\ifx\bfpage    \undefined \def\bfpage#1{#1}\fi
\ifx\blpage    \undefined \def\blpage #1{#1}\fi
\ifx\burl      \undefined \def\burl#1{#1}\fi
\ifx\href      \undefined \def\href#1#2{#2}\fi
\ifx\betal     \undefined \def\betal{et al.}\fi
\ifx\bctitle   \undefined \def\bctitle#1{#1}\fi
\ifx\beditor   \undefined \def\beditor#1{#1}\fi
\ifx\bbtitle   \undefined \def\bbtitle#1{\textit{#1}}\fi
\ifx\bedition  \undefined \def\bedition#1{#1}\fi
\ifx\bseriesno \undefined \def\bseriesno#1{\textbf{#1}}\fi
\ifx\blocation \undefined \def\blocation#1{#1}\fi
\ifx\bsertitle \undefined \def\bsertitle#1{\textit{#1}}\fi
\ifx\bsnm      \undefined \def\bsnm#1{#1}\fi
\ifx\bsuffix   \undefined \def\bsuffix#1{#1}\fi
\ifx\bparticle \undefined \def\bparticle#1{#1}\fi
\ifx\barticle  \undefined \def\barticle#1{}\fi
\ifx\binstitute  \undefined \def\binstitute#1{#1}\fi
\ifx\bpublisher  \undefined \def\bpublisher#1{#1}\fi
\ifx\doiurl    \undefined \def\doiurl#1{\href{#1}{DOI}}\fi
\makeatletter
\def\safeHref#1#2#3{\in@{http}{#2}\ifin@\href{#2}{#3}\else\href{#1#2}{#3}\fi}
\makeatother
\ifx\adsurl    \undefined
  \def\adsurl#1{\safeHref{https://ui.adsabs.harvard.edu/abs/}{#1}{ADS}}\fi
\ifx\arxivurl  \undefined
  \def\arxivurl#1{\safeHref{http://arxiv.org/abs/}{#1}{arXiv}}\fi
\ifx\botherref \undefined \def\botherref#1{}\fi
\ifx\url       \undefined \def\url#1{#1}\fi
\ifx\bchapter  \undefined \def\bchapter#1{}\fi
\ifx\bbook     \undefined \def\bbook#1{}\fi
\ifx\bcomment  \undefined \def\bcomment#1{#1}\fi
\ifx\oauthor   \undefined \def\oauthor#1{#1}\fi
\ifx\citeauthoryear \undefined\def \citeauthoryear#1{#1}\fi
\def\endbibitem {}
\ifx\bconflocation  \undefined \def\bconflocation#1{#1} \fi

\bibitem[\protect\citeauthoryear{{Andretta} and
  {Jones}}{1997}]{1997ApJ...489..375A}
\begin{barticle}
\bauthor{\bsnm{{Andretta}}, \binits{V.}},
\bauthor{\bsnm{{Jones}}, \binits{H.P.}}:
\byear{1997},
\batitle{{On the Role of the Solar Corona and Transition Region in the
  Excitation of the Spectrum of Neutral Helium}}.
\bjtitle{\apj}
\bvolume{489},
\bfpage{375}.
\doiurl{https://doi.org/10.1086/304760}.
\adsurl{1997ApJ...489..375A}.
\end{barticle}
\endbibitem

\bibitem[\protect\citeauthoryear{{Arge} and
  {Pizzo}}{2000}]{2000JGR...10510465A}
\begin{barticle}
\bauthor{\bsnm{{Arge}}, \binits{C.N.}},
\bauthor{\bsnm{{Pizzo}}, \binits{V.J.}}:
\byear{2000},
\batitle{{Improvement in the prediction of solar wind conditions using
  near-real time solar magnetic field updates}}.
\bjtitle{\jgr}
\bvolume{105},
\bfpage{10465}.
\doiurl{https://doi.org/10.1029/1999JA000262}.
\adsurl{2000JGR...10510465A}.
\end{barticle}
\endbibitem

\bibitem[\protect\citeauthoryear{{Arge} et~al.}{2004}]{2004JASTP..66.1295A}
\begin{barticle}
\bauthor{\bsnm{{Arge}}, \binits{C.N.}},
\bauthor{\bsnm{{Luhmann}}, \binits{J.G.}},
\bauthor{\bsnm{{Odstrcil}}, \binits{D.}},
\bauthor{\bsnm{{Schrijver}}, \binits{C.J.}},
\bauthor{\bsnm{{Li}}, \binits{Y.}}:
\byear{2004},
\batitle{{Stream structure and coronal sources of the solar wind during the May
  12th, 1997 CME}}.
\bjtitle{Journal of Atmospheric and Solar-Terrestrial Physics}
\bvolume{66},
\bfpage{1295}.
\doiurl{https://doi.org/10.1016/j.jastp.2004.03.018}.
\adsurl{2004JASTP..66.1295A}.
\end{barticle}
\endbibitem

\bibitem[\protect\citeauthoryear{{Arge} et~al.}{2024}]{2024ApJ...964..115A}
\begin{barticle}
\bauthor{\bsnm{{Arge}}, \binits{C.N.}},
\bauthor{\bsnm{{Leisner}}, \binits{A.}},
\bauthor{\bsnm{{Antiochos}}, \binits{S.K.}},
\bauthor{\bsnm{{Wallace}}, \binits{S.}},
\bauthor{\bsnm{{Henney}}, \binits{C.J.}}:
\byear{2024},
\batitle{{Proposed Resolution to the Solar Open Magnetic Flux Problem}}.
\bjtitle{\apj}
\bvolume{964},
\bfpage{115}.
\doiurl{https://doi.org/10.3847/1538-4357/ad20e2}.
\adsurl{2024ApJ...964..115A}.
\end{barticle}
\endbibitem

\bibitem[\protect\citeauthoryear{{Astropy Collaboration}
  et~al.}{2022}]{2022ApJ...935..167A}
\begin{barticle}
\bauthor{\bsnm{{Astropy Collaboration}}},
\bauthor{\bsnm{{Price-Whelan}}, \binits{A.M.}},
\bauthor{\bsnm{{Lim}}, \binits{P.L.}},
\bauthor{\bsnm{{Earl}}, \binits{N.}},
\bauthor{\bsnm{{Starkman}}, \binits{N.}},
\bauthor{\bsnm{{Bradley}}, \binits{L.}},
\bauthor{\bsnm{{Shupe}}, \binits{D.L.}},
\bauthor{\bsnm{{Patil}}, \binits{A.A.}},
\bauthor{\bsnm{{Corrales}}, \binits{L.}},
\bauthor{\bsnm{{Brasseur}}, \binits{C.E.}},
\bauthor{\bsnm{{N{\"o}the}}, \binits{M.}},
\bauthor{\bsnm{{Donath}}, \binits{A.}},
\bauthor{\bsnm{{Tollerud}}, \binits{E.}},
\bauthor{\bsnm{{Morris}}, \binits{B.M.}},
\bauthor{\bsnm{{Ginsburg}}, \binits{A.}},
\bauthor{\bsnm{{Vaher}}, \binits{E.}},
\bauthor{\bsnm{{Weaver}}, \binits{B.A.}},
\bauthor{\bsnm{{Tocknell}}, \binits{J.}},
\bauthor{\bsnm{{Jamieson}}, \binits{W.}},
\bauthor{\bsnm{{van Kerkwijk}}, \binits{M.H.}},
\bauthor{\bsnm{{Robitaille}}, \binits{T.P.}},
\bauthor{\bsnm{{Merry}}, \binits{B.}},
\bauthor{\bsnm{{Bachetti}}, \binits{M.}},
\bauthor{\bsnm{{G{\"u}nther}}, \binits{H.M.}},
\bauthor{\bsnm{{Aldcroft}}, \binits{T.L.}},
\bauthor{\bsnm{{Alvarado-Montes}}, \binits{J.A.}},
\bauthor{\bsnm{{Archibald}}, \binits{A.M.}},
\bauthor{\bsnm{{B{\'o}di}}, \binits{A.}},
\bauthor{\bsnm{{Bapat}}, \binits{S.}},
\bauthor{\bsnm{{Barentsen}}, \binits{G.}},
\bauthor{\bsnm{{Baz{\'a}n}}, \binits{J.}},
\bauthor{\bsnm{{Biswas}}, \binits{M.}},
\bauthor{\bsnm{{Boquien}}, \binits{M.}},
\bauthor{\bsnm{{Burke}}, \binits{D.J.}},
\bauthor{\bsnm{{Cara}}, \binits{D.}},
\bauthor{\bsnm{{Cara}}, \binits{M.}},
\bauthor{\bsnm{{Conroy}}, \binits{K.E.}},
\bauthor{\bsnm{{Conseil}}, \binits{S.}},
\bauthor{\bsnm{{Craig}}, \binits{M.W.}},
\bauthor{\bsnm{{Cross}}, \binits{R.M.}},
\bauthor{\bsnm{{Cruz}}, \binits{K.L.}},
\bauthor{\bsnm{{D'Eugenio}}, \binits{F.}},
\bauthor{\bsnm{{Dencheva}}, \binits{N.}},
\bauthor{\bsnm{{Devillepoix}}, \binits{H.A.R.}},
\bauthor{\bsnm{{Dietrich}}, \binits{J.P.}},
\bauthor{\bsnm{{Eigenbrot}}, \binits{A.D.}},
\bauthor{\bsnm{{Erben}}, \binits{T.}},
\bauthor{\bsnm{{Ferreira}}, \binits{L.}},
\bauthor{\bsnm{{Foreman-Mackey}}, \binits{D.}},
\bauthor{\bsnm{{Fox}}, \binits{R.}},
\bauthor{\bsnm{{Freij}}, \binits{N.}},
\bauthor{\bsnm{{Garg}}, \binits{S.}},
\bauthor{\bsnm{{Geda}}, \binits{R.}},
\bauthor{\bsnm{{Glattly}}, \binits{L.}},
\bauthor{\bsnm{{Gondhalekar}}, \binits{Y.}},
\bauthor{\bsnm{{Gordon}}, \binits{K.D.}},
\bauthor{\bsnm{{Grant}}, \binits{D.}},
\bauthor{\bsnm{{Greenfield}}, \binits{P.}},
\bauthor{\bsnm{{Groener}}, \binits{A.M.}},
\bauthor{\bsnm{{Guest}}, \binits{S.}},
\bauthor{\bsnm{{Gurovich}}, \binits{S.}},
\bauthor{\bsnm{{Handberg}}, \binits{R.}},
\bauthor{\bsnm{{Hart}}, \binits{A.}},
\bauthor{\bsnm{{Hatfield-Dodds}}, \binits{Z.}},
\bauthor{\bsnm{{Homeier}}, \binits{D.}},
\bauthor{\bsnm{{Hosseinzadeh}}, \binits{G.}},
\bauthor{\bsnm{{Jenness}}, \binits{T.}},
\bauthor{\bsnm{{Jones}}, \binits{C.K.}},
\bauthor{\bsnm{{Joseph}}, \binits{P.}},
\bauthor{\bsnm{{Kalmbach}}, \binits{J.B.}},
\bauthor{\bsnm{{Karamehmetoglu}}, \binits{E.}},
\bauthor{\bsnm{{Ka{\l}uszy{\'n}ski}}, \binits{M.}},
\bauthor{\bsnm{{Kelley}}, \binits{M.S.P.}},
\bauthor{\bsnm{{Kern}}, \binits{N.}},
\bauthor{\bsnm{{Kerzendorf}}, \binits{W.E.}},
\bauthor{\bsnm{{Koch}}, \binits{E.W.}},
\bauthor{\bsnm{{Kulumani}}, \binits{S.}},
\bauthor{\bsnm{{Lee}}, \binits{A.}},
\bauthor{\bsnm{{Ly}}, \binits{C.}},
\bauthor{\bsnm{{Ma}}, \binits{Z.}},
\bauthor{\bsnm{{MacBride}}, \binits{C.}},
\bauthor{\bsnm{{Maljaars}}, \binits{J.M.}},
\bauthor{\bsnm{{Muna}}, \binits{D.}},
\bauthor{\bsnm{{Murphy}}, \binits{N.A.}},
\bauthor{\bsnm{{Norman}}, \binits{H.}},
\bauthor{\bsnm{{O'Steen}}, \binits{R.}},
\bauthor{\bsnm{{Oman}}, \binits{K.A.}},
\bauthor{\bsnm{{Pacifici}}, \binits{C.}},
\bauthor{\bsnm{{Pascual}}, \binits{S.}},
\bauthor{\bsnm{{Pascual-Granado}}, \binits{J.}},
\bauthor{\bsnm{{Patil}}, \binits{R.R.}},
\bauthor{\bsnm{{Perren}}, \binits{G.I.}},
\bauthor{\bsnm{{Pickering}}, \binits{T.E.}},
\bauthor{\bsnm{{Rastogi}}, \binits{T.}},
\bauthor{\bsnm{{Roulston}}, \binits{B.R.}},
\bauthor{\bsnm{{Ryan}}, \binits{D.F.}},
\bauthor{\bsnm{{Rykoff}}, \binits{E.S.}},
\bauthor{\bsnm{{Sabater}}, \binits{J.}},
\bauthor{\bsnm{{Sakurikar}}, \binits{P.}},
\bauthor{\bsnm{{Salgado}}, \binits{J.}},
\bauthor{\bsnm{{Sanghi}}, \binits{A.}},
\bauthor{\bsnm{{Saunders}}, \binits{N.}},
\bauthor{\bsnm{{Savchenko}}, \binits{V.}},
\bauthor{\bsnm{{Schwardt}}, \binits{L.}},
\bauthor{\bsnm{{Seifert-Eckert}}, \binits{M.}},
\bauthor{\bsnm{{Shih}}, \binits{A.Y.}},
\bauthor{\bsnm{{Jain}}, \binits{A.S.}},
\bauthor{\bsnm{{Shukla}}, \binits{G.}},
\bauthor{\bsnm{{Sick}}, \binits{J.}},
\bauthor{\bsnm{{Simpson}}, \binits{C.}},
\bauthor{\bsnm{{Singanamalla}}, \binits{S.}},
\bauthor{\bsnm{{Singer}}, \binits{L.P.}},
\bauthor{\bsnm{{Singhal}}, \binits{J.}},
\bauthor{\bsnm{{Sinha}}, \binits{M.}},
\bauthor{\bsnm{{Sip{\H{o}}cz}}, \binits{B.M.}},
\bauthor{\bsnm{{Spitler}}, \binits{L.R.}},
\bauthor{\bsnm{{Stansby}}, \binits{D.}},
\bauthor{\bsnm{{Streicher}}, \binits{O.}},
\bauthor{\bsnm{{{\v{S}}umak}}, \binits{J.}},
\bauthor{\bsnm{{Swinbank}}, \binits{J.D.}},
\bauthor{\bsnm{{Taranu}}, \binits{D.S.}},
\bauthor{\bsnm{{Tewary}}, \binits{N.}},
\bauthor{\bsnm{{Tremblay}}, \binits{G.R.}},
\bauthor{\bsnm{{de Val-Borro}}, \binits{M.}},
\bauthor{\bsnm{{Van Kooten}}, \binits{S.J.}},
\bauthor{\bsnm{{Vasovi{\'c}}}, \binits{Z.}},
\bauthor{\bsnm{{Verma}}, \binits{S.}},
\bauthor{\bsnm{{de Miranda Cardoso}}, \binits{J.V.}},
\bauthor{\bsnm{{Williams}}, \binits{P.K.G.}},
\bauthor{\bsnm{{Wilson}}, \binits{T.J.}},
\bauthor{\bsnm{{Winkel}}, \binits{B.}},
\bauthor{\bsnm{{Wood-Vasey}}, \binits{W.M.}},
\bauthor{\bsnm{{Xue}}, \binits{R.}},
\bauthor{\bsnm{{Yoachim}}, \binits{P.}},
\bauthor{\bsnm{{Zhang}}, \binits{C.}},
\bauthor{\bsnm{{Zonca}}, \binits{A.}},
\bauthor{\bsnm{{Astropy Project Contributors}}}:
\byear{2022},
\batitle{{The Astropy Project: Sustaining and Growing a Community-oriented
  Open-source Project and the Latest Major Release (v5.0) of the Core
  Package}}.
\bjtitle{\apj}
\bvolume{935},
\bfpage{167}.
\doiurl{https://doi.org/10.3847/1538-4357/ac7c74}.
\adsurl{2022ApJ...935..167A}.
\end{barticle}
\endbibitem

\bibitem[\protect\citeauthoryear{Barra, Delouille, and
  Hochedez}{2007}]{2007Barra}
\begin{bchapter}
\bauthor{\bsnm{Barra}, \binits{V.}},
\bauthor{\bsnm{Delouille}, \binits{V.}},
\bauthor{\bsnm{Hochedez}, \binits{J.-F.}}:
\byear{2007},
\bctitle{Segmentation of Extreme Ultraviolet Solar Images using a Multispectral
  Data Fusion Process}.
In: \bbtitle{2007 IEEE International Fuzzy Systems Conference},
\bfpage{1}.
\doiurl{https://doi.org/10.1109/FUZZY.2007.4295367}.
\end{bchapter}
\endbibitem

\bibitem[\protect\citeauthoryear{{Boucheron}, {Valluri}, and
  {McAteer}}{2016}]{2016SoPh..291.2353B}
\begin{barticle}
\bauthor{\bsnm{{Boucheron}}, \binits{L.E.}},
\bauthor{\bsnm{{Valluri}}, \binits{M.}},
\bauthor{\bsnm{{McAteer}}, \binits{R.T.J.}}:
\byear{2016},
\batitle{{Segmentation of Coronal Holes Using Active Contours Without Edges}}.
\bjtitle{\solphys}
\bvolume{291},
\bfpage{2353}.
\doiurl{https://doi.org/10.1007/s11207-016-0985-z}.
\adsurl{2016SoPh..291.2353B}.
\end{barticle}
\endbibitem

\bibitem[\protect\citeauthoryear{{Braj{\v{s}}a}
  et~al.}{1996}]{1996SoPh..163...79B}
\begin{barticle}
\bauthor{\bsnm{{Braj{\v{s}}a}}, \binits{R.}},
\bauthor{\bsnm{{Pohjolainen}}, \binits{S.}},
\bauthor{\bsnm{{Ru{\v{z}}djak}}, \binits{V.}},
\bauthor{\bsnm{{Sakurai}}, \binits{T.}},
\bauthor{\bsnm{{Urpo}}, \binits{S.}},
\bauthor{\bsnm{{Vr{\v{s}}nak}}, \binits{B.}},
\bauthor{\bsnm{{W{\"o}hl}}, \binits{H.}}:
\byear{1996},
\batitle{{Helium 10830 {\r{A}} measurements of the Sun}}.
\bjtitle{\solphys}
\bvolume{163},
\bfpage{79}.
\doiurl{https://doi.org/10.1007/BF00165457}.
\adsurl{1996SoPh..163...79B}.
\end{barticle}
\endbibitem

\bibitem[\protect\citeauthoryear{Brunton and Kutz}{2022}]{Brunton_Kutz_2022}
\begin{bbook}
\bauthor{\bsnm{Brunton}, \binits{S.L.}},
\bauthor{\bsnm{Kutz}, \binits{J.N.}}:
\byear{2022},
\bbtitle{Data-Driven Science and Engineering: Machine Learning, Dynamical
  Systems, and Control},
\bedition{2}nd edn.
\bpublisher{Cambridge University Press}.
\end{bbook}
\endbibitem

\bibitem[\protect\citeauthoryear{{Caplan}, {Downs}, and
  {Linker}}{2016}]{2016ApJ...823...53C}
\begin{barticle}
\bauthor{\bsnm{{Caplan}}, \binits{R.M.}},
\bauthor{\bsnm{{Downs}}, \binits{C.}},
\bauthor{\bsnm{{Linker}}, \binits{J.A.}}:
\byear{2016},
\batitle{{Synchronic Coronal Hole Mapping Using Multi-instrument EUV Images:
  Data Preparation and Detection Method}}.
\bjtitle{\apj}
\bvolume{823},
\bfpage{53}.
\doiurl{https://doi.org/10.3847/0004-637X/823/1/53}.
\adsurl{2016ApJ...823...53C}.
\end{barticle}
\endbibitem

\bibitem[\protect\citeauthoryear{{Caplan} et~al.}{2023}]{2023ApJ...958...43C}
\begin{barticle}
\bauthor{\bsnm{{Caplan}}, \binits{R.M.}},
\bauthor{\bsnm{{Mason}}, \binits{E.I.}},
\bauthor{\bsnm{{Downs}}, \binits{C.}},
\bauthor{\bsnm{{Linker}}, \binits{J.A.}}:
\byear{2023},
\batitle{{Improving Coronal Hole Detections and Open Flux Estimates}}.
\bjtitle{\apj}
\bvolume{958},
\bfpage{43}.
\doiurl{https://doi.org/10.3847/1538-4357/ad01b6}.
\adsurl{2023ApJ...958...43C}.
\end{barticle}
\endbibitem

\bibitem[\protect\citeauthoryear{{Cranmer}}{2009}]{2009LRSP....6....3C}
\begin{barticle}
\bauthor{\bsnm{{Cranmer}}, \binits{S.R.}}:
\byear{2009},
\batitle{{Coronal Holes}}.
\bjtitle{Living Reviews in Solar Physics}
\bvolume{6},
\bfpage{3}.
\doiurl{https://doi.org/10.12942/lrsp-2009-3}.
\adsurl{2009LRSP....6....3C}.
\end{barticle}
\endbibitem

\bibitem[\protect\citeauthoryear{{Czolbe} et~al.}{2021}]{2021arXiv210316265C}
\begin{botherref}
\oauthor{\bsnm{{Czolbe}}, \binits{S.}},
\oauthor{\bsnm{{Arnavaz}}, \binits{K.}},
\oauthor{\bsnm{{Krause}}, \binits{O.}},
\oauthor{\bsnm{{Feragen}}, \binits{A.}}:
2021,
{Is segmentation uncertainty useful?}
\textit{arXiv e-prints},
arXiv:2103.16265.
\doiurl{https://doi.org/10.48550/arXiv.2103.16265}.
\adsurl{2021arXiv210316265C}.
\end{botherref}
\endbibitem

\bibitem[\protect\citeauthoryear{{de Toma} and
  {Arge}}{2005}]{2005ASPC..346..251T}
\begin{bchapter}
\bauthor{\bsnm{{de Toma}}, \binits{G.}},
\bauthor{\bsnm{{Arge}}, \binits{C.N.}}:
\byear{2005},
\bctitle{{Multi--wavelength Observations of Coronal Holes}}.
In: \beditor{\bsnm{{Sankarasubramanian}}, \binits{K.}},
\beditor{\bsnm{{Penn}}, \binits{M.}},
\beditor{\bsnm{{Pevtsov}}, \binits{A.}} (eds.)
\bbtitle{Large-scale Structures and their Role in Solar Activity},
\bsertitle{Astronomical Society of the Pacific Conference Series}
\bseriesno{346},
\bfpage{251}.
\adsurl{2005ASPC..346..251T}.
\end{bchapter}
\endbibitem

\bibitem[\protect\citeauthoryear{{de Wijn} et~al.}{2023}]{2023BAAS...55c.088D}
\begin{bchapter}
\bauthor{\bsnm{{de Wijn}}, \binits{A.}},
\bauthor{\bsnm{{Tomczyk}}, \binits{S.}},
\bauthor{\bsnm{{Burkepile}}, \binits{J.}},
\bauthor{\bsnm{{Casini}}, \binits{R.}},
\bauthor{\bsnm{{de Toma}}, \binits{G.}}:
\byear{2023},
\bctitle{{COSMO ChroMag: The Chromosphere and Prominence Magnetometer}}.
In: \bbtitle{Bulletin of the American Astronomical Society}
\bseriesno{55},
\bfpage{088}.
\doiurl{https://doi.org/10.3847/25c2cfeb.86f73330}.
\adsurl{2023BAAS...55c.088D}.
\end{bchapter}
\endbibitem

\bibitem[\protect\citeauthoryear{{Garton}, {Gallagher}, and
  {Murray}}{2018}]{2018JSWSC...8A...2G}
\begin{barticle}
\bauthor{\bsnm{{Garton}}, \binits{T.M.}},
\bauthor{\bsnm{{Gallagher}}, \binits{P.T.}},
\bauthor{\bsnm{{Murray}}, \binits{S.A.}}:
\byear{2018},
\batitle{{Automated coronal hole identification via multi-thermal intensity
  segmentation}}.
\bjtitle{Journal of Space Weather and Space Climate}
\bvolume{8},
\bfpage{A02}.
\doiurl{https://doi.org/10.1051/swsc/2017039}.
\adsurl{2018JSWSC...8A...2G}.
\end{barticle}
\endbibitem

\bibitem[\protect\citeauthoryear{Gonzalez and
  Woods}{2018}]{gonzalez2018digital}
\begin{bbook}
\bauthor{\bsnm{Gonzalez}, \binits{R.C.}},
\bauthor{\bsnm{Woods}, \binits{R.E.}}:
\byear{2018},
\bbtitle{Digital Image Processing},
\bpublisher{Pearson}.
\bisbn{9780133356724}.
\burl{https://books.google.com/books?id=0F05vgAACAAJ}.
\end{bbook}
\endbibitem

\bibitem[\protect\citeauthoryear{{Grajeda} et~al.}{2023}]{2023SoPh..298..133G}
\begin{barticle}
\bauthor{\bsnm{{Grajeda}}, \binits{J.A.}},
\bauthor{\bsnm{{Boucheron}}, \binits{L.E.}},
\bauthor{\bsnm{{Kirk}}, \binits{M.S.}},
\bauthor{\bsnm{{Leisner}}, \binits{A.}},
\bauthor{\bsnm{{Arge}}, \binits{C.N.}}:
\byear{2023},
\batitle{{Quantifying the Consistency and Characterizing the Confidence of
  Coronal Holes Detected by Active Contours Without Edges (ACWE)}}.
\bjtitle{\solphys}
\bvolume{298},
\bfpage{133}.
\doiurl{https://doi.org/10.1007/s11207-023-02228-0}.
\adsurl{2023SoPh..298..133G}.
\end{barticle}
\endbibitem

\bibitem[\protect\citeauthoryear{{Hamada} et~al.}{2018}]{2018SoPh..293...71H}
\begin{barticle}
\bauthor{\bsnm{{Hamada}}, \binits{A.}},
\bauthor{\bsnm{{Asikainen}}, \binits{T.}},
\bauthor{\bsnm{{Virtanen}}, \binits{I.}},
\bauthor{\bsnm{{Mursula}}, \binits{K.}}:
\byear{2018},
\batitle{{Automated Identification of Coronal Holes from Synoptic EUV Maps}}.
\bjtitle{\solphys}
\bvolume{293},
\bfpage{71}.
\doiurl{https://doi.org/10.1007/s11207-018-1289-2}.
\adsurl{2018SoPh..293...71H}.
\end{barticle}
\endbibitem

\bibitem[\protect\citeauthoryear{Harris et~al.}{2020}]{harris2020array}
\begin{barticle}
\bauthor{\bsnm{Harris}, \binits{C.R.}},
\bauthor{\bsnm{Millman}, \binits{K.J.}},
\bauthor{\bparticle{van~der} \bsnm{Walt}, \binits{S.J.}},
\bauthor{\bsnm{Gommers}, \binits{R.}},
\bauthor{\bsnm{Virtanen}, \binits{P.}},
\bauthor{\bsnm{Cournapeau}, \binits{D.}},
\bauthor{\bsnm{Wieser}, \binits{E.}},
\bauthor{\bsnm{Taylor}, \binits{J.}},
\bauthor{\bsnm{Berg}, \binits{S.}},
\bauthor{\bsnm{Smith}, \binits{N.J.}},
\bauthor{\bsnm{Kern}, \binits{R.}},
\bauthor{\bsnm{Picus}, \binits{M.}},
\bauthor{\bsnm{Hoyer}, \binits{S.}},
\bauthor{\bparticle{van} \bsnm{Kerkwijk}, \binits{M.H.}},
\bauthor{\bsnm{Brett}, \binits{M.}},
\bauthor{\bsnm{Haldane}, \binits{A.}},
\bauthor{\bparticle{del} \bsnm{R{\'{i}}o}, \binits{J.F.}},
\bauthor{\bsnm{Wiebe}, \binits{M.}},
\bauthor{\bsnm{Peterson}, \binits{P.}},
\bauthor{\bsnm{G{\'{e}}rard-Marchant}, \binits{P.}},
\bauthor{\bsnm{Sheppard}, \binits{K.}},
\bauthor{\bsnm{Reddy}, \binits{T.}},
\bauthor{\bsnm{Weckesser}, \binits{W.}},
\bauthor{\bsnm{Abbasi}, \binits{H.}},
\bauthor{\bsnm{Gohlke}, \binits{C.}},
\bauthor{\bsnm{Oliphant}, \binits{T.E.}}:
\byear{2020},
\batitle{Array programming with {NumPy}}.
\bjtitle{Nature}
\bvolume{585},
\bfpage{357}.
\doiurl{https://doi.org/10.1038/s41586-020-2649-2}.
\burl{https://doi.org/10.1038/s41586-020-2649-2}.
\end{barticle}
\endbibitem

\bibitem[\protect\citeauthoryear{{Harvey} and
  {Recely}}{2002}]{2002SoPh..211...31H}
\begin{barticle}
\bauthor{\bsnm{{Harvey}}, \binits{K.L.}},
\bauthor{\bsnm{{Recely}}, \binits{F.}}:
\byear{2002},
\batitle{{Polar Coronal Holes During Cycles 22 and 23}}.
\bjtitle{\solphys}
\bvolume{211},
\bfpage{31}.
\doiurl{https://doi.org/10.1023/A:1022469023581}.
\adsurl{2002SoPh..211...31H}.
\end{barticle}
\endbibitem

\bibitem[\protect\citeauthoryear{{Heinemann}
  et~al.}{2019}]{2019SoPh..294..144H}
\begin{barticle}
\bauthor{\bsnm{{Heinemann}}, \binits{S.G.}},
\bauthor{\bsnm{{Temmer}}, \binits{M.}},
\bauthor{\bsnm{{Heinemann}}, \binits{N.}},
\bauthor{\bsnm{{Dissauer}}, \binits{K.}},
\bauthor{\bsnm{{Samara}}, \binits{E.}},
\bauthor{\bsnm{{Jer{\v{c}}i{\'c}}}, \binits{V.}},
\bauthor{\bsnm{{Hofmeister}}, \binits{S.J.}},
\bauthor{\bsnm{{Veronig}}, \binits{A.M.}}:
\byear{2019},
\batitle{{Statistical Analysis and Catalog of Non-polar Coronal Holes Covering
  the SDO-Era Using CATCH}}.
\bjtitle{\solphys}
\bvolume{294},
\bfpage{144}.
\doiurl{https://doi.org/10.1007/s11207-019-1539-y}.
\adsurl{2019SoPh..294..144H}.
\end{barticle}
\endbibitem

\bibitem[\protect\citeauthoryear{{Henney} and
  {Harvey}}{2005}]{2005ASPC..346..261H}
\begin{bchapter}
\bauthor{\bsnm{{Henney}}, \binits{C.J.}},
\bauthor{\bsnm{{Harvey}}, \binits{J.W.}}:
\byear{2005},
\bctitle{{Automated Coronal Hole Detection using He 1083 nm Spectroheliograms
  and Photospheric Magnetograms}}.
In: \beditor{\bsnm{{Sankarasubramanian}}, \binits{K.}},
\beditor{\bsnm{{Penn}}, \binits{M.}},
\beditor{\bsnm{{Pevtsov}}, \binits{A.}} (eds.)
\bbtitle{Large-scale Structures and their Role in Solar Activity},
\bsertitle{Astronomical Society of the Pacific Conference Series}
\bseriesno{346},
\bfpage{261}.
\doiurl{https://doi.org/10.48550/arXiv.astro-ph/0701122}.
\adsurl{2005ASPC..346..261H}.
\end{bchapter}
\endbibitem

\bibitem[\protect\citeauthoryear{{Hofmeister}
  et~al.}{2017}]{2017ApJ...835..268H}
\begin{barticle}
\bauthor{\bsnm{{Hofmeister}}, \binits{S.J.}},
\bauthor{\bsnm{{Veronig}}, \binits{A.}},
\bauthor{\bsnm{{Reiss}}, \binits{M.A.}},
\bauthor{\bsnm{{Temmer}}, \binits{M.}},
\bauthor{\bsnm{{Vennerstrom}}, \binits{S.}},
\bauthor{\bsnm{{Vr{\v{s}}nak}}, \binits{B.}},
\bauthor{\bsnm{{Heber}}, \binits{B.}}:
\byear{2017},
\batitle{{Characteristics of Low-latitude Coronal Holes near the Maximum of
  Solar Cycle 24}}.
\bjtitle{\apj}
\bvolume{835},
\bfpage{268}.
\doiurl{https://doi.org/10.3847/1538-4357/835/2/268}.
\adsurl{2017ApJ...835..268H}.
\end{barticle}
\endbibitem

\bibitem[\protect\citeauthoryear{{Hofmeister}
  et~al.}{2019}]{2019A&A...629A..22H}
\begin{barticle}
\bauthor{\bsnm{{Hofmeister}}, \binits{S.J.}},
\bauthor{\bsnm{{Utz}}, \binits{D.}},
\bauthor{\bsnm{{Heinemann}}, \binits{S.G.}},
\bauthor{\bsnm{{Veronig}}, \binits{A.}},
\bauthor{\bsnm{{Temmer}}, \binits{M.}}:
\byear{2019},
\batitle{{Photospheric magnetic structure of coronal holes}}.
\bjtitle{\aap}
\bvolume{629},
\bfpage{A22}.
\doiurl{https://doi.org/10.1051/0004-6361/201935918}.
\adsurl{2019A&A...629A..22H}.
\end{barticle}
\endbibitem

\bibitem[\protect\citeauthoryear{{Hong} et~al.}{2014}]{2014AGUFMSH21A4089H}
\begin{bchapter}
\bauthor{\bsnm{{Hong}}, \binits{S.}},
\bauthor{\bsnm{{Kim}}, \binits{J.}},
\bauthor{\bsnm{{Han}}, \binits{J.}},
\bauthor{\bsnm{{Kim}}, \binits{Y.}}:
\byear{2014},
\bctitle{{The Automatic Solar Synoptic Analyzer and Solar Wind Prediction}}.
In: \bbtitle{AGU Fall Meeting Abstracts}
\bseriesno{2014},
\bfpage{SH21A}.
\adsurl{2014AGUFMSH21A4089H}.
\end{bchapter}
\endbibitem

\bibitem[\protect\citeauthoryear{Hunter}{2007}]{Hunter:2007}
\begin{barticle}
\bauthor{\bsnm{Hunter}, \binits{J.D.}}:
\byear{2007},
\batitle{Matplotlib: A 2D graphics environment}.
\bjtitle{Computing in Science \& Engineering}
\bvolume{9},
\bfpage{90}.
\doiurl{https://doi.org/10.1109/MCSE.2007.55}.
\end{barticle}
\endbibitem

\bibitem[\protect\citeauthoryear{{Illarionov} and
  {Tlatov}}{2018}]{2018MNRAS.481.5014I}
\begin{barticle}
\bauthor{\bsnm{{Illarionov}}, \binits{E.A.}},
\bauthor{\bsnm{{Tlatov}}, \binits{A.G.}}:
\byear{2018},
\batitle{{Segmentation of coronal holes in solar disc images with a
  convolutional neural network}}.
\bjtitle{\mnras}
\bvolume{481},
\bfpage{5014}.
\doiurl{https://doi.org/10.1093/mnras/sty2628}.
\adsurl{2018MNRAS.481.5014I}.
\end{barticle}
\endbibitem

\bibitem[\protect\citeauthoryear{{Issan} et~al.}{2023}]{2023SpWea..2103555I}
\begin{barticle}
\bauthor{\bsnm{{Issan}}, \binits{O.}},
\bauthor{\bsnm{{Riley}}, \binits{P.}},
\bauthor{\bsnm{{Camporeale}}, \binits{E.}},
\bauthor{\bsnm{{Kramer}}, \binits{B.}}:
\byear{2023},
\batitle{{Bayesian Inference and Global Sensitivity Analysis for Ambient Solar
  Wind Prediction}}.
\bjtitle{Space Weather}
\bvolume{21},
\bfpage{e2023SW003555}.
\doiurl{https://doi.org/10.1029/2023SW003555}.
\adsurl{2023SpWea..2103555I}.
\end{barticle}
\endbibitem

\bibitem[\protect\citeauthoryear{{Jarolim} et~al.}{2021}]{2021A&A...652A..13J}
\begin{barticle}
\bauthor{\bsnm{{Jarolim}}, \binits{R.}},
\bauthor{\bsnm{{Veronig}}, \binits{A.M.}},
\bauthor{\bsnm{{Hofmeister}}, \binits{S.}},
\bauthor{\bsnm{{Heinemann}}, \binits{S.G.}},
\bauthor{\bsnm{{Temmer}}, \binits{M.}},
\bauthor{\bsnm{{Podladchikova}}, \binits{T.}},
\bauthor{\bsnm{{Dissauer}}, \binits{K.}}:
\byear{2021},
\batitle{{Multi-channel coronal hole detection with convolutional neural
  networks}}.
\bjtitle{\aap}
\bvolume{652},
\bfpage{A13}.
\doiurl{https://doi.org/10.1051/0004-6361/202140640}.
\adsurl{2021A&A...652A..13J}.
\end{barticle}
\endbibitem

\bibitem[\protect\citeauthoryear{{Jatla}}{2022}]{2022arXiv220710070J}
\begin{botherref}
\oauthor{\bsnm{{Jatla}}, \binits{V.}}:
2022,
{Automatic Segmentation of Coronal Holes in Solar Images and Solar Prediction
  Map Classification}.
\textit{arXiv e-prints},
arXiv:2207.10070.
\doiurl{https://doi.org/10.48550/arXiv.2207.10070}.
\adsurl{2022arXiv220710070J}.
\end{botherref}
\endbibitem

\bibitem[\protect\citeauthoryear{{Jones} et~al.}{2005}]{2005AGUSMSP51A..02J}
\begin{bchapter}
\bauthor{\bsnm{{Jones}}, \binits{H.P.}},
\bauthor{\bsnm{{Malanushenko}}, \binits{O.V.}},
\bauthor{\bsnm{{Harvey}}, \binits{J.W.}},
\bauthor{\bsnm{{Henney}}, \binits{C.J.}},
\bauthor{\bsnm{{Keller}}, \binits{C.U.}}:
\byear{2005},
\bctitle{{Reduction of SOLIS/Vector Spectromagnetograph He I 1083 nm
  Observations}}.
In: \bbtitle{AGU Spring Meeting Abstracts}
\bseriesno{2005},
\bfpage{SP51A}.
\adsurl{2005AGUSMSP51A..02J}.
\end{bchapter}
\endbibitem

\bibitem[\protect\citeauthoryear{{Jones}, {Uritsky}, and
  {Davila}}{2017}]{2017ApJ...844...93J}
\begin{barticle}
\bauthor{\bsnm{{Jones}}, \binits{S.I.}},
\bauthor{\bsnm{{Uritsky}}, \binits{V.}},
\bauthor{\bsnm{{Davila}}, \binits{J.M.}}:
\byear{2017},
\batitle{{Image-optimized Coronal Magnetic Field Models}}.
\bjtitle{\apj}
\bvolume{844},
\bfpage{93}.
\doiurl{https://doi.org/10.3847/1538-4357/aa7b7a}.
\adsurl{2017ApJ...844...93J}.
\end{barticle}
\endbibitem

\bibitem[\protect\citeauthoryear{{Kaiser} et~al.}{2008}]{2008SSRv..136....5K}
\begin{barticle}
\bauthor{\bsnm{{Kaiser}}, \binits{M.L.}},
\bauthor{\bsnm{{Kucera}}, \binits{T.A.}},
\bauthor{\bsnm{{Davila}}, \binits{J.M.}},
\bauthor{\bsnm{{St. Cyr}}, \binits{O.C.}},
\bauthor{\bsnm{{Guhathakurta}}, \binits{M.}},
\bauthor{\bsnm{{Christian}}, \binits{E.}}:
\byear{2008},
\batitle{{The STEREO Mission: An Introduction}}.
\bjtitle{\ssr}
\bvolume{136},
\bfpage{5}.
\doiurl{https://doi.org/10.1007/s11214-007-9277-0}.
\adsurl{2008SSRv..136....5K}.
\end{barticle}
\endbibitem

\bibitem[\protect\citeauthoryear{{Keller}, {Harvey}, and {Solis
  Team}}{2003}]{2003ASPC..307...13K}
\begin{bchapter}
\bauthor{\bsnm{{Keller}}, \binits{C.U.}},
\bauthor{\bsnm{{Harvey}}, \binits{J.W.}},
\bauthor{\bsnm{{Solis Team}}}:
\byear{2003},
\bctitle{{The SOLIS Vector-Spectromagnetograph}}.
In: \beditor{\bsnm{{Trujillo-Bueno}}, \binits{J.}},
\beditor{\bsnm{{Sanchez Almeida}}, \binits{J.}} (eds.)
\bbtitle{Solar Polarization},
\bsertitle{Astronomical Society of the Pacific Conference Series}
\bseriesno{307},
\bfpage{13}.
\adsurl{2003ASPC..307...13K}.
\end{bchapter}
\endbibitem

\bibitem[\protect\citeauthoryear{{Kirk} et~al.}{2009}]{2009SoPh..257...99K}
\begin{barticle}
\bauthor{\bsnm{{Kirk}}, \binits{M.S.}},
\bauthor{\bsnm{{Pesnell}}, \binits{W.D.}},
\bauthor{\bsnm{{Young}}, \binits{C.A.}},
\bauthor{\bsnm{{Hess Webber}}, \binits{S.A.}}:
\byear{2009},
\batitle{{Automated detection of EUV Polar Coronal Holes during Solar Cycle
  23}}.
\bjtitle{\solphys}
\bvolume{257},
\bfpage{99}.
\doiurl{https://doi.org/10.1007/s11207-009-9369-y}.
\adsurl{2009SoPh..257...99K}.
\end{barticle}
\endbibitem

\bibitem[\protect\citeauthoryear{{Ko} et~al.}{2014}]{2014ApJ...787..121K}
\begin{barticle}
\bauthor{\bsnm{{Ko}}, \binits{Y.-K.}},
\bauthor{\bsnm{{Muglach}}, \binits{K.}},
\bauthor{\bsnm{{Wang}}, \binits{Y.-M.}},
\bauthor{\bsnm{{Young}}, \binits{P.R.}},
\bauthor{\bsnm{{Lepri}}, \binits{S.T.}}:
\byear{2014},
\batitle{{Temporal Evolution of Solar Wind Ion Composition and their Source
  Coronal Holes during the Declining Phase of Cycle 23. I. Low-latitude
  Extension of Polar Coronal Holes}}.
\bjtitle{\apj}
\bvolume{787},
\bfpage{121}.
\doiurl{https://doi.org/10.1088/0004-637X/787/2/121}.
\adsurl{2014ApJ...787..121K}.
\end{barticle}
\endbibitem

\bibitem[\protect\citeauthoryear{{Krista} and
  {Gallagher}}{2009}]{2009SoPh..256...87K}
\begin{barticle}
\bauthor{\bsnm{{Krista}}, \binits{L.D.}},
\bauthor{\bsnm{{Gallagher}}, \binits{P.T.}}:
\byear{2009},
\batitle{{Automated Coronal Hole Detection Using Local Intensity Thresholding
  Techniques}}.
\bjtitle{\solphys}
\bvolume{256},
\bfpage{87}.
\doiurl{https://doi.org/10.1007/s11207-009-9357-2}.
\adsurl{2009SoPh..256...87K}.
\end{barticle}
\endbibitem

\bibitem[\protect\citeauthoryear{Landeros et~al.}{2024}]{my_zenodo}
\begin{botherref}
\oauthor{\bsnm{Landeros}, \binits{J.}},
\oauthor{\bsnm{Kirk}, \binits{M.S.}},
\oauthor{\bsnm{Arge}, \binits{C.N.}},
\oauthor{\bsnm{Boucheron}, \binits{L.E.}},
\oauthor{\bsnm{Zhang}, \binits{J.}},
\oauthor{\bsnm{Uritsky}, \binits{V.M.}},
\oauthor{\bsnm{Grajeda}, \binits{J.A.}},
\oauthor{\bsnm{Dupertuis}, \binits{M.}}:
2024,
\textit{Magnetic Field-Constrained Ensemble Image Segmentation of Coronal Holes
  in Chromospheric Observations},
Zenodo.
\doiurl{https://doi.org/10.5281/zenodo.14402981}.
\url{https://doi.org/10.5281/zenodo.14402981}.
\end{botherref}
\endbibitem

\bibitem[\protect\citeauthoryear{{Linker} et~al.}{2017}]{2017ApJ...848...70L}
\begin{barticle}
\bauthor{\bsnm{{Linker}}, \binits{J.A.}},
\bauthor{\bsnm{{Caplan}}, \binits{R.M.}},
\bauthor{\bsnm{{Downs}}, \binits{C.}},
\bauthor{\bsnm{{Riley}}, \binits{P.}},
\bauthor{\bsnm{{Mikic}}, \binits{Z.}},
\bauthor{\bsnm{{Lionello}}, \binits{R.}},
\bauthor{\bsnm{{Henney}}, \binits{C.J.}},
\bauthor{\bsnm{{Arge}}, \binits{C.N.}},
\bauthor{\bsnm{{Liu}}, \binits{Y.}},
\bauthor{\bsnm{{Derosa}}, \binits{M.L.}},
\bauthor{\bsnm{{Yeates}}, \binits{A.}},
\bauthor{\bsnm{{Owens}}, \binits{M.J.}}:
\byear{2017},
\batitle{{The Open Flux Problem}}.
\bjtitle{\apj}
\bvolume{848},
\bfpage{70}.
\doiurl{https://doi.org/10.3847/1538-4357/aa8a70}.
\adsurl{2017ApJ...848...70L}.
\end{barticle}
\endbibitem

\bibitem[\protect\citeauthoryear{{Linker} et~al.}{2021}]{2021ApJ...918...21L}
\begin{barticle}
\bauthor{\bsnm{{Linker}}, \binits{J.A.}},
\bauthor{\bsnm{{Heinemann}}, \binits{S.G.}},
\bauthor{\bsnm{{Temmer}}, \binits{M.}},
\bauthor{\bsnm{{Owens}}, \binits{M.J.}},
\bauthor{\bsnm{{Caplan}}, \binits{R.M.}},
\bauthor{\bsnm{{Arge}}, \binits{C.N.}},
\bauthor{\bsnm{{Asvestari}}, \binits{E.}},
\bauthor{\bsnm{{Delouille}}, \binits{V.}},
\bauthor{\bsnm{{Downs}}, \binits{C.}},
\bauthor{\bsnm{{Hofmeister}}, \binits{S.J.}},
\bauthor{\bsnm{{Jebaraj}}, \binits{I.C.}},
\bauthor{\bsnm{{Madjarska}}, \binits{M.S.}},
\bauthor{\bsnm{{Pinto}}, \binits{R.F.}},
\bauthor{\bsnm{{Pomoell}}, \binits{J.}},
\bauthor{\bsnm{{Samara}}, \binits{E.}},
\bauthor{\bsnm{{Scolini}}, \binits{C.}},
\bauthor{\bsnm{{Vr{\v{s}}nak}}, \binits{B.}}:
\byear{2021},
\batitle{{Coronal Hole Detection and Open Magnetic Flux}}.
\bjtitle{\apj}
\bvolume{918},
\bfpage{21}.
\doiurl{https://doi.org/10.3847/1538-4357/ac090a}.
\adsurl{2021ApJ...918...21L}.
\end{barticle}
\endbibitem

\bibitem[\protect\citeauthoryear{{Lionello} et~al.}{2006}]{2006ApJ...642L..69L}
\begin{barticle}
\bauthor{\bsnm{{Lionello}}, \binits{R.}},
\bauthor{\bsnm{{Linker}}, \binits{J.A.}},
\bauthor{\bsnm{{Miki{\'c}}}, \binits{Z.}},
\bauthor{\bsnm{{Riley}}, \binits{P.}}:
\byear{2006},
\batitle{{The Latitudinal Excursion of Coronal Magnetic Field Lines in Response
  to Differential Rotation: MHD Simulations}}.
\bjtitle{\apjl}
\bvolume{642},
\bfpage{L69}.
\doiurl{https://doi.org/10.1086/504289}.
\adsurl{2006ApJ...642L..69L}.
\end{barticle}
\endbibitem

\bibitem[\protect\citeauthoryear{{Livingston}
  et~al.}{1976}]{1976ApOpt..15...33L}
\begin{barticle}
\bauthor{\bsnm{{Livingston}}, \binits{W.C.}},
\bauthor{\bsnm{{Harvey}}, \binits{J.}},
\bauthor{\bsnm{{Pierce}}, \binits{A.K.}},
\bauthor{\bsnm{{Schrage}}, \binits{D.}},
\bauthor{\bsnm{{Gillespie}}, \binits{B.}},
\bauthor{\bsnm{{Simmons}}, \binits{J.}},
\bauthor{\bsnm{{Slaughter}}, \binits{C.}}:
\byear{1976},
\batitle{{Kitt Peak 60-cm vacuum telescope}}.
\bjtitle{\ao}
\bvolume{15},
\bfpage{33}.
\doiurl{https://doi.org/10.1364/AO.15.000033}.
\adsurl{1976ApOpt..15...33L}.
\end{barticle}
\endbibitem

\bibitem[\protect\citeauthoryear{{Lowder}, {Qiu}, and
  {Leamon}}{2017}]{2017SoPh..292...18L}
\begin{barticle}
\bauthor{\bsnm{{Lowder}}, \binits{C.}},
\bauthor{\bsnm{{Qiu}}, \binits{J.}},
\bauthor{\bsnm{{Leamon}}, \binits{R.}}:
\byear{2017},
\batitle{{Coronal Holes and Open Magnetic Flux over Cycles 23 and 24}}.
\bjtitle{\solphys}
\bvolume{292},
\bfpage{18}.
\doiurl{https://doi.org/10.1007/s11207-016-1041-8}.
\adsurl{2017SoPh..292...18L}.
\end{barticle}
\endbibitem

\bibitem[\protect\citeauthoryear{{Lowder} et~al.}{2014}]{2014ApJ...783..142L}
\begin{barticle}
\bauthor{\bsnm{{Lowder}}, \binits{C.}},
\bauthor{\bsnm{{Qiu}}, \binits{J.}},
\bauthor{\bsnm{{Leamon}}, \binits{R.}},
\bauthor{\bsnm{{Liu}}, \binits{Y.}}:
\byear{2014},
\batitle{{Measurements of EUV Coronal Holes and Open Magnetic Flux}}.
\bjtitle{\apj}
\bvolume{783},
\bfpage{142}.
\doiurl{https://doi.org/10.1088/0004-637X/783/2/142}.
\adsurl{2014ApJ...783..142L}.
\end{barticle}
\endbibitem

\bibitem[\protect\citeauthoryear{{Mackay} and
  {Yeates}}{2012}]{2012LRSP....9....6M}
\begin{barticle}
\bauthor{\bsnm{{Mackay}}, \binits{D.H.}},
\bauthor{\bsnm{{Yeates}}, \binits{A.R.}}:
\byear{2012},
\batitle{{The Sun's Global Photospheric and Coronal Magnetic Fields:
  Observations and Models}}.
\bjtitle{Living Reviews in Solar Physics}
\bvolume{9},
\bfpage{6}.
\doiurl{https://doi.org/10.12942/lrsp-2012-6}.
\adsurl{2012LRSP....9....6M}.
\end{barticle}
\endbibitem

\bibitem[\protect\citeauthoryear{{Mason} and
  {Uritsky}}{2022}]{2022ApJ...937L..19M}
\begin{barticle}
\bauthor{\bsnm{{Mason}}, \binits{E.I.}},
\bauthor{\bsnm{{Uritsky}}, \binits{V.M.}}:
\byear{2022},
\batitle{{Statistical Evidence for Small-scale Interchange Reconnection at a
  Coronal Hole Boundary}}.
\bjtitle{\apjl}
\bvolume{937},
\bfpage{L19}.
\doiurl{https://doi.org/10.3847/2041-8213/ac9124}.
\adsurl{2022ApJ...937L..19M}.
\end{barticle}
\endbibitem

\bibitem[\protect\citeauthoryear{{Mogilevsky}, {Obridko}, and
  {Shilova}}{1997}]{1997SoPh..176..107M}
\begin{barticle}
\bauthor{\bsnm{{Mogilevsky}}, \binits{E.I.}},
\bauthor{\bsnm{{Obridko}}, \binits{V.N.}},
\bauthor{\bsnm{{Shilova}}, \binits{N.S.}}:
\byear{1997},
\batitle{{Large-Scale Magnetic Field Structures and Coronal Holes on the Sun}}.
\bjtitle{\solphys}
\bvolume{176},
\bfpage{107}.
\doiurl{https://doi.org/10.1023/A:1004908014970}.
\adsurl{1997SoPh..176..107M}.
\end{barticle}
\endbibitem

\bibitem[\protect\citeauthoryear{Nayar}{2022}]{CAVE_0396}
\begin{bchapter}
\bauthor{\bsnm{Nayar}, \binits{S.K.}}:
\byear{2022},
\bctitle{{E}dge {D}etection}.
In: \bbtitle{Monograph FPCV-2-1, First Principles of Computer Vision},
\bconflocation{Columbia University, New York}.
\end{bchapter}
\endbibitem

\bibitem[\protect\citeauthoryear{NSO}{2016}]{2016NSO}
\begin{botherref}
\oauthor{\bsnm{NSO}}:
2016,
\textit{SOLIS VSM 10830 Coronal Holes image maps}.
\url{https://nispdata.nso.edu/webProdDesc2/presenter.php?file=solis_vsm_10830_Coronal_Holes_image_maps.html&echoExact=0&name=SOLIS\%20VSM\%2010830\%20Coronal\%20Holes\%20image\%20maps}.
\end{botherref}
\endbibitem

\bibitem[\protect\citeauthoryear{Park}{1997}]{ParkDECOMPOSITIONOS}
\begin{bchapter}
\bauthor{\bsnm{Park}, \binits{H.}}:
\byear{1997},
\bctitle{DECOMPOSITION OF STRUCTURING ELEMENTS FOR OPTIMAL IMPLEMENTATION OF
  MORPHOLOGICAL OPERATIONS}.
In: \bbtitle{???}
\burl{https://api.semanticscholar.org/CorpusID:17506327}.
\end{bchapter}
\endbibitem

\bibitem[\protect\citeauthoryear{{Pevtsov} et~al.}{2023}]{2023BAAS...55c.320P}
\begin{bchapter}
\bauthor{\bsnm{{Pevtsov}}, \binits{A.}},
\bauthor{\bsnm{{Martinez-Pillet}}, \binits{V.}},
\bauthor{\bsnm{{Gilbert}}, \binits{H.}},
\bauthor{\bsnm{{de Wijn}}, \binits{A.}},
\bauthor{\bsnm{{Roth}}, \binits{M.}},
\bauthor{\bsnm{{Gosain}}, \binits{S.}},
\bauthor{\bsnm{{Upton}}, \binits{L.}},
\bauthor{\bsnm{{Katsukawa}}, \binits{Y.}},
\bauthor{\bsnm{{Burkepile}}, \binits{J.}},
\bauthor{\bsnm{{Zhang}}, \binits{J.}},
\bauthor{\bsnm{{Reardon}}, \binits{K.}},
\bauthor{\bsnm{{Bertello}}, \binits{L.}},
\bauthor{\bsnm{{Jain}}, \binits{K.}},
\bauthor{\bsnm{{Tripathy}}, \binits{S.}},
\bauthor{\bsnm{{Leka}}, \binits{K.D.}}:
\byear{2023},
\bctitle{{ngGONG {\textemdash} Future Ground-based Facilities for Research in
  Heliophysics and Space Weather Operational Forecast}}.
In: \bbtitle{Bulletin of the American Astronomical Society}
\bseriesno{55},
\bfpage{320}.
\doiurl{https://doi.org/10.3847/25c2cfeb.612b0ef9}.
\adsurl{2023BAAS...55c.320P}.
\end{bchapter}
\endbibitem

\bibitem[\protect\citeauthoryear{{Pozhalova}}{1987}]{1987SvAL...13..255P}
\begin{barticle}
\bauthor{\bsnm{{Pozhalova}}, \binits{Z.A.}}:
\byear{1987},
\batitle{{The Helium 10830A Line as an Indicator of Coronal Holes - Theory}}.
\bjtitle{Soviet Astronomy Letters}
\bvolume{13},
\bfpage{255}.
\adsurl{1987SvAL...13..255P}.
\end{barticle}
\endbibitem

\bibitem[\protect\citeauthoryear{{Reiss} et~al.}{2015}]{2015JSWSC...5A..23R}
\begin{barticle}
\bauthor{\bsnm{{Reiss}}, \binits{M.A.}},
\bauthor{\bsnm{{Hofmeister}}, \binits{S.J.}},
\bauthor{\bsnm{{De Visscher}}, \binits{R.}},
\bauthor{\bsnm{{Temmer}}, \binits{M.}},
\bauthor{\bsnm{{Veronig}}, \binits{A.M.}},
\bauthor{\bsnm{{Delouille}}, \binits{V.}},
\bauthor{\bsnm{{Mampaey}}, \binits{B.}},
\bauthor{\bsnm{{Ahammer}}, \binits{H.}}:
\byear{2015},
\batitle{{Improvements on coronal hole detection in SDO/AIA images using
  supervised classification}}.
\bjtitle{Journal of Space Weather and Space Climate}
\bvolume{5},
\bfpage{A23}.
\doiurl{https://doi.org/10.1051/swsc/2015025}.
\adsurl{2015JSWSC...5A..23R}.
\end{barticle}
\endbibitem

\bibitem[\protect\citeauthoryear{{Reiss} et~al.}{2020}]{2020ApJ...891..165R}
\begin{barticle}
\bauthor{\bsnm{{Reiss}}, \binits{M.A.}},
\bauthor{\bsnm{{MacNeice}}, \binits{P.J.}},
\bauthor{\bsnm{{Muglach}}, \binits{K.}},
\bauthor{\bsnm{{Arge}}, \binits{C.N.}},
\bauthor{\bsnm{{M{\"o}stl}}, \binits{C.}},
\bauthor{\bsnm{{Riley}}, \binits{P.}},
\bauthor{\bsnm{{Hinterreiter}}, \binits{J.}},
\bauthor{\bsnm{{Bailey}}, \binits{R.L.}},
\bauthor{\bsnm{{Weiss}}, \binits{A.J.}},
\bauthor{\bsnm{{Owens}}, \binits{M.J.}},
\bauthor{\bsnm{{Amerstorfer}}, \binits{T.}},
\bauthor{\bsnm{{Amerstorfer}}, \binits{U.}}:
\byear{2020},
\batitle{{Forecasting the Ambient Solar Wind with Numerical Models. II. An
  Adaptive Prediction System for Specifying Solar Wind Speed near the Sun}}.
\bjtitle{\apj}
\bvolume{891},
\bfpage{165}.
\doiurl{https://doi.org/10.3847/1538-4357/ab78a0}.
\adsurl{2020ApJ...891..165R}.
\end{barticle}
\endbibitem

\bibitem[\protect\citeauthoryear{{Reiss} et~al.}{2021}]{2021ApJ...913...28R}
\begin{barticle}
\bauthor{\bsnm{{Reiss}}, \binits{M.A.}},
\bauthor{\bsnm{{Muglach}}, \binits{K.}},
\bauthor{\bsnm{{M{\"o}stl}}, \binits{C.}},
\bauthor{\bsnm{{Arge}}, \binits{C.N.}},
\bauthor{\bsnm{{Bailey}}, \binits{R.}},
\bauthor{\bsnm{{Delouille}}, \binits{V.}},
\bauthor{\bsnm{{Garton}}, \binits{T.M.}},
\bauthor{\bsnm{{Hamada}}, \binits{A.}},
\bauthor{\bsnm{{Hofmeister}}, \binits{S.}},
\bauthor{\bsnm{{Illarionov}}, \binits{E.}},
\bauthor{\bsnm{{Jarolim}}, \binits{R.}},
\bauthor{\bsnm{{Kirk}}, \binits{M.S.F.}},
\bauthor{\bsnm{{Kosovichev}}, \binits{A.}},
\bauthor{\bsnm{{Krista}}, \binits{L.}},
\bauthor{\bsnm{{Lee}}, \binits{S.}},
\bauthor{\bsnm{{Lowder}}, \binits{C.}},
\bauthor{\bsnm{{MacNeice}}, \binits{P.J.}},
\bauthor{\bsnm{{Veronig}}, \binits{A.}},
\bauthor{\bsnm{{Cospar Iswat Coronal Hole Boundary Working Team}}}:
\byear{2021},
\batitle{{The Observational Uncertainty of Coronal Hole Boundaries in Automated
  Detection Schemes}}.
\bjtitle{\apj}
\bvolume{913},
\bfpage{28}.
\doiurl{https://doi.org/10.3847/1538-4357/abf2c8}.
\adsurl{2021ApJ...913...28R}.
\end{barticle}
\endbibitem

\bibitem[\protect\citeauthoryear{{Reiss} et~al.}{2024}]{2024ApJS..271....6R}
\begin{barticle}
\bauthor{\bsnm{{Reiss}}, \binits{M.A.}},
\bauthor{\bsnm{{Muglach}}, \binits{K.}},
\bauthor{\bsnm{{Mason}}, \binits{E.}},
\bauthor{\bsnm{{Davies}}, \binits{E.E.}},
\bauthor{\bsnm{{Chakraborty}}, \binits{S.}},
\bauthor{\bsnm{{Delouille}}, \binits{V.}},
\bauthor{\bsnm{{Downs}}, \binits{C.}},
\bauthor{\bsnm{{Garton}}, \binits{T.M.}},
\bauthor{\bsnm{{Grajeda}}, \binits{J.A.}},
\bauthor{\bsnm{{Hamada}}, \binits{A.}},
\bauthor{\bsnm{{Heinemann}}, \binits{S.G.}},
\bauthor{\bsnm{{Hofmeister}}, \binits{S.}},
\bauthor{\bsnm{{Illarionov}}, \binits{E.}},
\bauthor{\bsnm{{Jarolim}}, \binits{R.}},
\bauthor{\bsnm{{Krista}}, \binits{L.}},
\bauthor{\bsnm{{Lowder}}, \binits{C.}},
\bauthor{\bsnm{{Verwichte}}, \binits{E.}},
\bauthor{\bsnm{{Arge}}, \binits{C.N.}},
\bauthor{\bsnm{{Boucheron}}, \binits{L.E.}},
\bauthor{\bsnm{{Foullon}}, \binits{C.}},
\bauthor{\bsnm{{Kirk}}, \binits{M.S.}},
\bauthor{\bsnm{{Kosovichev}}, \binits{A.}},
\bauthor{\bsnm{{Leisner}}, \binits{A.}},
\bauthor{\bsnm{{M{\"o}stl}}, \binits{C.}},
\bauthor{\bsnm{{Turtle}}, \binits{J.}},
\bauthor{\bsnm{{Veronig}}, \binits{A.}}:
\byear{2024},
\batitle{{A Community Data Set for Comparing Automated Coronal Hole Detection
  Schemes}}.
\bjtitle{\apjs}
\bvolume{271},
\bfpage{6}.
\doiurl{https://doi.org/10.3847/1538-4365/ad1408}.
\adsurl{2024ApJS..271....6R}.
\end{barticle}
\endbibitem

\bibitem[\protect\citeauthoryear{{Rotter} et~al.}{2012}]{2012SoPh..281..793R}
\begin{barticle}
\bauthor{\bsnm{{Rotter}}, \binits{T.}},
\bauthor{\bsnm{{Veronig}}, \binits{A.M.}},
\bauthor{\bsnm{{Temmer}}, \binits{M.}},
\bauthor{\bsnm{{Vr{\v{s}}nak}}, \binits{B.}}:
\byear{2012},
\batitle{{Relation Between Coronal Hole Areas on the Sun and the Solar Wind
  Parameters at 1 AU}}.
\bjtitle{\solphys}
\bvolume{281},
\bfpage{793}.
\doiurl{https://doi.org/10.1007/s11207-012-0101-y}.
\adsurl{2012SoPh..281..793R}.
\end{barticle}
\endbibitem

\bibitem[\protect\citeauthoryear{Rutten}{2003}]{2003Rutten}
\begin{botherref}
\oauthor{\bsnm{Rutten}, \binits{R.J.}}:
2003,
\textit{Radiative Transfer in Stellar Atmospheres},
Utrecht University lecture notes.
\url{https://robrutten.nl/rrweb/rjr-pubs/2003rtsa.book.....R.pdf}.
\end{botherref}
\endbibitem

\bibitem[\protect\citeauthoryear{Samara et~al.}{2024}]{2024Samara}
\begin{barticle}
\bauthor{\bsnm{Samara}, \binits{E.}},
\bauthor{\bsnm{Arge}, \binits{C.N.}},
\bauthor{\bsnm{Pinto}, \binits{R.F.}},
\bauthor{\bsnm{Magdalenić}, \binits{J.}},
\bauthor{\bsnm{Wijsen}, \binits{N.}},
\bauthor{\bsnm{Stevens}, \binits{M.L.}},
\bauthor{\bsnm{Rodriguez}, \binits{L.}},
\bauthor{\bsnm{Poedts}, \binits{S.}}:
\byear{2024},
\batitle{Calibrating the WSA Model in EUHFORIA Based on Parker Solar Probe
  Observations}.
\bjtitle{The Astrophysical Journal}
\bvolume{971},
\bfpage{83}.
\doiurl{https://doi.org/10.3847/1538-4357/ad53c6}.
\burl{https://dx.doi.org/10.3847/1538-4357/ad53c6}.
\end{barticle}
\endbibitem

\bibitem[\protect\citeauthoryear{{Scholl} and
  {Habbal}}{2008}]{2008SoPh..248..425S}
\begin{barticle}
\bauthor{\bsnm{{Scholl}}, \binits{I.F.}},
\bauthor{\bsnm{{Habbal}}, \binits{S.R.}}:
\byear{2008},
\batitle{{Automatic Detection and Classification of Coronal Holes and Filaments
  Based on EUV and Magnetogram Observations of the Solar Disk}}.
\bjtitle{\solphys}
\bvolume{248},
\bfpage{425}.
\doiurl{https://doi.org/10.1007/s11207-007-9075-6}.
\adsurl{2008SoPh..248..425S}.
\end{barticle}
\endbibitem

\bibitem[\protect\citeauthoryear{{The SunPy Community}
  et~al.}{2020}]{sunpy_community2020}
\begin{barticle}
\bauthor{\bsnm{{The SunPy Community}}},
\bauthor{\bsnm{Barnes}, \binits{W.T.}},
\bauthor{\bsnm{Bobra}, \binits{M.G.}},
\bauthor{\bsnm{Christe}, \binits{S.D.}},
\bauthor{\bsnm{Freij}, \binits{N.}},
\bauthor{\bsnm{Hayes}, \binits{L.A.}},
\bauthor{\bsnm{Ireland}, \binits{J.}},
\bauthor{\bsnm{Mumford}, \binits{S.}},
\bauthor{\bsnm{Perez-Suarez}, \binits{D.}},
\bauthor{\bsnm{Ryan}, \binits{D.F.}},
\bauthor{\bsnm{Shih}, \binits{A.Y.}},
\bauthor{\bsnm{Chanda}, \binits{P.}},
\bauthor{\bsnm{Glogowski}, \binits{K.}},
\bauthor{\bsnm{Hewett}, \binits{R.}},
\bauthor{\bsnm{Hughitt}, \binits{V.K.}},
\bauthor{\bsnm{Hill}, \binits{A.}},
\bauthor{\bsnm{Hiware}, \binits{K.}},
\bauthor{\bsnm{Inglis}, \binits{A.}},
\bauthor{\bsnm{Kirk}, \binits{M.S.F.}},
\bauthor{\bsnm{Konge}, \binits{S.}},
\bauthor{\bsnm{Mason}, \binits{J.P.}},
\bauthor{\bsnm{Maloney}, \binits{S.A.}},
\bauthor{\bsnm{Murray}, \binits{S.A.}},
\bauthor{\bsnm{Panda}, \binits{A.}},
\bauthor{\bsnm{Park}, \binits{J.}},
\bauthor{\bsnm{Pereira}, \binits{T.M.D.}},
\bauthor{\bsnm{Reardon}, \binits{K.}},
\bauthor{\bsnm{Savage}, \binits{S.}},
\bauthor{\bsnm{Sipőcz}, \binits{B.M.}},
\bauthor{\bsnm{Stansby}, \binits{D.}},
\bauthor{\bsnm{Jain}, \binits{Y.}},
\bauthor{\bsnm{Taylor}, \binits{G.}},
\bauthor{\bsnm{Yadav}, \binits{T.}},
\bauthor{\bsnm{Rajul}},
\bauthor{\bsnm{Dang}, \binits{T.K.}}:
\byear{2020},
\batitle{The SunPy Project: Open Source Development and Status of the Version
  1.0 Core Package}.
\bjtitle{The Astrophysical Journal}
\bvolume{890}.
\doiurl{https://doi.org/10.3847/1538-4357/ab4f7a}.
\burl{https://iopscience.iop.org/article/10.3847/1538-4357/ab4f7a}.
\end{barticle}
\endbibitem

\bibitem[\protect\citeauthoryear{{Tlatov}, {Tavastsherna}, and
  {Vasil'eva}}{2014}]{2014SoPh..289.1349T}
\begin{barticle}
\bauthor{\bsnm{{Tlatov}}, \binits{A.}},
\bauthor{\bsnm{{Tavastsherna}}, \binits{K.}},
\bauthor{\bsnm{{Vasil'eva}}, \binits{V.}}:
\byear{2014},
\batitle{{Coronal Holes in Solar Cycles 21 to 23}}.
\bjtitle{\solphys}
\bvolume{289},
\bfpage{1349}.
\doiurl{https://doi.org/10.1007/s11207-013-0387-4}.
\adsurl{2014SoPh..289.1349T}.
\end{barticle}
\endbibitem

\bibitem[\protect\citeauthoryear{van~der Walt et~al.}{2014}]{scikit-image}
\begin{barticle}
\bauthor{\bparticle{van~der} \bsnm{Walt}, \binits{S.}},
\bauthor{\bsnm{{S}ch\"onberger}, \binits{J.L.}},
\bauthor{\bsnm{{Nunez-Iglesias}}, \binits{J.}},
\bauthor{\bsnm{{B}oulogne}, \binits{F.}},
\bauthor{\bsnm{{W}arner}, \binits{J.D.}},
\bauthor{\bsnm{{Y}ager}, \binits{N.}},
\bauthor{\bsnm{{G}ouillart}, \binits{E.}},
\bauthor{\bsnm{{Y}u}, \binits{T.}},
\bauthor{\bparticle{the~scikit{-}image} \bsnm{contributors}}:
\byear{2014},
\batitle{scikit-image: image processing in {P}ython}.
\bjtitle{PeerJ}
\bvolume{2},
\bfpage{e453}.
\doiurl{https://doi.org/10.7717/peerj.453}.
\burl{https://doi.org/10.7717/peerj.453}.
\end{barticle}
\endbibitem

\bibitem[\protect\citeauthoryear{{Verbeeck} et~al.}{2014}]{2014A&A...561A..29V}
\begin{barticle}
\bauthor{\bsnm{{Verbeeck}}, \binits{C.}},
\bauthor{\bsnm{{Delouille}}, \binits{V.}},
\bauthor{\bsnm{{Mampaey}}, \binits{B.}},
\bauthor{\bsnm{{De Visscher}}, \binits{R.}}:
\byear{2014},
\batitle{{The SPoCA-suite: Software for extraction, characterization, and
  tracking of active regions and coronal holes on EUV images}}.
\bjtitle{\aap}
\bvolume{561},
\bfpage{A29}.
\doiurl{https://doi.org/10.1051/0004-6361/201321243}.
\adsurl{2014A&A...561A..29V}.
\end{barticle}
\endbibitem

\bibitem[\protect\citeauthoryear{{Wallace} et~al.}{2019}]{2019SoPh..294...19W}
\begin{barticle}
\bauthor{\bsnm{{Wallace}}, \binits{S.}},
\bauthor{\bsnm{{Arge}}, \binits{C.N.}},
\bauthor{\bsnm{{Pattichis}}, \binits{M.}},
\bauthor{\bsnm{{Hock-Mysliwiec}}, \binits{R.A.}},
\bauthor{\bsnm{{Henney}}, \binits{C.J.}}:
\byear{2019},
\batitle{{Estimating Total Open Heliospheric Magnetic Flux}}.
\bjtitle{\solphys}
\bvolume{294},
\bfpage{19}.
\doiurl{https://doi.org/10.1007/s11207-019-1402-1}.
\adsurl{2019SoPh..294...19W}.
\end{barticle}
\endbibitem

\bibitem[\protect\citeauthoryear{{Wuelser} et~al.}{2004}]{2004SPIE.5171..111W}
\begin{bchapter}
\bauthor{\bsnm{{Wuelser}}, \binits{J.-P.}},
\bauthor{\bsnm{{Lemen}}, \binits{J.R.}},
\bauthor{\bsnm{{Tarbell}}, \binits{T.D.}},
\bauthor{\bsnm{{Wolfson}}, \binits{C.J.}},
\bauthor{\bsnm{{Cannon}}, \binits{J.C.}},
\bauthor{\bsnm{{Carpenter}}, \binits{B.A.}},
\bauthor{\bsnm{{Duncan}}, \binits{D.W.}},
\bauthor{\bsnm{{Gradwohl}}, \binits{G.S.}},
\bauthor{\bsnm{{Meyer}}, \binits{S.B.}},
\bauthor{\bsnm{{Moore}}, \binits{A.S.}},
\bauthor{\bsnm{{Navarro}}, \binits{R.L.}},
\bauthor{\bsnm{{Pearson}}, \binits{J.D.}},
\bauthor{\bsnm{{Rossi}}, \binits{G.R.}},
\bauthor{\bsnm{{Springer}}, \binits{L.A.}},
\bauthor{\bsnm{{Howard}}, \binits{R.A.}},
\bauthor{\bsnm{{Moses}}, \binits{J.D.}},
\bauthor{\bsnm{{Newmark}}, \binits{J.S.}},
\bauthor{\bsnm{{Delaboudiniere}}, \binits{J.-P.}},
\bauthor{\bsnm{{Artzner}}, \binits{G.E.}},
\bauthor{\bsnm{{Auchere}}, \binits{F.}},
\bauthor{\bsnm{{Bougnet}}, \binits{M.}},
\bauthor{\bsnm{{Bouyries}}, \binits{P.}},
\bauthor{\bsnm{{Bridou}}, \binits{F.}},
\bauthor{\bsnm{{Clotaire}}, \binits{J.-Y.}},
\bauthor{\bsnm{{Colas}}, \binits{G.}},
\bauthor{\bsnm{{Delmotte}}, \binits{F.}},
\bauthor{\bsnm{{Jerome}}, \binits{A.}},
\bauthor{\bsnm{{Lamare}}, \binits{M.}},
\bauthor{\bsnm{{Mercier}}, \binits{R.}},
\bauthor{\bsnm{{Mullot}}, \binits{M.}},
\bauthor{\bsnm{{Ravet}}, \binits{M.-F.}},
\bauthor{\bsnm{{Song}}, \binits{X.}},
\bauthor{\bsnm{{Bothmer}}, \binits{V.}},
\bauthor{\bsnm{{Deutsch}}, \binits{W.}}:
\byear{2004},
\bctitle{{EUVI: the STEREO-SECCHI extreme ultraviolet imager}}.
In: \beditor{\bsnm{{Fineschi}}, \binits{S.}},
\beditor{\bsnm{{Gummin}}, \binits{M.A.}} (eds.)
\bbtitle{Telescopes and Instrumentation for Solar Astrophysics},
\bsertitle{Society of Photo-Optical Instrumentation Engineers (SPIE) Conference
  Series}
\bseriesno{5171},
\bfpage{111}.
\doiurl{https://doi.org/10.1117/12.506877}.
\adsurl{2004SPIE.5171..111W}.
\end{bchapter}
\endbibitem

\bibitem[\protect\citeauthoryear{Zeng, Zhang, and Genderen}{2006}]{2006Zeng}
\begin{bchapter}
\bauthor{\bsnm{Zeng}, \binits{Y.}},
\bauthor{\bsnm{Zhang}, \binits{J.}},
\bauthor{\bsnm{Genderen}, \binits{J.L.V.}}:
\byear{2006},
\bctitle{Comparison and analysis of remote sensing data fusion techniques at
  feature and decision levels}.
In: \bbtitle{International Society for Photogrammetry and Remote Sensing}.
\burl{https://api.semanticscholar.org/CorpusID:16968639}.
\end{bchapter}
\endbibitem

\bibitem[\protect\citeauthoryear{{Zhang}}{2010}]{2010IJIDF...1....5Z}
\begin{barticle}
\bauthor{\bsnm{{Zhang}}, \binits{J.}}:
\byear{2010},
\batitle{{Multi-source remote sensing data fusion: status and trends}}.
\bjtitle{International Journal of Image and Data Fusion}
\bvolume{1},
\bfpage{5}.
\doiurl{https://doi.org/10.1080/19479830903561035}.
\adsurl{2010IJIDF...1....5Z}.
\end{barticle}
\endbibitem

\bibitem[\protect\citeauthoryear{Zhou}{2012}]{2012Zhou}
\begin{botherref}
\oauthor{\bsnm{Zhou}, \binits{Z.-H.}}:
2012,
\textit{Ensemble Methods: Foundations and Algorithms},
Chapman and Hall/CRC.
\doiurl{https://doi.org/10.1201/b12207}.
\end{botherref}
\endbibitem

\bibitem[\protect\citeauthoryear{{Zirin}}{1975}]{1975ApJ...199L..63Z}
\begin{barticle}
\bauthor{\bsnm{{Zirin}}, \binits{H.}}:
\byear{1975},
\batitle{{The helium chromosphere, coronal holes, and stellar X-rays.}}
\bjtitle{\apjl}
\bvolume{199},
\bfpage{L63}.
\doiurl{https://doi.org/10.1086/181849}.
\adsurl{1975ApJ...199L..63Z}.
\end{barticle}
\endbibitem

\end{thebibliography}

     % Checking: look if the file containing the ``\bibitem'' exits
     %           so check if the .bbl file exist (bibTeX compilation)
\IfFileExists{\jobname.bbl}{} {\typeout{}
\typeout{****************************************************}
\typeout{****************************************************}
\typeout{** Please run "bibtex \jobname" to obtain} \typeout{**
the bibliography and then re-run LaTeX} \typeout{** twice to fix
the references !}
\typeout{****************************************************}
\typeout{****************************************************}
\typeout{}}

\end{document}